\documentclass[11pt]{article}
\usepackage{graphicx}
\usepackage[usenames,dvipsnames]{color}
\usepackage{url}
\usepackage[colorlinks = true,
            linkcolor = blue,
            urlcolor  = blue,
            citecolor = blue,
            anchorcolor = blue]{hyperref}

\usepackage{lineno,hyperref,color}
\modulolinenumbers[5]
\usepackage{todonotes}
\usepackage{ulem}
\usepackage{booktabs}
\usepackage{float}
\usepackage{wrapfig}

\usepackage{appendix}
\usepackage{pdfpages}
\usepackage{hyperref}
\usepackage{geometry}                
\usepackage[parfill]{parskip}    
\usepackage{graphicx, subfigure}
\usepackage{amssymb}
\usepackage{amsmath}
\usepackage{epstopdf}
\usepackage{array}
\usepackage{lineno}
\usepackage{scrextend}

\def\Title#1{\begin{center} {\LARGE #1 } \end{center}}
\def\Author#1{\begin{center}{ \sc #1} \end{center}}

\newenvironment{Abstract}{\begin{quotation} \begin{center}
                       ABSTRACT
     \end{center}\bigskip  }{\end{quotation}}

\def\beq{\begin{equation}}
\def\eeq#1{\label{#1}\end{equation}}
\def\eeqn{\end{equation}}

\newenvironment{Eqnarray}%
   {\arraycolsep 0.14em\begin{eqnarray}}{\end{eqnarray}}
\def\beqa{\begin{Eqnarray}}
\def\eeqa#1{\label{#1}\end{Eqnarray}}
\def\eeqan{\end{Eqnarray}}

\let\bar=\overbar

\def\lsim{\mathrel{\raise.3ex\hbox{$<$\kern-.75em\lower1ex\hbox{$\sim$}}}}
\def\gsim{\mathrel{\raise.3ex\hbox{$>$\kern-.75em\lower1ex\hbox{$\sim$}}}}

\def\del{\partial}
\def\Dslash{\not{\hbox{\kern-4pt $D$}}}
\def\dslash{\not{\hbox{\kern-2pt $\del$}}}
\def\pslash{\not{\hbox{\kern-2pt $p$}}}
\def\ETmiss{\not{\hbox{\kern-4pt $E$}}_T}

\def\Dlr{\mathrel{\raise1.5ex\hbox{$\leftrightarrow$\kern-1em\lower1.5ex\hbox{$D$}}}}

\def\MSB{{\bar{M \kern -2pt S}}}
\def\msb{{\bar{\scriptsize M \kern -1pt S}}}

\def\drb{{\bar{\scriptsize D \kern -1pt R}}}

\def\GeV{{\rm GeV}}

\input{defs}

\newcommand\snowmass{\begin{center}\rule[-0.2in]{\hsize}{0.01in}\\\rule{\hsize}{0.01in}\\
\vskip 0.1in Submitted to the  Proceedings of the US Community Study\\ 
on the Future of Particle Physics (Snowmass 2021)\\ 
\rule{\hsize}{0.01in}\\\rule[+0.2in]{\hsize}{0.01in} \end{center}}

\begin{document}


\Title{Dual-Readout Calorimetry for Future Experiments Probing Fundamental Physics}

\bigskip 
\Author{CERN, Switzerland}{L.~Pezzotti}
\Author{Caltech, CA, USA}{Harvey~Newman}
\Author{Fermi National Accelerator Laboratory, IL, USA}{J.~Freeman, J.~Hirschauer}
\Author{INFN Pavia, Italy}{R.~Ferrari, G.~Gaudio, G.~Polesello}
\Author{INFN Milano \& University of Insubria, Como, Italy}{R.~Santoro}
\Author{INFN \& University of Milano-Bicocca, Italy}{M.~Lucchini}
\Author{INFN \& University of Roma La Sapienza, Italy}{S.~Giagu}
\Author{Istituto Nazionale di Fisica Nucleare, Sezione di Pisa: Pisa, Italy}{F.~Bedeschi}
\Author{Kyungpook National University, S. Korea}{Sehwook Lee }
\Author{Massachusetts Institute of Technology, Cambridge, MA, USA}{P.~Harris}
\Author{ Princeton University, NJ, USA}{C.~Tully}
\Author{Purdue University, IN, USA}{A.~Jung}
\Author{Texas Tech University, TX, USA}{Nural Akchurin }
\Author{U. Maryland, MD, USA}{A.~Belloni, S.~Eno}
\Author{University of Michigan, MI, USA}{J.~Qian, B.~Zhou, J.~Zhu}
\Author{University of Seoul, S. Korea}{Jason Sang Hun Lee}
\Author{University of Sussex, UK}{I.~Vivarelli}
\Author{University of Virginia, USA}{R.~Hirosky}
\Author{Yonsei University, S. Korea}{Hwidong Yoo}

\medskip

\medskip

 \begin{Abstract}
\noindent In this White Paper for the 2021 Snowmass process, we detail the status and prospects for dual-readout calorimetry.
While all calorimeters allow estimation of energy depositions in their active material, dual-readout calorimeters aim to provide additional information on the light produced in the sensitive media via, for example, wavelength and polarization, and/or a precision timing measurements, allowing an estimation of the shower-by-shower particle content. Utilizing this knowledge of the shower particle content may allow unprecedented energy resolution for hadronic particles and jets, and new types of particle flow algorithms.
We also discuss the impact continued development of this kind of calorimetry could have on precision on Higgs boson property measurements at future colliders.
\end{Abstract}

\snowmass

\def\thefootnote{\fnsymbol{footnote}}
\setcounter{footnote}{0}
%


\section{Introduction}
Calorimeters are essential elements in most experiments designed to study fundamental physics.
Advances in materials, photodetectors, electronics, and algorithmic tools such as machine learning allow the improvement of current calorimeter designs and the development of new types of calorimeters that will make future measurements more precise and cost effective. Additional information on the future of calorimetery can be found in Ref.~\cite{Aleksa:2021ztd}. 

The United States Department of Energy (DOE) and the National Science Foundation (NSF)
in a ``Basic Research Needs" (BRN) report~\cite{brn}
 have recently defined ambitious goals for the next generation of calorimetry,
 driven by our fundamental physics objectives.
 In this white paper, we will discuss how calorimeter improvements, via the development of an innovative maximal use of information to measure the substructure of showers, can improve particle resolutions and identification and enhance our physics capabilities.
  The {\bf ``dual-readout"} calorimeter, pioneered
 by the RD52/DREAM/IDEA collaborations, is the progenitor of this 
 calorimetry thrust.
 
 This White Paper is organized as follows. In Section~\ref{sec:physics} we review the physics needs identified in the BRN and their calorimetry requirements.  We argue that
 improvements beyond the current requirements will improve the precision of the measurements, allowing  deeper probes of fundamental physics. 
We also discuss the complementarity of the leading thrusts in precision calorimetry:
high granularity and dual readout.
 In Section~\ref{sec:dualreview} we review the concept of dual readout and discuss briefly possible extensions.  In Section~\ref{sec:currentstatus} we describe the current status of dual-readout calorimeters.  In Section~\ref{sec:future} we describe a program of US R\&D, in conjunction with our international partners, needed to provide detectors that maximize the precision of future physics measurements.

\section{Calorimetry needs for future fundamental physics studies}
\label{sec:physics}
\begin{figure}[hbtp]
\centering
\resizebox{0.90\textwidth}{!}{\includegraphics{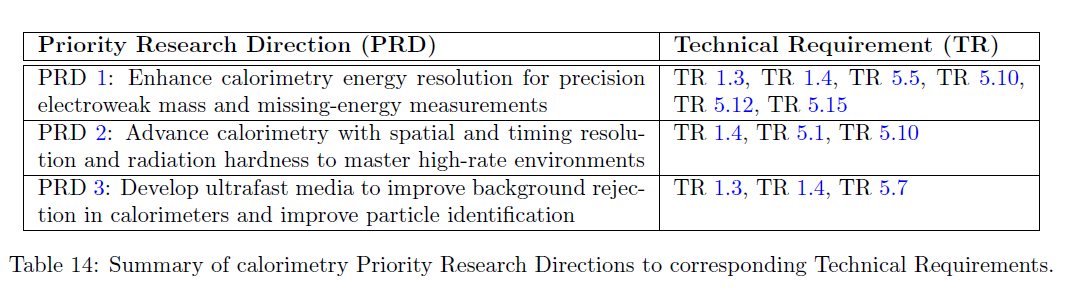}}
\caption{ 
Table 14 from Ref.~\cite{brn} giving the US priorities in calorimetry. 
} \label{fig:table14}
\end{figure}

The calorimetry requirements given in Table 14 of the DOE/NSF BRN~\cite{brn} are 
 reproduced in Fig.~\ref{fig:table14}.
 Similar sets of requirements, designed specifically for future circular electron-positron colliders, are discussed in Refs.~\cite{Aleksa:2021ztd, Azzi:2021ylt}.
 The table links the calorimeter needs to physics goals via technical requirements (TRs).
 TRs 1.1-1.5 relate to measuring Higgs boson properties with sub-percent precision and self-coupling with 5\% precision, its connections to dark matter, and new particles and phenomena at the multi-TeV scale.
 TR 5.5 and 5.10 are related to flavor physics (searches for new physics through rare flavor interactions, tests of the CKM quark mixing matrix description, studies of lepton flavor universality).  
 TR 5.12 is related to charged lepton flavor violation studies.  The Priority Research Directions (PRDs) 2 and 3 refer primarily to operation in hadron colliders, while PRD 1 is much more focused on electron-positron colliders. The latter are more likely to be available for experiments in the medium term future; this paper will then mainly discuss requirements for future accelerator of this type. 
 
	The associated production of a Z and a Higgs boson in $e^+e^-$ collisions at 240 GeV is one of the key processes to be studied in detail at future colliders. Over 90\% of the possible final states contain hadronic jets. One third has two jets, while the rest have four or six.
	It is therefore straightforward that good jet resolution is a very important feature of any detector to be deployed at these accelerators. In particular this generates strong requirements on the performance of the calorimeters. Jet resolution is relevant in many specific analyses, for instance: 1) Higgs recoil mass from Z decaying to two jets, to be distinguished from hadronic decays of pair-produced $W$'s or $Z$'s; 2) classification of specific Higgs decay final states, in particular separation of $H\rightarrow W^+W^-$ from $H\rightarrow Z\,Z$ when $W$'s or $Z$'s decay to two jets~\cite{RuanZhu}. At higher energies the separation between the vector boson fusion process $H\nu\nu$ from $ZH$ associated production  with $Z\rightarrow\nu\nu$, which is important in the determination of the Higgs width, also relies on good jet energy resolution~\cite{Azzi:2021ylt}.
	
	Typical ElectroMagnetic (EM) resolutions in high-granularity particle flow calorimeters are $\sim$15\%/$\sqrt{E}$ and close to
$\sim$10\%/$\sqrt{E}$ in Liquid Argon or dual-readout sampling calorimeters respectively~\cite{Aleksa:2021ztd}. The hadronic jet resolution is not significantly affected by improving this level of precision, although it has been shown that associating non-prompt photons to $\pi^0s$ and therefore to jets is strongly improved with state-of-the-art EM resolutions from homogeneous calorimetry with EM resolutions in the 3-5\%/$\sqrt{E}$ range~\cite{Lucchini:2020bac}. 
Sampling calorimeter resolutions are sufficient to detect $H\rightarrow\gamma\gamma$ decays and measure the Higgs to two photon coupling, although with smaller accuracy than HL-LHC due to the lower statistics at FCC-ee. Improving the EM resolution has a small effect on the coupling measurement~\cite{An:2018dwb}. So neither the jet resolution nor the $H\rightarrow\gamma\gamma$ measurements are strong drivers for pushing the EM resolution to state-of-the-art levels, that would require the use of crystals and therefore potentially increasing the overall detector cost.

On the other hand, improvements in EM resolution would be very helpful in studies of several processes at the Z pole. Heavy flavor physics would strongly benefit from a high precision and high granularity EM calorimeter, especially for the identification and measurement of $\pi^0$'s and thus the improved reconstruction of the  many final states that include them. An example is the measurement of CP violation in the decay $B_s\rightarrow D_s K$~\cite{AleksanFCCweek20, Aleksan:2021gii}. Measurements based on radiative return also require very good EM resolution. In particular the coupling of the $Z$ boson to electron neutrinos, which is a unique opportunity enabled by a future $e^+e^-$ collider~\cite{Aleksan:2019erl},
and also the high-dimension anomaly detection via radiative return described in Ref.~\cite{Gorisek2016} (see especially Sec.~5).
The search for rare processes such as $\tau\rightarrow \mu\gamma$ or $Z\rightarrow \tau e$ would also profit from state-of-the-art electromagnetic resolutions~\cite{Dam:2021ibi, Dam:2018rfz}.  
In Ref.~\cite{Dam:2021pnx}, 
the author shows that the background for the rare flavor-violating decay
$\tau \rightarrow \mu \gamma$ in 1.6\% of the full FCC-ee statistics 
 scales linearly (or slightly stronger) in the photon energy resolution.
	
The requirements can then be summarized as follows:
\begin{itemize}
    \item Jet resolution: 4\%  jet energy resolution for jets from W/Z/Higgs bosons~\cite{THOMSON200925,POSCHL2020162234}.
    \item High granularity: EM cells of 0.5 $\times$ 0.5\,cm$^2$ to better resolve $\gamma$'s from $\pi^0$ decay.
    \item EM resolution: minimally for Higgs physics $\sigma/E = 10\% / \sqrt{E} + 1\%$. For $Z$ pole physics, typical crystal EM calorimeter resolutions.
\end{itemize}

It is possible now to match these requirements with dual-readout calorimetry and to even exceed them with a crystal-based EM calorimeter front section and through a new type of particle flow algorithm (Sec.~\ref{sec:lucchini}).

Precision calorimetry is challenging. Currently there are two main thrusts: high granularity and dual readout.  It is important that both of these options be vigorously pursued.  Especially, if a circular collider such as FCC-ee or the muon collider is built, it is optimal if the detectors
at the (potentially up to 4) interaction regions have different technologies and thus different systematics.

\begin{figure}[hbtp]
\centering
\resizebox{0.8\textwidth}{!}{\includegraphics{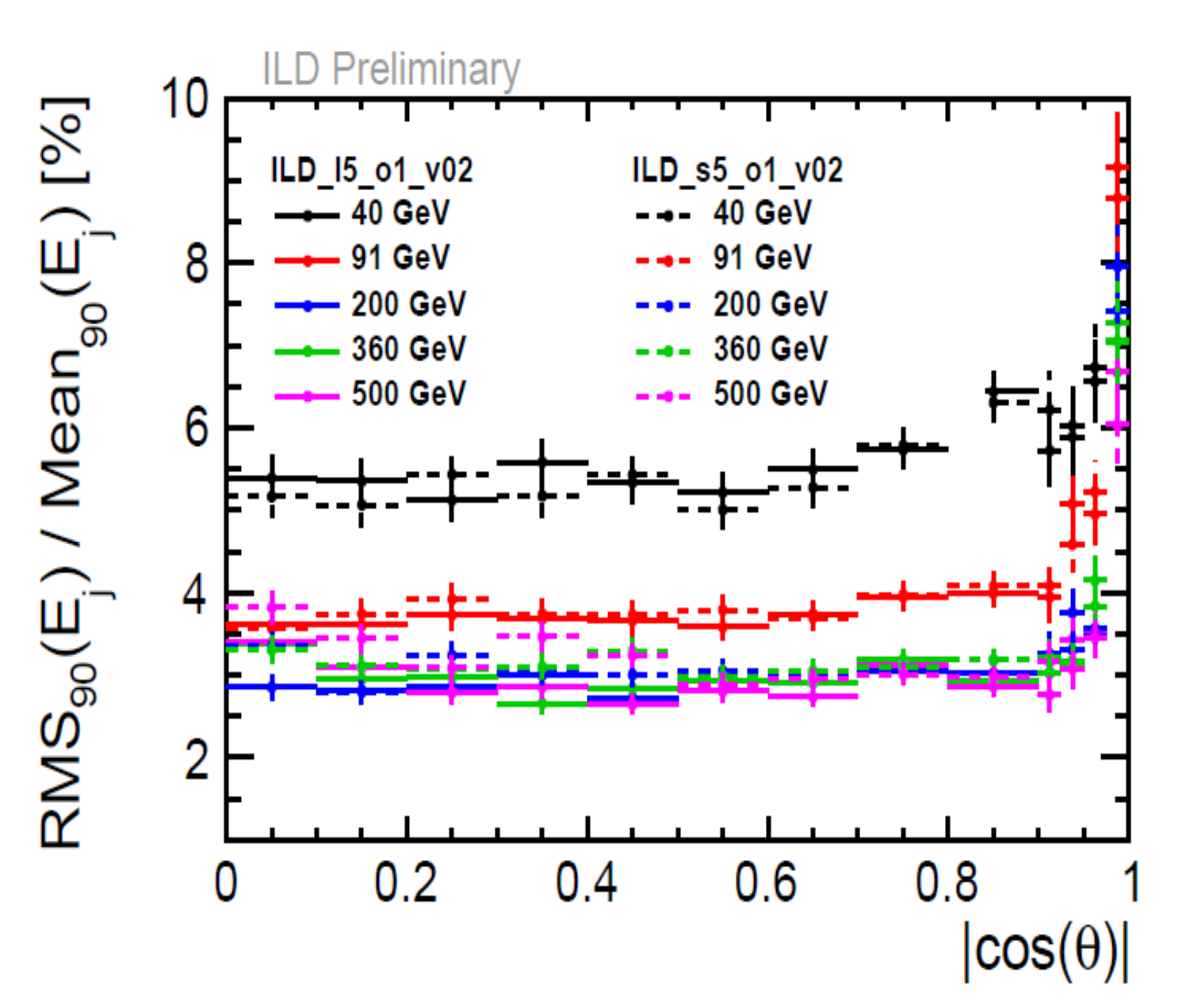}}
\caption{ 
Typical jet energy resolution at a Higgs factory from a highly granular calorimeter in conjunction with particle flow algorithms (specifically, for the  ILD high granularity calorimeter, from Ref.~\cite{theildcollaboration2019ild}). 
}\label{fig:siDjet}
\end{figure}

\begin{figure}[hbtp]
\centering
\resizebox{0.90\textwidth}{!}{\includegraphics{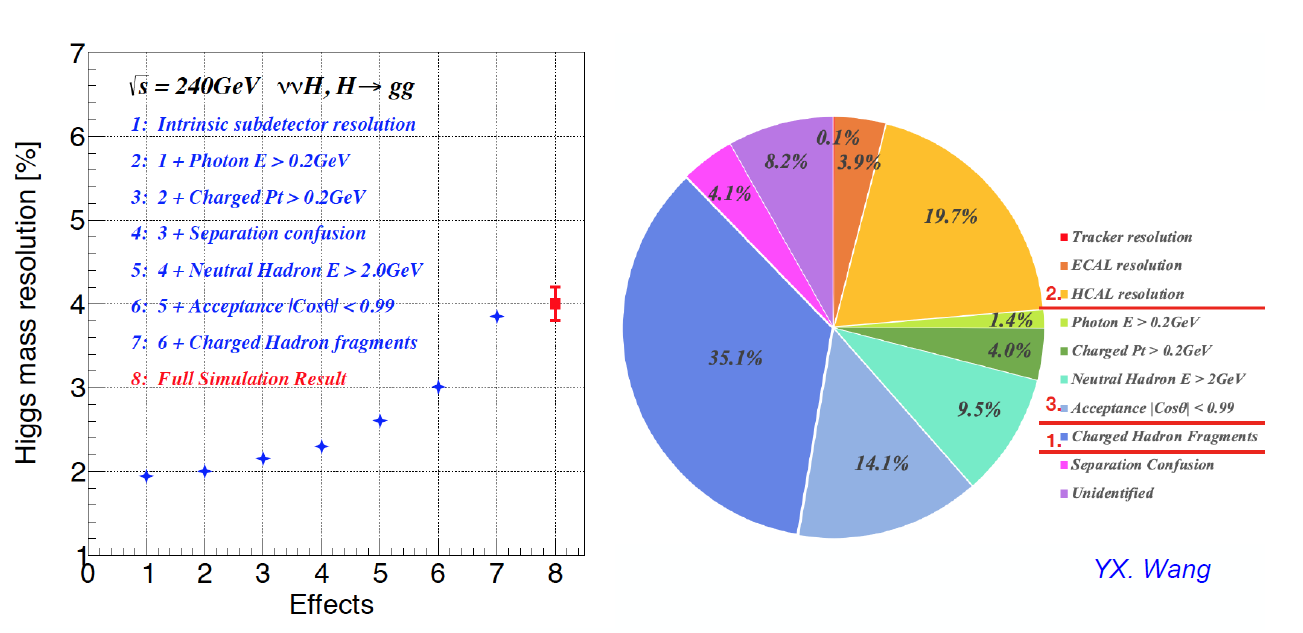}}
\caption{ 
Reproduction of slide 17 of Ref.~\cite{manqislides}, showing the breakdown of the contributions to the jet energy resolution for a detector with a high granularity calorimeter designed for the CEPC.
}\label{fig:favoritemanqi}
\end{figure}

High granularity calorimetry currently has the largest world-wide community, lead by the CALICE community and
the CMS HGCAL~\cite{hgcaltdr} community.  The strategy is to use the calorimeter
as little as possible and to obtain
excellent jet resolutions by  using the tracker for measurement of electrically charged particles to 0.3\% or better in a momentum range of 0.3 to 30\,GeV, and then use high granularity and pattern recognition to remove their calorimetric energy deposits, leaving those from photons and neutral hadrons~\cite{THOMSON200925,POSCHL2020162234}. 
The pattern recognition requirements favor sampling calorimeters made of materials with small \moliere radius, and both transverse and longitudinal segmentation.  However,
since the active material typically has a larger \moliere radius, the desire for compact showers
prefers low sampling fractions.  The large channel count requires substantial electronics embedded in the volume of the calorimeter,
again diluting the \moliere radius and increasing the complexity of cabling and cooling.
The result is modest photon (15 -20 \%/$\sqrt{E}$) and neutral hadron (45\% - 150\%$/ \sqrt{E}$) energy resolutions.  
However, since the calorimeter is only used for photons and neutral hadrons,
current high granularity calorimeters using  these ``particle flow" algorithms~\cite{THOMSON200925} typically have jet energy resolutions with a stochastic term of 30\%$/ \sqrt{E}$, largely determined by the neutral particle calorimeter-based measurements.
Unfortunately, the resolutions are not Gaussian, and the energy resolution imbalance between tracker and calorimeter measurements for charged hadrons is so large that the correct assignment of showers to tracks is limited.  
Figure~\ref{fig:siDjet} shows a typical expected jet resolution from Ref.~\cite{theildcollaboration2019ild} (for the ILC ILD detector). Note that due to non-gaussian tails, the ``rms'' is calculated by truncating the distribution to include 90\% of the area.
No operating experiment currently has this kind of calorimeter, although the CMS detector at CERN's LHC will have one (the HGCAL~\cite{hgcaltdr}) at the 
start of the LHC HL-LHC run (around 2029).
One of the systematic uncertainties associated with jet reconstructions using this type of calorimetry, is due to the splitting of hadronic showers in the pattern recognition.
A fast simulation study of
Ref.~\cite{manqislides} slide 17, reproduced in Fig.~\ref{fig:favoritemanqi}, shows  that this splitting is currently
the largest contribution to the resolutions of jets in high granularity calorimeters.
 When showers are split, and only the part most closely
 matched to a track removed from the jet energy calculation, the split energy is a noise term similar to pileup in pp collisions.
Due to sizable uncertainties in
the low-energy nuclear physics modeling in GEANT4~\cite{Howard_2020}, these systematic uncertainties cannot be understood until a lepton collider Higgs factory is built and operating (although the CALICE collaboration has done much groundbreaking work using test beams to reduce these systematic uncertainties for single particles in a detector of limited size and the work will certainly benefit from ML techniques such as discussed in Ref.~\cite{AkchurinNN}).  
For high energy jets (order 100\,GeV), the individual showers in the jet will overlap even for materials with the smallest \moliere radius, leading to the so-called
confusion term~\cite{THOMSON200925}.  Particle flow reconstruction becomes less effective at these energies.
It would be sensible, given the high cost and potential physics impact of these machines, to develop another form of calorimetry with different systematic uncertainties.

Another path, allowing  better resolutions and certainly having very different systematic uncertainties, is to improve the calorimetry resolution for
all particles.  
Dual-readout calorimetry uses additional information, beyond simply the total energy deposition, to separately identify the contribution due to relativistic particles (mainly electrons and positrons). Either two different sensitive media are used (scintillating and \v{C}erenkov media) or the light pulse is integrated over different wavelength regions and/or time intervals. Neutrons can also be separately measured (triple readout) through delayed signal integration (or late pulse shape sampling).
The differences in response between the relativistic (EM) and non-relativistic (hadronic) components are the leading driver of hadronic resolution for most calorimeters, therefore estimating the fraction of the shower associated with each sub-component can be used to reduce these fluctuations.  The jet reconstruction in this type
of calorimeter would be more driven by the calorimeter (although as discussed in Sec.~\ref{sec:lucchini}, novel particle flow algorithms
can improve the resolutions of dual-readout calorimeters as well).  Without particle flow, excellent jet resolutions are obtained
without the split/merge systematic.  With particle flow, the systematics will be very different, as the particle flow algorithms
are very different (see Sec.~\ref{sec:lucchini}).  The electronics can be entirely at the front or back of the calorimeter,
maintaining the compactness of the shower and simplifying all services.

 \section{Short review of the dual-readout concept}
\label{sec:dualreview}

In this section, we briefly review why ancillary information can improve shower resolutions. 
Hadronic showers involve inelastic interactions of hadrons with nuclei.    The breakup of the nuclei consumes some of the kinetic energy of the incident particle, and produces many different types of secondary particles, including charged and neutral pions, heavy-ion fragments, strange mesons, photons, etc.  The signal generated by a hadronic shower  is lower in general compared to that of an electron/photon (``EM'' shower) of the same energy due to {\it invisible energy losses} from e.g. nuclear binding energies, neutrinos, and particles with small inelastic cross sections such as neutrons escaping the detector or depositing energy outside of the sampling window. A variety of other effects also contribute to the lower detector response for hadronic showers. The amount of missing (invisible) energy is correlated with different observables such as the number of inelastic nuclear collisions or the electromagnetic energy fraction. Maximal information calorimetry finds measurable quantities that are correlated with the missing energy to make shower-by-shower corrections.

\begin{figure}[hbtp]
\centering
\resizebox{0.48\textwidth}{!}{\includegraphics{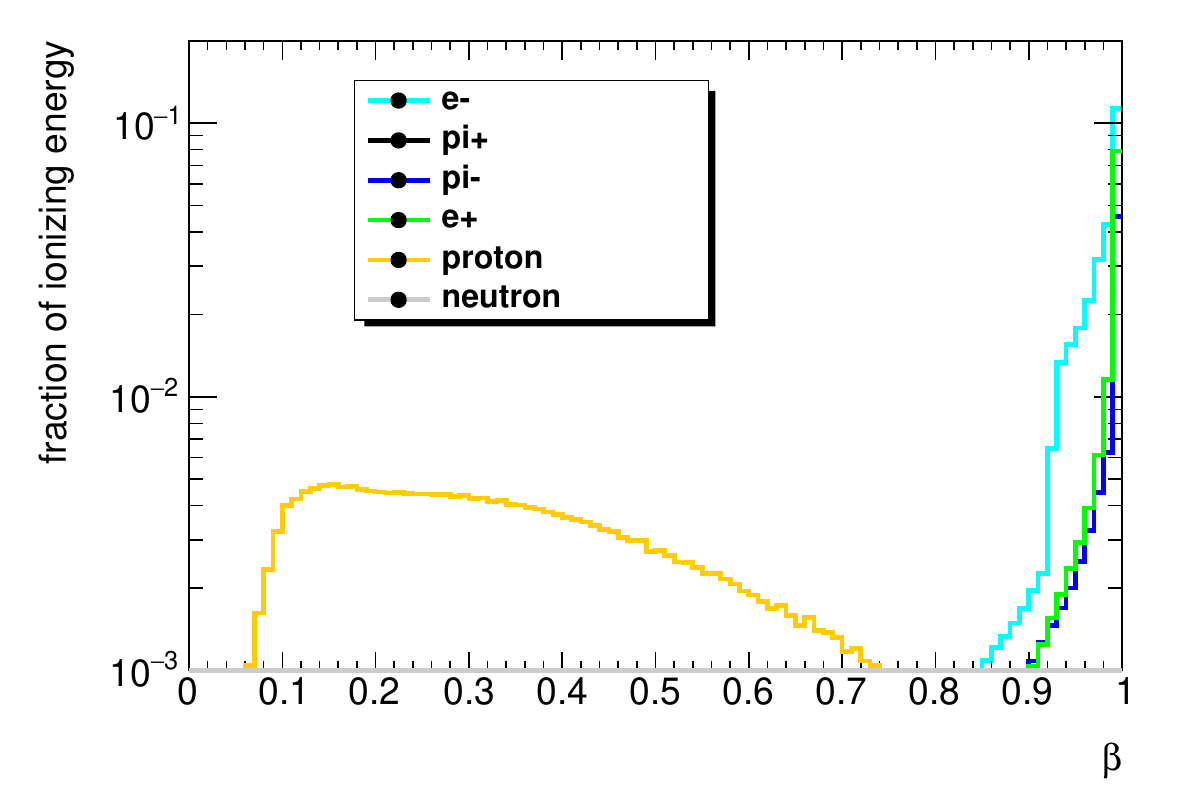}}
\resizebox{0.48\textwidth}{!}{\includegraphics{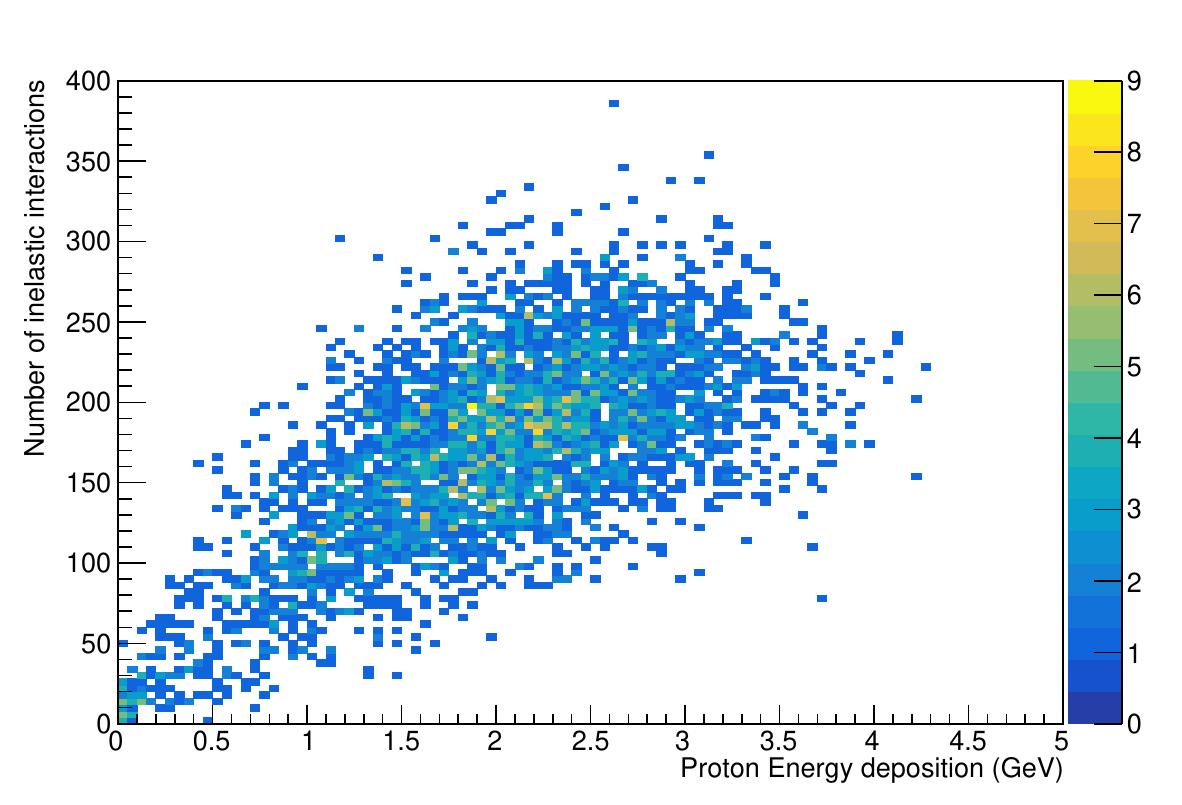}}
\caption{ [left] Fraction of total ionizing energy deposit in a large block of PbWO$_4$ versus particle velocity ($\beta$) by particle type  in a GEANT4 simulation of a  shower initiated by a high energy charged pion. [right] Correlation between ionizing energy deposition by protons and the number of inelastic interactions in a GEANT4-simulated  pion shower in a large block of PbWO$_4$. }
\label{fig:beta}
\end{figure}

A example of the use of additional information is dual-readout calorimetry, 
which is used to estimate the fraction of the shower particles that are relativistic.    Figure~\ref{fig:beta} [left] shows the fractional contribution to the ionizing energy deposition as a function of particle velocity and type for a shower initiated by a high-energy charged pion in a large block of PbWO$_4$.
The energy depositions by non-relativistic charged particles are dominated by protons.  Figure~\ref{fig:beta} [right] shows the correlation between the number of nuclear breakups (which is correlated with the missing energy) and the energy deposition by protons.  An estimation of the fraction of the energy due to non-relativistic particles gives an estimation of the missing energy. It should also be stressed that the complementary information (i.e. an estimation of the fraction of the energy due to relativistic particles) is equivalent for this purpose.

\begin{figure}[hbtp]
\centering
\resizebox{0.48\textwidth}{!}{\includegraphics{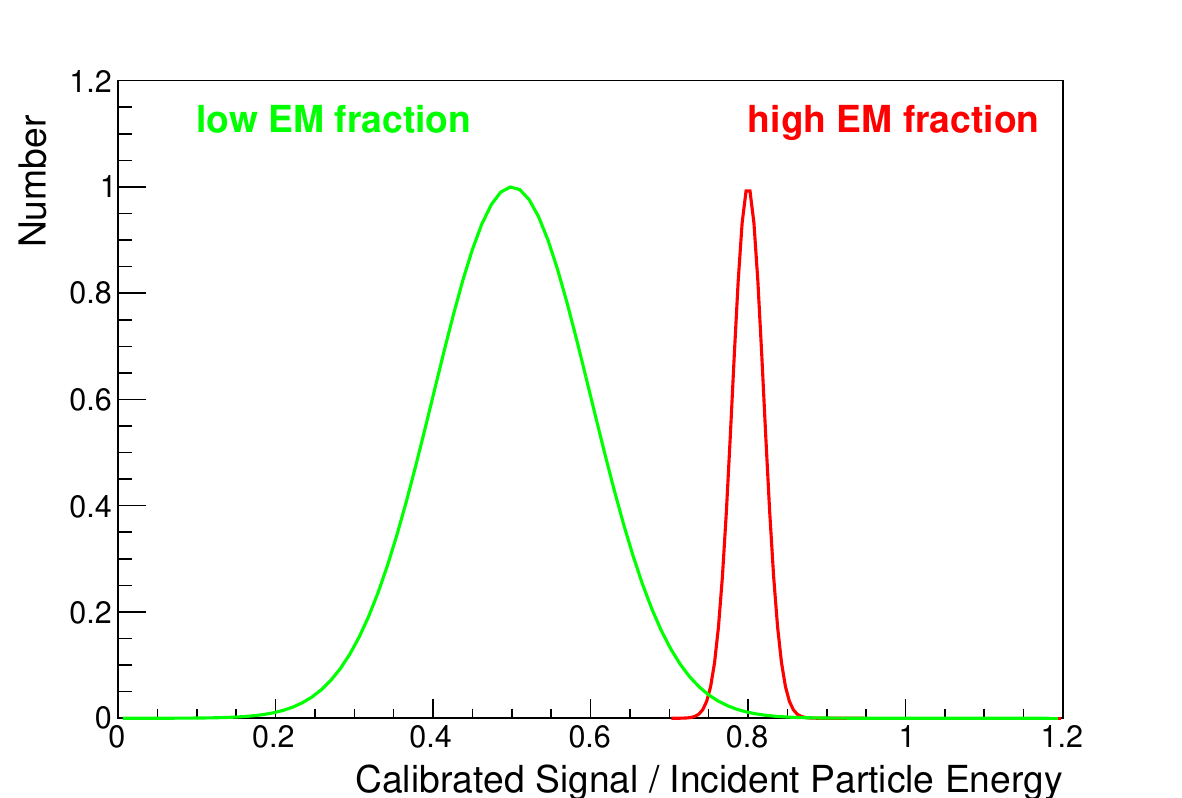}}
\resizebox{0.48\textwidth}{!}{\includegraphics{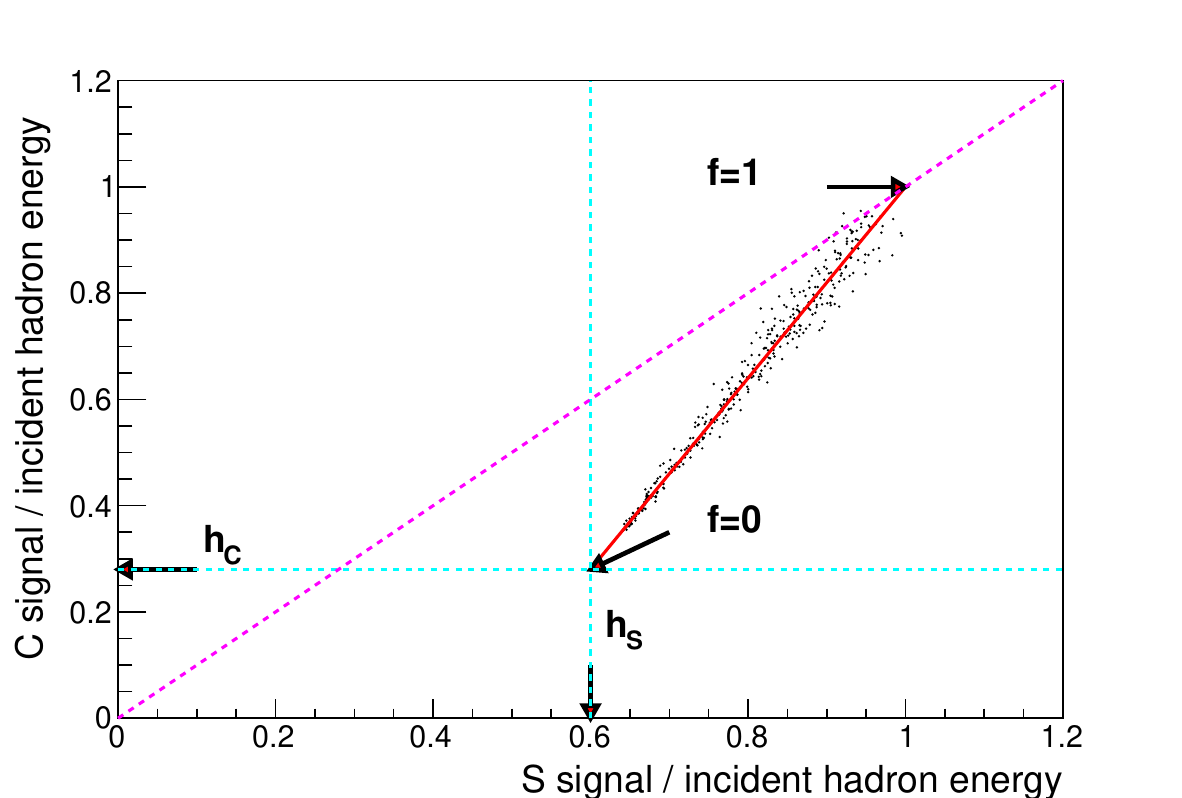}}
\caption{ [left] Inspired by Ref.~\cite{Lee:2017xss}, an illustration of why fluctuations in the subcomponents of a shower dominate the resolutions of hadronic calorimeters with high sampling fraction. The two Gaussians represent the spread in the measured signal for a fixed incident particle energy for two different values of the ``electromagnetic'' fraction $f$.
[right] An illustration of how two measurements, one with sensitivity to all components of the shower (``S'') and another with reduced sensitivity to the non-EM portion of the shower (``C'') can be used to correct for fluctuations in the energy scale due to fluctuations in the subcomponent composition.  The points represent a pair of ``C'' and ``S'' measurements from an individual hadronic shower.  For the meanings of the text, see Eqn.~\ref{eqn:dual}.
The dual-readout correction moves points along the red line towards the purple dashed line, reducing their spread.
}\label{fig:wigmans1}
\end{figure}

The correction process is illustrated
in Fig.~\ref{fig:wigmans1} [left], following the description detailed in Fig. 2 of Ref.~\cite{Lee:2017xss}.
As shown in the cartoon, the average response to a purely EM shower is calibrated to  1.0 (i.e. the calorimeter is calibrated only ``once'' at the EM scale). Showers with lower relativistic content will have a response that is lower than unity mainly due to their larger missing energy.
Fluctuations in the electromagnetic fraction will lead to fluctuations in response that limit the resolution.  

The correction formula when using only one additional measurement is simple, and is shown graphically in Fig. 7 of Ref.~\cite{Lee:2017xss} (Figure~\ref{fig:wigmans1} [right] here).  Given two signals ``S'' and ``C'' with different sensitivities to the non-EM part of the hadronic shower:
 \begin{equation}\label{eqn:basiceqn}
\begin{aligned}
     S & = k_S \cdot E [f + h_S(1-f)] \\
     C & = k_C \cdot E [f + h_C(1-f)],
     \end{aligned}
 \end{equation}
 where $E$ is the energy of the incident particle (assuming full shower containment),
 $h_S$ ($h_C$) is the response of the ``S'' (``C'') signal to the non-EM energy relative to that of the EM energy, 
 $f$ is the fraction of the shower that is EM,
 and $k_S $ and $k_C$ are the conversion factors 
 (with units $GeV^{-1}$) for EM showers, which by definition have f=1.
 In ``classical" dual readout, ``S'' is light from scintillation  and ``C'' is light from \v{C}erenkov radiation,
    and thus $k_S$ and $k_C$ contain the e.g. light production, detection, and 
  identification efficiency. $k_S$ and $k_C$ set the size of the counting uncertainty due to photostatistics.
  Both $h_S$ and $h_C$ are usually less than one.
  In an ideal dual-readout calorimeter, $h_S$ is one, and  $h_C$ is zero (in this case, S directly provides E and C/S provides f). A compensating calorimeter is one for which $h_S$ is one.
 
 Rearrangement yields:
\begin{equation}\label{eqn:dual}      E  = \frac{S/k_S ~-~ \chi \cdot C/k_C}{1~-~\chi}
\end{equation}
 where:
\begin{equation}\label{eqn:chi}
     \chi  = \frac{1-h_S}{1-h_C}
\end{equation}
 
 The dependence on $f$ is eliminated, and the contribution to the resolution due to the spread in $f$ is removed.  In Fig.~\ref{fig:wigmans1} [right], which shows the correlation between the two measurements, applying this equation is algebraically equivalent to  projecting the points, each of which represents a measurement of an individual shower,
 along the red line (whose extent corresponds to different values of $f$) to the purple dashed line in the neighborhood of $C=S=1$,
 removing the spread along the red line due to the variation in $f$.

While this section uses the example of detecting of both \v{C}erenkov and scintillation light, complementary  corrections can be made using timing windows to select late deposits from protons produced via neutron interactions~\cite{7454834}.

\section{Status of dual-readout calorimetry}
\label{sec:currentstatus}
A comprehensive review of the history of dual-readout calorimetry can be found in Ref.~\cite{Lee:2017xss}. Briefly, in 1983
Paul Mockett (U. Washington) proposed a dual-readout concept which recognized measuring the two main components in a shower with multiple sampling media to improve hadron calorimetry~\cite{mockett}. This approach was subsequently
taken up by Richard Wigmans, Vladimir Nagaslaev, and Alan Sill of Texas Tech University~\cite{firstdual} and 
implemented by the DREAM Collaboration led by Richard Wigmans (Texas Tech University)~\cite{Lee:2017xss}.
Research on dual readout continued with the DREAM/RD52 collaboration, studying both dual readout in homogeneous scintillating crystals, and in a spaghetti-type calorimeter with two types of fibers: scintillating and clear (for \v{C}erenkov light production), demonstrating that unprecedented hadronic resolutions could be achieved.
Recently, by exploiting the advancements in solid stated sensors, they also proved that individual readout of the fibers with Silicon PhotoMultipliers (SiPMs) opens the way for resolving the shower structure with unprecedented granularity~\cite{Antonello2018}. The extensive amount and the high quality of the information also allows the development of new types of particle-flow algorithms for identifying and resolving highly complex final states.

Current research on dual-readout calorimetry concentrates on high-resolution fiber-sampling calorimeters and is carried out primarily by the IDEA collaboration, a consortium of institutions from Italy, S. Korea, UK, Chile, and the US.
The status of their work is described in Section~\ref{sec:idea}.
Recently, inspired by past work on homogeneous calorimeters done by the RD52 collaboration and by Ref.~\cite{Lucchini:2020bac}, a group of US physicists has received initial seed money to revive studies of dual readout in homogeneous calorimeters based on scintillating crystals.  This proposed work is described
in Sec~\ref{sec:calvision}.
Work~\cite{lucchinipf} on novel particle-flow algorithms by physicists at University of Milano-Bicocca, Sussex, CERN, and Princeton for this type of calorimeter appeared in 2021 and is summarized in Sec~\ref{sec:lucchini}.

\subsection{The IDEA detector and its spaghetti calorimeter}
\label{sec:idea}
\begin{figure}[!htbp]
\centering
\includegraphics[width=1.00\linewidth]{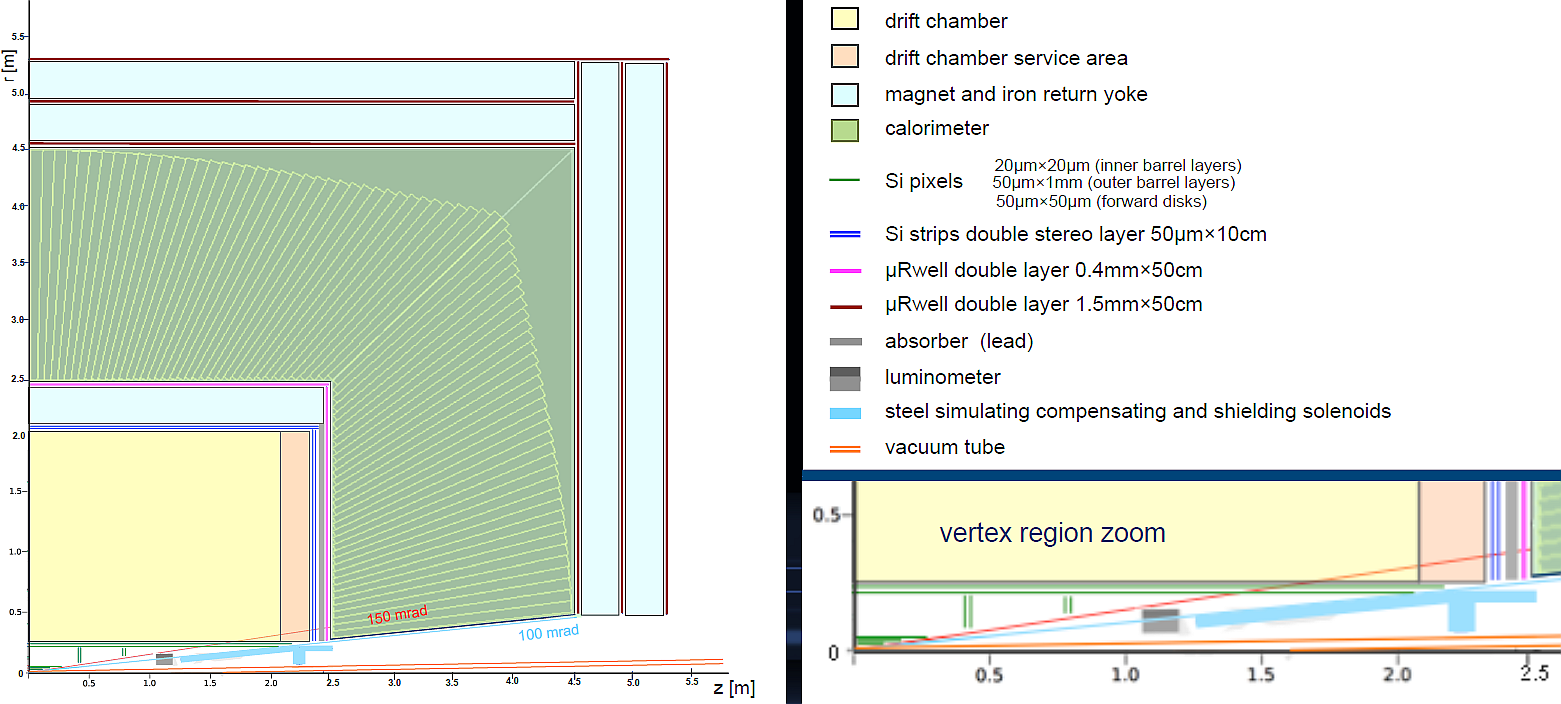}
\caption{Cross section of the proposed layout for the IDEA detector concept.}
\label{fig:IDEAdet}
\end{figure}

IDEA (Innovative Detector for an Electron-positron Accelerator) is an innovative general-purpose detector concept~\cite{Bedeschi:2021nln}, designed to study electron-positron collisions in a wide energy range provided by a very large circular lepton collider, with a typical circumference of 100 km. IDEA is a hermetic detector, geometrically subdivided in a cylindrical barrel region and closed at the extremities by two endcaps, as can be seen in Figure~\ref{fig:IDEAdet}.
The detector is composed of the following subdetectors in order of increasing distance from the primary vertex: a central tracking system, a solenoid, a calorimeter, and a muon detection system.  The calorimeter is discussed in Section~\ref{sec:ideacalorimeter}.

The central tracking system is composed of a vertex detector made of silicon pixel and strip detectors~\cite{andreazzaPED, andreazzaHVmaps, Pancheri:2019qto} and a large drift chamber with a 2 m outer radius~\cite{Tassielli:2021rjk}, providing more than 100 measurements along the track of every charged particle. The vertex detector measures tracks of charged particles with very high precision, of the order of 3\,$\mu m$ in the innermost layers, and is able
to reconstruct secondary vertices, originating for example from decays of heavy flavor hadrons, with exquisite precision. The drift chamber also offers outstanding particle-identification performance using the cluster counting technology. The central tracker is then completed by a silicon wrapper, made of silicon detectors, that surrounds the drift chamber. 

The central tracker is surrounded by a large but thin and low-material-budget solenoidal magnet that provides a 2 T magnetic field~\cite{tenkateECFA}. The solenoid is followed by the preshower detector. The preshower is used to identify and measure electromagnetic showers that originate in the material of the solenoid before reaching the calorimeter. The preshower is constituted of a large array of a novel type of micro-pattern gas detectors, the $\mu$-RWELL~\cite{Bencivenni:2017wee, Morello:2017edb}. The preshower provides a spatial resolution for charged particles of about 60-70\,$\mu m$. In the baseline design, the preshower is followed by a dual-readout (DR) fiber-sampling calorimeter. An option with a high-resolution (and dual-readout) EM crystal calorimeter is also being studied.

The last element is the muon detection system. The muon detector also uses the $\mu$-RWELL technology, but with a coarser strip pitch. It is subdivided into three stations at increasing distance from the vertex, located within the iron return yoke that closes the magnetic field. Each muon station can provide a space point with a spatial resolution of about 400\,$\mu m$ in the plane perpendicular to the particle direction. Combining the three stations enables standalone tracking of charged particles at 5-6\,m from the vertex. The precision achieved
also allows identification of secondary vertices that could be produced by long-lived particles.

The IDEA detector concept has been considered by both the FCC-ee collider and the CEPC collider and is described in detail in both Conceptual Design Reports~\cite{CEPCStudyGroup:2018ghi,Benedikt:2651299}.

\subsubsection{The spaghetti DR calorimeter}
\label{sec:ideacalorimeter}

The IDEA calorimetric section adopts a dual-readout longitudinally unsegmented fully projective fiber calorimeter. Following the ideas expressed in~\cite{richardbook} on high-resolution compensating fiber calorimeters, and the experimental work on the SPACAL~\cite{spacalyellowreport} and the RD52 calorimeters~\cite{Lee:2017xss}, the geometry is dictated as follows.
The segmentation of the calorimeter is chosen in such a way that the volume in which a shower develops always corresponds to a small number of cells, and most of the energy is deposited in a single cell. This greatly simplifies the calibration procedures, as each cell response can be equalized as a single entity. This type of segmentation can be achieved, with a tower-based structure:
the detector volumes are divided into towers, and the signal produced in each tower is integrated over the tower length.
Active volumes are optical fibers running parallel to the tower axis. As demonstrated in~\cite{Antonello2018}, the possibility of reading out each fiber independently with a dedicated SiPM frees the position and angular measurements from the limitations due to the tower structure, and potentially leads to an unprecedented spatial and angular resolution. This also opens the possibility of separating two particles showering inside a single tower on the basis of the pattern of the signals produced. 

The main quantities provided by a calorimeter are the momentum vector, the energy, and the identity of the incident particles. 
Following this reasoning, the IDEA calorimeter is composed of $36$ rotations around the beam axis corresponding to a segmentation of $\Delta\phi = 10.0^\circ$. Each rotation is referred to as a slice. Each slice contains both barrel and endcap towers with a $\theta$ coverage down to $\simeq0.1$ rad. At present, the towers are $2$\,m long but a solution with 
$2.5$\,m long capillary tubes is also being considered.
The calorimeter barrel-region of a single slice is made of $80$ towers while each slice endcap-region consists of $35$ towers. Each tower has a $\theta$ segmentation of $\Delta\theta=1.125^\circ$. The barrel geometry is perfectly symmetrical with respect to the plane perpendicular to the beam passing through the IP, \textit{i.e.} the first right tower is identical to the first left tower, and so on. Each slice is identical so that each tower (both endcap and barrel ones) is repeated $2\times36=72$ times.

The calorimeter active elements are scintillating (polystyrene) and clear-plastic (PolyMethyl MethAcrylate, PMMA) fibers embedded in copper, in a chess-board like geometry. In order to minimize the sampling fluctuations, each fiber is $1$\,mm thick (core + cladding), separated from its closest neighbors by $0.5$\,mm of absorber material. Copper was chosen for its $e/mip$ response ratio of about 0.85, which guarantees a better linear response to low-energy charged hadrons compared to high-Z materials.

\begin{figure}
\begin{center}
\includegraphics[scale=0.3]{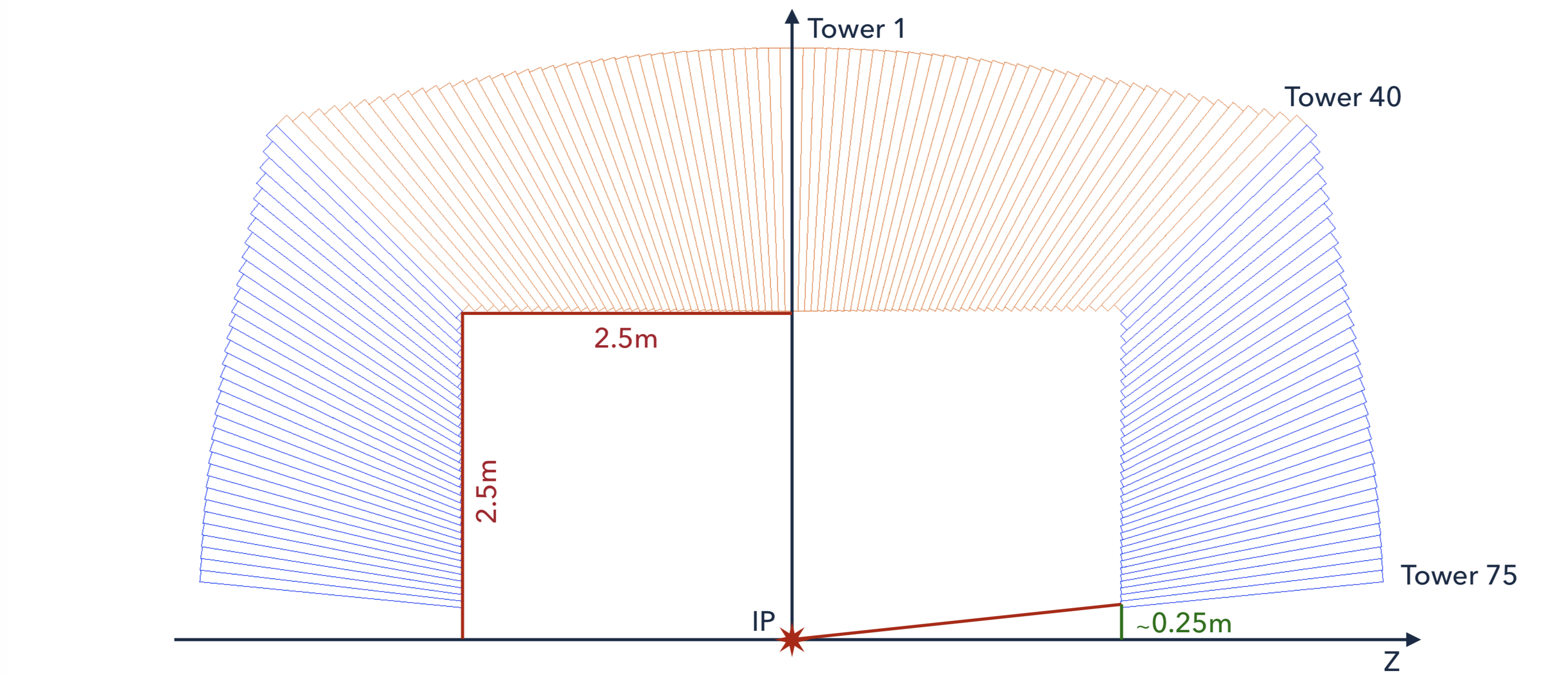}
\caption{Sketch of a single slice of the IDEA calorimeter.}
\label{IDEA_DR_slice}
\end{center}
\end{figure}

\begin{figure}
\begin{center}
\includegraphics[scale=0.25]{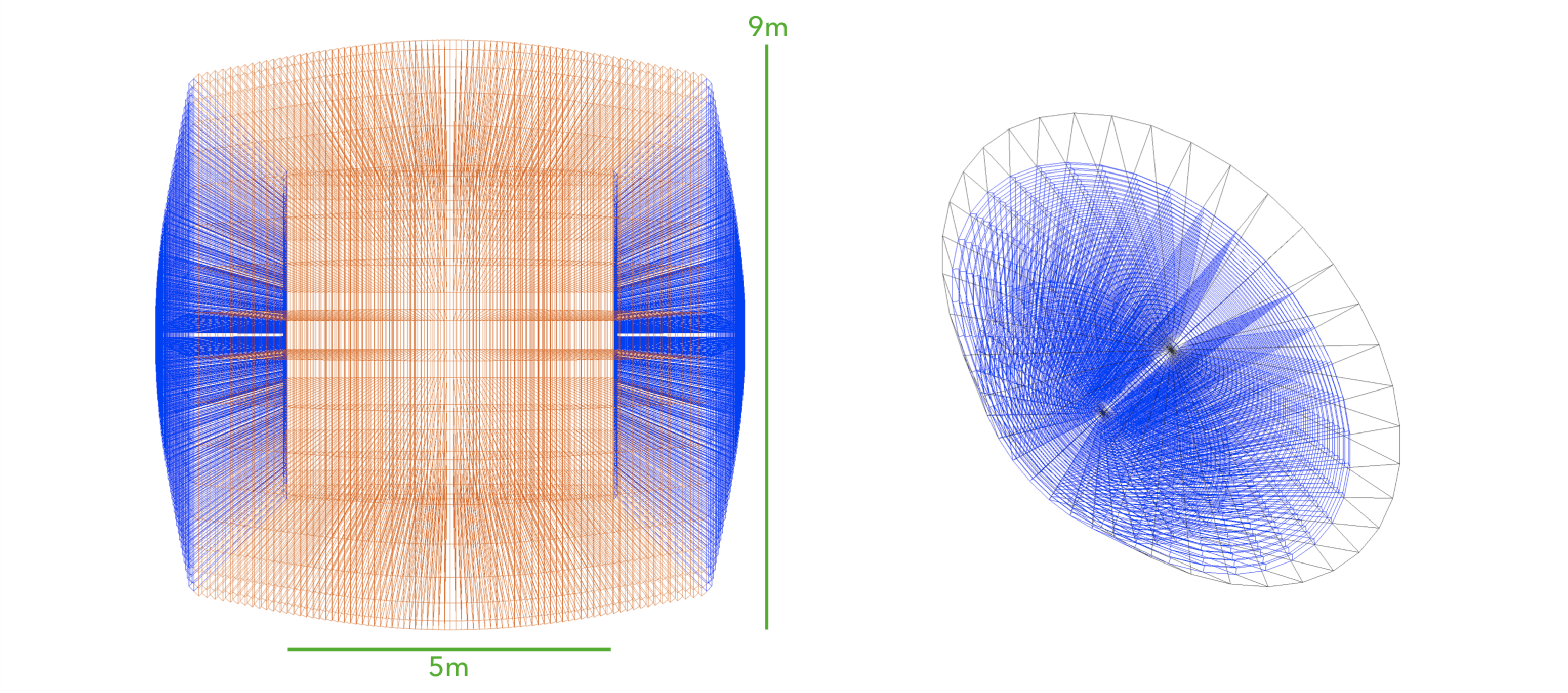}
\caption{Sketch of the IDEA calorimeter (left) and endcap geometry (right).}
\label{IDEA_DR_geometry}
\end{center}
\end{figure}

This complex geometry has been implemented in a program based on the GEANT4 simulation toolkit~\cite{Agostinelli:2002hh}, and all results in this section are based on such simulation. Figs.~\ref{IDEA_DR_slice},~\ref{IDEA_DR_geometry} show respectively a slice of the IDEA calorimeter and sketches of the full calorimeter and a single endcap. The simulation program is described in detail in~\cite{Pezzotti:2021jxh}. An adaptation of this detector was later used to combine dual-readout crystals and fibers in a hybrid segmented calorimeter: details are given in Section~\ref{sec:crystal_cal}.
In the simulation, the Birks-corrected scintillation light produced on a step-by-step basis is smeared by a Poissonian distribution according to the statistical nature of the light production mechanism. This approach correctly reproduces photo-statistics fluctuations in scintillation light production and allows  the implementation inside the simulation of the desired scintillating light yield (p.e./GeV). The simulation is tuned to $\simeq 400$ scintillation p.e. per GeV deposited at the electromagnetic scale, \textit{i.e.} by electrons hitting the face of the calorimeter.
Similarly for the \v{C}erenkov process, each light seed is smeared by a Poissonian distribution tuned to $\simeq 100$ p.e. per GeV deposited at the electromagnetic scale.\\

\subsection*{Electromagnetic showers performance}

\begin{figure}
\includegraphics[scale=0.4]{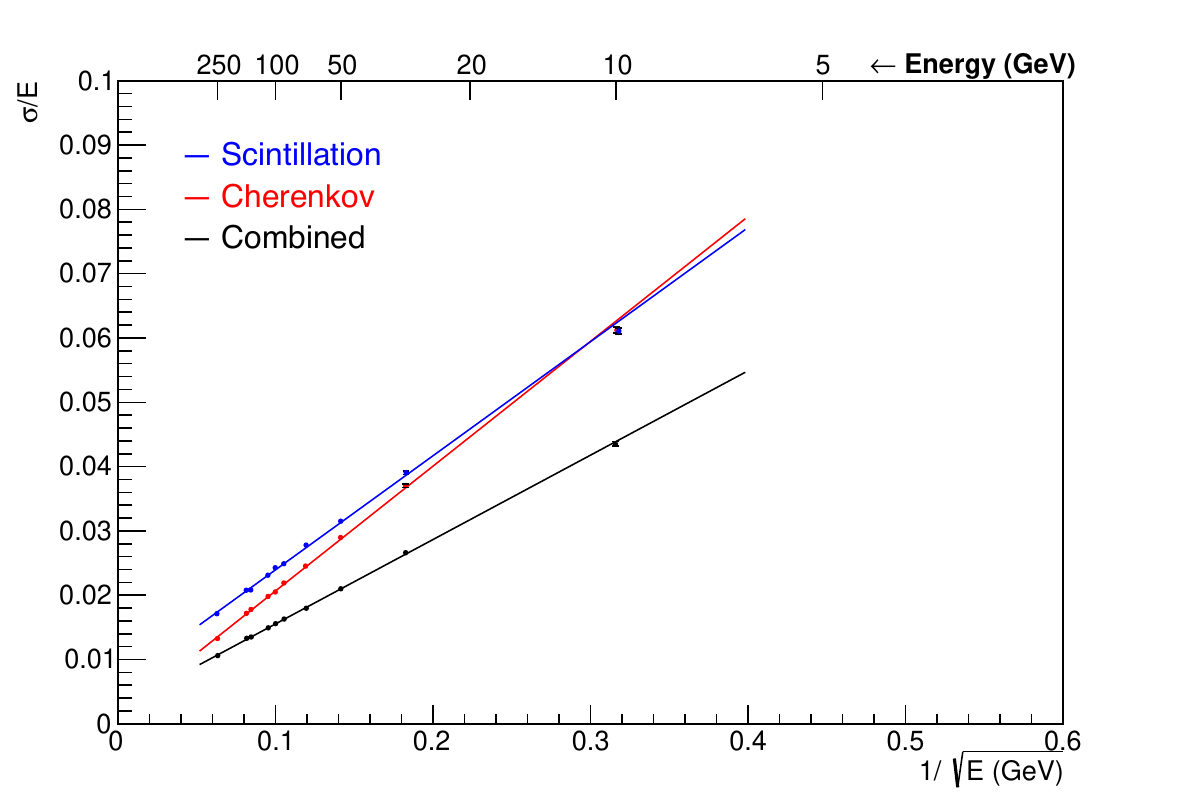}
\includegraphics[scale=0.4]{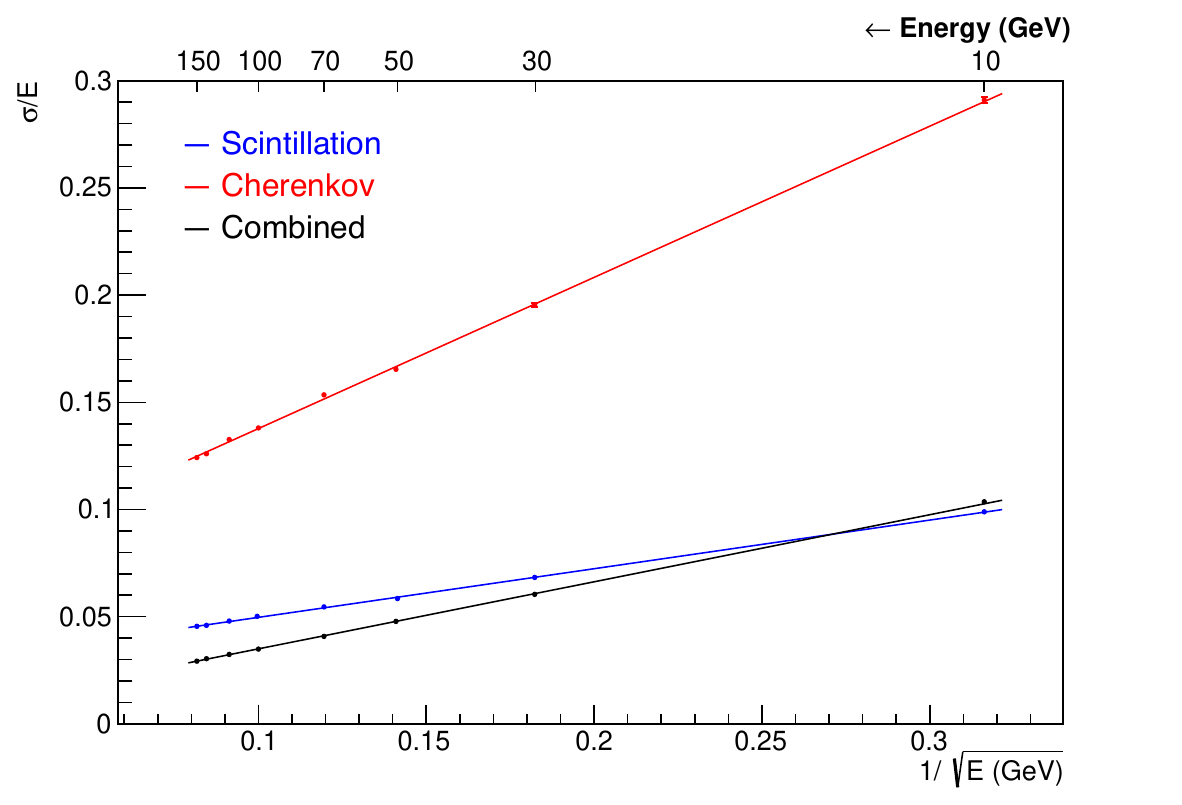}
\caption{Electromagnetic energy resolutions of the IDEA dual-readout calorimeter (left). Energy resolution for $\pi^-$s from $10$ to $150$ GeV (right).}
\label{drresolution}
\end{figure}

A dual-readout calorimeter samples electromagnetic showers independently with the scintillation and \v{C}erenkov signals. The two signals can be further recombined to get the best electromagnetic energy resolution possible.
By fitting $\sigma/E$ values obtained for electrons in the energy range 10-250\,GeV, as shown in Fig.~\ref{drresolution}(a) (blue and red lines for scintillation and \v{C}erenkov signals, respectively), we find that the energy resolutions are well represented by the expressions:

\begin{equation}
\frac{\sigma}{E}=\frac{17.7\%}{\sqrt{E}}+0.6\% \text{ or, } \frac{19.6\%}{\sqrt{E}}\oplus1.3\%
\label{eq2}
\end{equation}
for the scintillation signal, and
\begin{equation}
\frac{\sigma}{E}=\frac{19.4\%}{\sqrt{E}}+0.1\% \text{ or, } \frac{20.0}{\sqrt{E}}\oplus0.5\%
\label{eq3}
\end{equation}

for the \v{C}erenkov signal, with $E$ expressed in GeV units.
The constant term of the scintillation signal is larger due to the more collimated nature of ionizing energy deposition in electromagnetic showers with respect to the \v{C}erenkov light signal. This makes the scintillation signal more sensitive to the impact point of the primary electron on the calorimeter face (the closer to a fiber the higher the signal). The best way to combine signals is the one that maximizes the component with the minimum standard deviation, on an event-by-event basis,

\begin{equation}
E = \frac{E_s/\sigma_s^2+E_c/\sigma_c^2}{1/\sigma_s^2+1/\sigma_c^2}
\label{eq4}
\end{equation}

with $\sigma_s$ ($\sigma_c$) being calculated with Eqs.~\ref{eq2}~(\ref{eq3}).
The energy resolution obtained, shown in Fig.~\ref{drresolution}(a) (black line), is well fitted by, 

\begin{equation}
\frac{\sigma}{E}=\frac{13.0\%}{\sqrt{E}}+0.2\% \text{ or, } \frac{14.0\%}{\sqrt{E}}\oplus 0.6\%
\label{eq5}
\end{equation}

The same studies also yield a linearity within $\pm1\%$ for electrons in the energy range $10 - 250$ GeV.

\subsection*{Hadronic shower performance}
The hadron performance was studied by using the $\chi$ factor obtained from Geant4 as the one that correctly reconstructs on average the primary particle energy, $\chi=0.41$. The resolution dependency on the $\chi$ factor was also studied and is reported in Ref.~\cite{Pezzotti:2021jxh}.
Fig.~\ref{drresolution}(b) shows the energy resolution in the range 10-150 GeV (bottom, black line). It corresponds to
\begin{equation}
\frac{\sigma}{E}=\frac{31\%}{\sqrt{E}}+0.4\% \text{ or, } \frac{32\%}{\sqrt{E}}\oplus 1.3\%
\label{eq8}
\end{equation}
with $E$ expressed in GeV units.
As discussed in Ref.~\cite{Pezzotti:2021jxh} the simulation also indicates that the asymmetric shape for the energy distribution, affecting any non-compensating calorimeter, is largely corrected, leading to a Gaussian distribution for the hadronic signal. Moreover, we found a linearity for charged pion detection within $\pm 1\%$ in the energy range $10-150$\,GeV. 


\subsection{Jet performance}

\begin{figure}[p]
\includegraphics[scale=0.4]{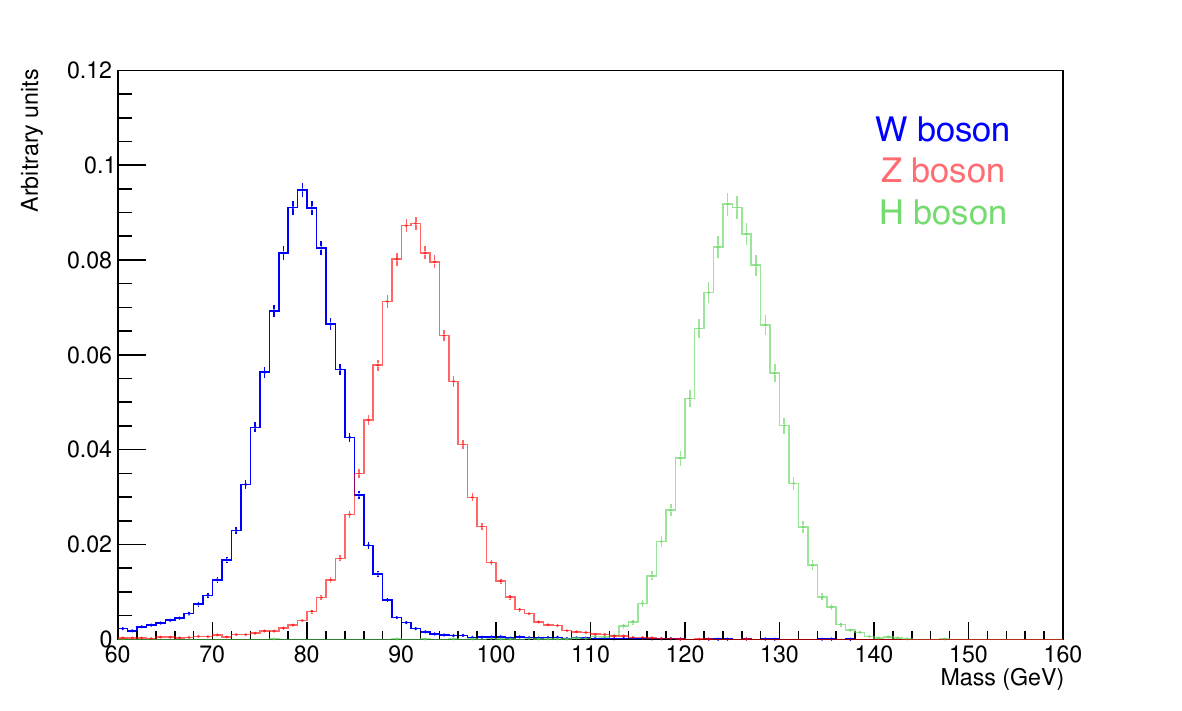}
\includegraphics[scale=0.4]{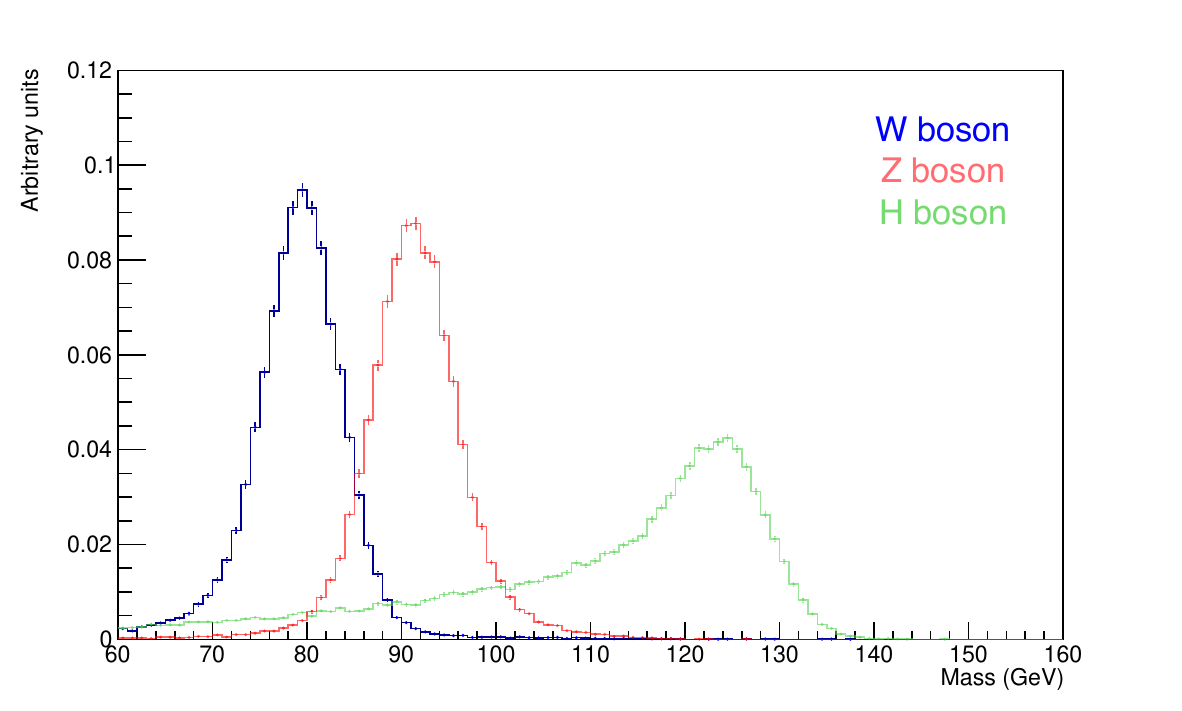}
\caption{Distribution of the reconstructed mass for three jet-jet resonances for the calo-only algorithm, excluding (left) or including (right) $b$ semileptonic decays.}
\label{2jbosondecays}
\end{figure}

Future circular electron-positron colliders will provide high statistics samples of $e^+e^-\rightarrow Z^{*}/\gamma \rightarrow jj$ events at several center-of-mass-energies ($E_{CM}$), resulting in events with two back-to-back jets with energy approximately equal to $E_{CM}/2 = E_{nom}$. These final states are used to study the IDEA calorimeter standalone performance in jet reconstruction. 
For each tower, we build a 4-vector with the direction obtained by linking the geometrical center of the tower to the IP and the energy measured as the signal calibrated at the electromagnetic scale, while the mass is set to zero. We neglect towers for which the energy deposition measured at the electromagnetic scale is below $10$ MeV. The 4-vectors are used as input to the FASTJET package~\cite{fastjetusermanual}, version 3.3.2. We use the generalized $k_t$ algorithm with $R=2\pi$ and $p=1$ and force the number of jets to be just two. Note that with these parameters the clustering sequence is identical to the one often referred to as the Durham algorithm. Details on the event selection and inter-calibration of the two signals are given in Ref.~\cite{Pezzotti:2021jxh}. We observe that for $\chi$ in the range $\chi=0.41-0.43$, the value yielding a linear response for single $\pi^-$, the calorimeter exhibits a similarly linear response for jet energy measurements, whereas it over(under)estimates the energy for higher(lower) values of $\chi$, respectively. 
To evaluate the energy resolution, we fit  the distribution of $(E_j^r - E_j^t)/E_j^t$, obtained with $\chi=0.43$ to a gaussian. The sigma of the fitted gaussian is taken as an estimator of the jet energy resolution. The measured relative resolution as a function of the reciprocal of the square root of the jet energy  suggests a resolution compatible with
\begin{equation}
    \frac{\sigma}{E} = \frac{38\%}{\sqrt{E}}
\end{equation}
with $E$ expressed in GeV units. As reported in Section~\ref{sec:crystal_cal}, this is in good agreement with what was found in a hybrid dual-readout crystal/fiber calorimeter, \textit{i.e.} $\sigma/E=5\%$ for jet energies around $50$ GeV. 
Such a standalone energy resolution translates into peaks for the $Z$ and $W$  boson masses from 2-jets decays for which  the instrumental resolution is comparable to the natural width (details are given in \cite{Pezzotti:2021jxh}). Indeed, a good discrimination power between the peaks of the $W$ and the $Z$ can be observed in Fig.~\ref{2jbosondecays}. This is likely the most stringent requirement for hadronic calorimetry at future $e^+e^-$ colliders.

\subsection{The performance of an alternative geometry of the dual-readout calorimeter}

The calorimeter performance was assessed for electromagnetic particles, single hadrons, hadron jets and position measurement with the configuration that the tower length was 2.5\,m long.
This contains hadron showers above the 99\% level in the longitudinal direction.
The scintillation and \v{C}erenkov fibers are alternately interleaved in the copper. 
The diameter of both fibers is 1\,mm. 
A photosensor is attached at the end of each fiber and mimics the photon detection efficiency of the HAMAMATSU S13615-1025, which has a peak efficiency of 25\%.
The distance between the centers of the scintillation and \v{C}erenkov fibers is set to 1.5 mm, chosen in order to minimize the sampling fluctuations.
The sampling fraction of this structure for minimum ionizing particles is 4.6\%, and the sampling fluctuations, independently in the scintillation and in the \v{C}erenkov channel, amount to 12.6\%. 
The combination of both channels reduces it to 8.9\%.
In this type of calorimeter, the simulation of light produced in the scintillation and \v{C}erenkov fibers are crucial to predict the performance.
For the realistic simulations in GEANT4, all parameters for the two types of fibers were configured with information such as the light attenuation lengths, refractive indices, fiber geometries provided from KURARY (SCSF-78) and Mitsubishi (SK40).
In GEANT4 we simulated the reflection on the boundary surface, photon transportation, light absorption, numerical aperture, Poisson fluctuations, and all other physics processes related to optical photons.
Under this simulation environment, both scintillation and \v{C}erenkov channels of each tower were calibrated with electrons of 20\,GeV. As a result, both channels have the same response to electromagnetic particles.

\begin{figure}[hbtp]
\centering
\includegraphics[width=0.95\textwidth]{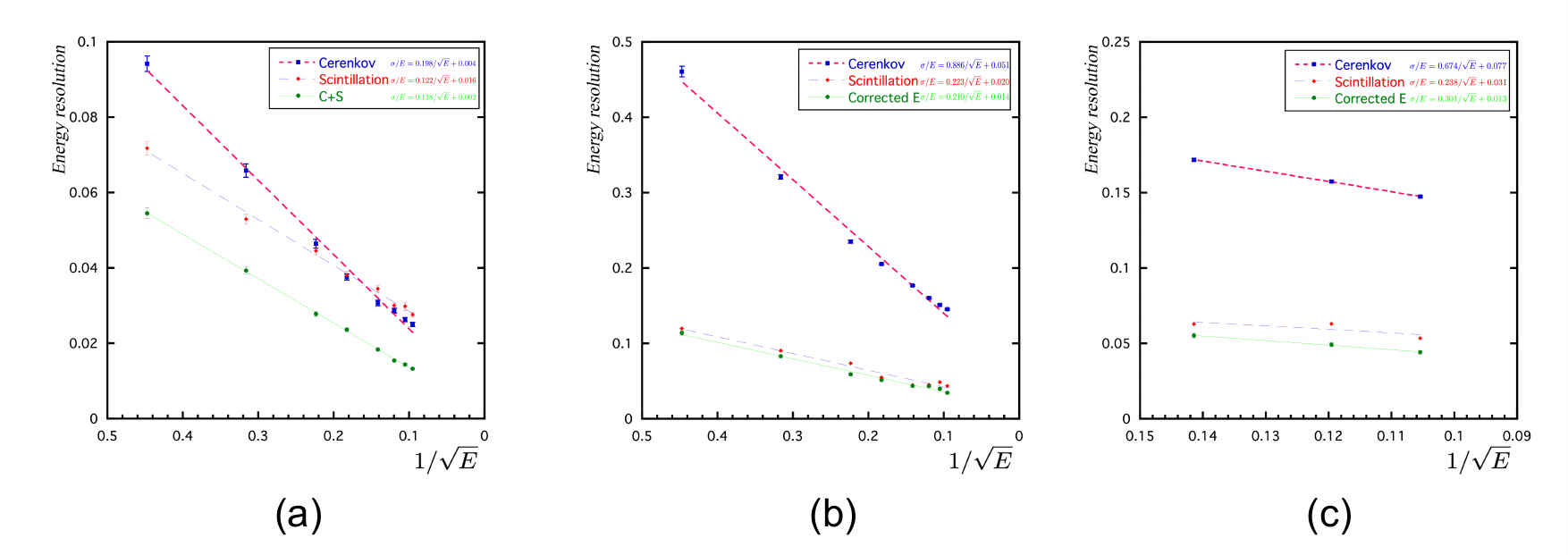}
\caption{The predicted energy resolutions for electrons (a), pions (b), and jets (c) as a function of $1/\sqrt{E}$ for the RD52 $4\pi$ dual-readout calorimeter.}
  \label{fig:E_res}
\end{figure}

Firstly, electrons of various energies were simulates in the calorimeter to derive the energy resolution for electromagnetic particles, as shown in Fig.~\ref{fig:E_res}(a). The stochastic term for the \v{C}erenkov channel is 19.8\% with a small constant term, whereas the constant term of the scintillation channel is 1.6\%, and its stochastic term is 12.2\%. 
In the combination of both channels, the fitting result suggests $11.8\%/\sqrt{E}$ with a constant term of 0.2\%. Combining the scintillation and \v{C}erenkov channels improves the energy resolution of the electromagnetic particles in both stochastic and constant terms. 
Secondly, the simulation studies for hadrons and jets show the essential property of the dual-readout calorimeter. 
According to the study about the limits of the hadronic energy resolution of Ref.~\cite{Lee:2017oye}, the dual-readout method for a calorimeter based on Cu may reach about $12\%/\sqrt{E}$. 
The hadronic performance of the calorimeter was estimated with pions of 5, 10, 20, 30, 50, 70, 90, and 110 GeV in energy.
Blue and red markers in Fig.~\ref{fig:E_res}(b) show the energy resolutions for the scintillation and \v{C}erenkov channels, which are $22.3\%/\sqrt{E}\oplus 2.0\%$ and $88.6\%/\sqrt{E}\oplus 5.1\%$ respectively. 
Correcting the pion energy using the dual-readout method formula of Eq.~\ref{eqn:dual}
yielded an energy resolution for pions of $21.9\%/\sqrt{E}\oplus 1.4\%$. 
In the Higgs factory, the jet energy resolution is crucial for studying hadronic decay modes of the Higgs boson without ambiguity among signals and backgrounds. 
Thus, we studied the precision of the calorimeter in the jet energy measurement. 
The particle gun of Pythia generated quark and anti-quark pair (a mixing of $u\bar{u}$ and $d\bar{d}$) events and hadronized them, the resulting cascade was used as input for the GEANT4 calorimeter simulation. 
The particle gun was adjusted for each quark to have energies of 50, 70, and 90 GeV. 
The anti-kt algorithm, with a cone radius of 0.8, clustered energy deposits to reconstruct jets. Fig.~\ref{fig:E_res} (c) refers to the jet energy resolutions, which show $67.4\%/\sqrt{E}\oplus 7.7\%$ for the scintillation and $23.8\%/\sqrt{E}\oplus 3.1\%$ for the \v{C}erenkov channel, respectively. 
The dual-readout correction improves the jet energy resolution to $30.1\%/\sqrt{E}\oplus 1.3\%$.

\begin{figure}[hbtp]
\centering
\includegraphics[width=0.4\textwidth]{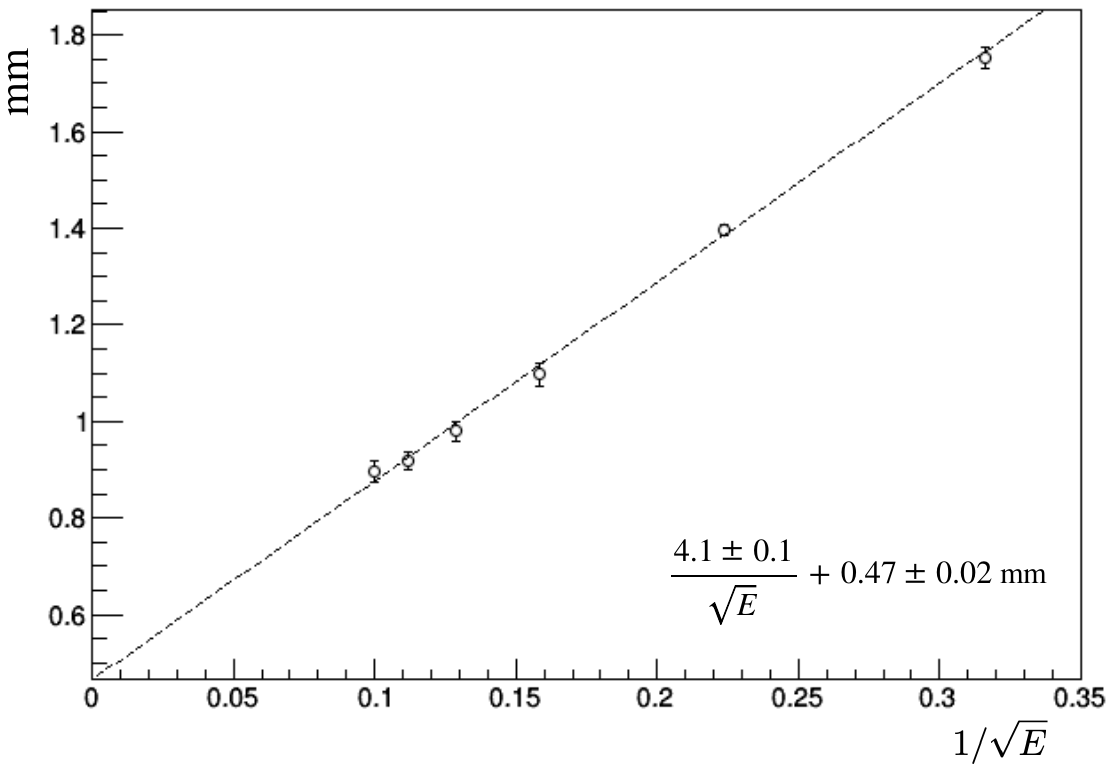}
\caption{The predicted position resolution for electrons as a function of $1/\sqrt{E}$ when the signal from each fiber is read out by one photosensor.}
  \label{fig:pos_res}
\end{figure}

In this design, one photosensor was connected to each fiber, exploiting the maximum possible granularity of the calorimeter. 
The position resolution was estimated by using a particle gun generated with a beam spot of 1 cm $\times$ 4 cm.
Electron beams with energies of 10, 20, 40, 60, 80, and 100 GeV were shot at the center of one tower, then gradually moved towards the neighbor tower. 
The position of the electrons was calculated using the center of gravity method. 
The position resolution was derived from the distribution drawn with the difference between the electron's position before entering the calorimeter and the reconstructed one.
Fig.~\ref{fig:pos_res} shows the position resolution of this detector for electromagnetic particles, estimated to be $4.1/\sqrt{E}\oplus 0.47~{\rm{mm}}$.
\begin{figure}[hbtp]
\centering
\includegraphics[width=1\textwidth]{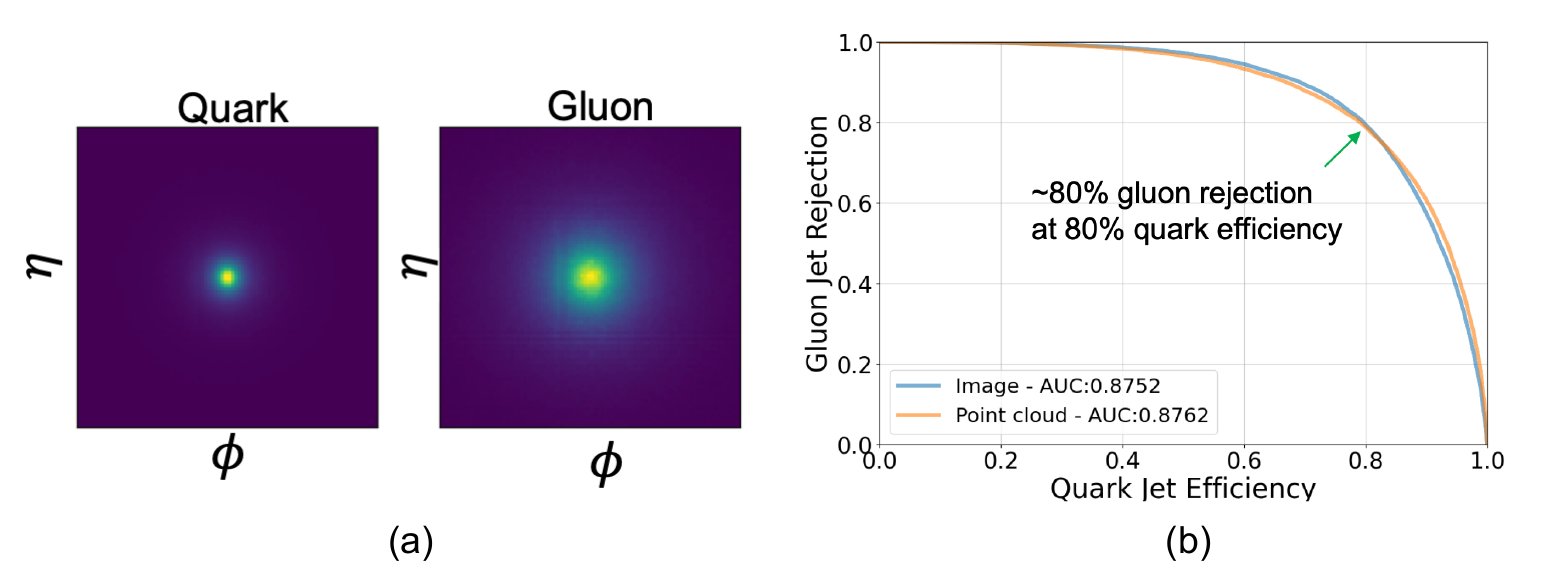}
\caption{Quark and gluon jet identification based on machine learning. Shown is the average image of quark and gluon jets reconstructed with the scintillation fibers (left), and the quark jet identification efficiency versus the gluon jet rejection (right).}
  \label{fig:I_ML}
\end{figure}

With a fiber-sampling structure, a dual-readout calorimeter can be regarded as a large 2D image sensor. 
Individual fibers read energy deposits produced by electromagnetic or hadron showers, and the different levels of the energy deposits can be converted to an image in the $\eta$-$\phi$ space as shown Fig.~\ref{fig:I_ML}. 
Therefore, we can expect electromagnetic particles, hadrons, and jets to produce different images and utilize machine learning algorithms to classify them. 
We tested how well this technique discriminates between quark and gluon jets. 
The particle gun in Pythia generated back-to-back di-jet events of quarks and gluons which were given as input to GEANT4 to simulate the detector response.
Fig.~\ref{fig:I_ML}(a) shows the images of the average energy deposited by a quark and gluon jet measured with the scintillation fibers. 
Due to the larger color factor, gluon jets have larger particle multiplicities which lead to broader energy deposits than quark jets in the $\eta$-$\phi$ plane. 
The ROC curve in Fig.~\ref{fig:I_ML}(b) shows the quark jet identification efficiency versus the gluon jet rejection power. 
For example, the image classification through machine learning can identify quark jets with a 80\% efficiency with a 20\% of gluon jet misidentification.

\subsection{Dual readout with crystal calorimeters}
\label{sec:calvision}
\label{sec:crystal_cal}

Recently a consortium of US Universities and National laboratories (Oak Ridge, FNAL, Argonne, Maryland, Princeton, Virginia, Texas Tech, Caltech, Michigan, MIT, Purdue) has revived the RD52 research on dual readout of homogeneous crystals.  A crystal EM calorimeter  used in conjunction with a spaghetti-type dual-readout hadron calorimeter  could have both excellent resolution for electrons and photons and state-of-the-art hadron and jet resolutions.  The proposal was inspired by Ref.~\cite{Lucchini:2020bac}, which proposed a detector called {\it SCEPCAL}.   Such a calorimeter could be built on the time scales of future Higgs factories such as ILC, FCC-ee, and the muon collider. Collaboration members are also interested in longer term dual-readout R\&D, which will be discussed in Sections \ref{sec:moreInfo} and \ref{sec:bluesky}.

Homogeneous dual-readout scintillating crystal calorimetry was abandoned by the DREAM/ RD52 collaboration due to limitations on the available instrumentation at the time.  The readout was assumed to be via (expensive) photomultipliers (PMT).  Because of the cost, the calorimeter would have to have long crystals with single or, at best, twofold readout. PMTs at that time had good quantum efficiency over a only limited range of wavelengths, with poor sensitivity in the red.  Because of this, the efforts focused on \v{C}erenkov light with UV wavelengths below the scintillation peak where self-absorption, especially in long crystals, is large. 
In order to limit contamination from the large scintillator signal in the \v{C}erenkov region, the scintillator light yield had to be reduced to a level that adversely affected the EM resolution\footnote{For an EM calorimeter, the scintillation light needs to produce O(1k) photoelectrons per GeV to retain the excellent resolutions for photons and electrons.}.  As a result, at that time, the EM performance of the combined crystal and spaghetti calorimeter was no better than that of the spaghetti calorimeter alone.

\begin{figure}[hbtp]
\centering
\includegraphics[width=0.53\textwidth]{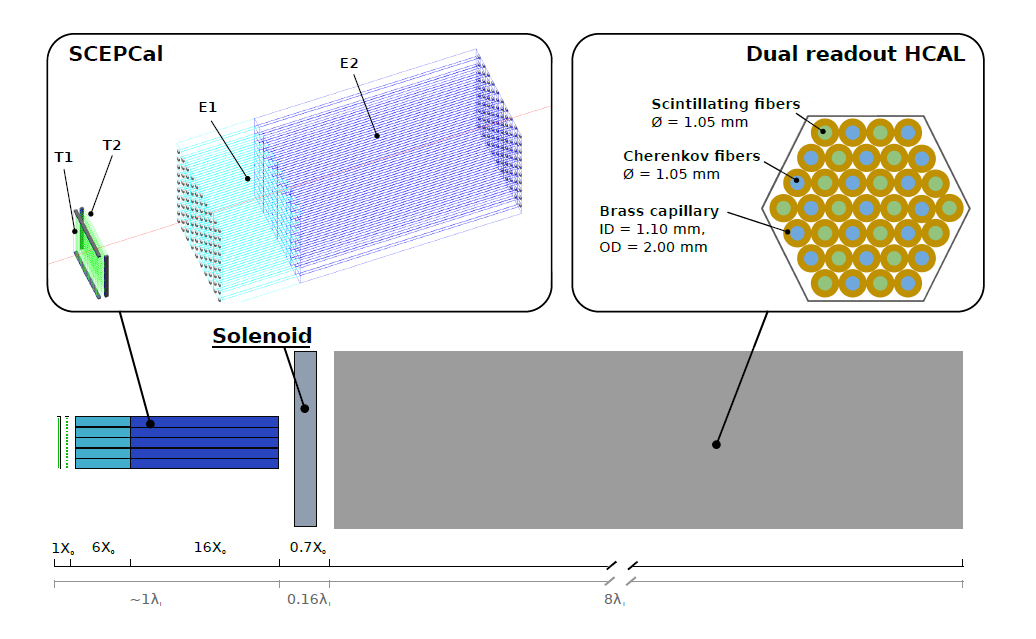}
\includegraphics[width=0.45\textwidth]{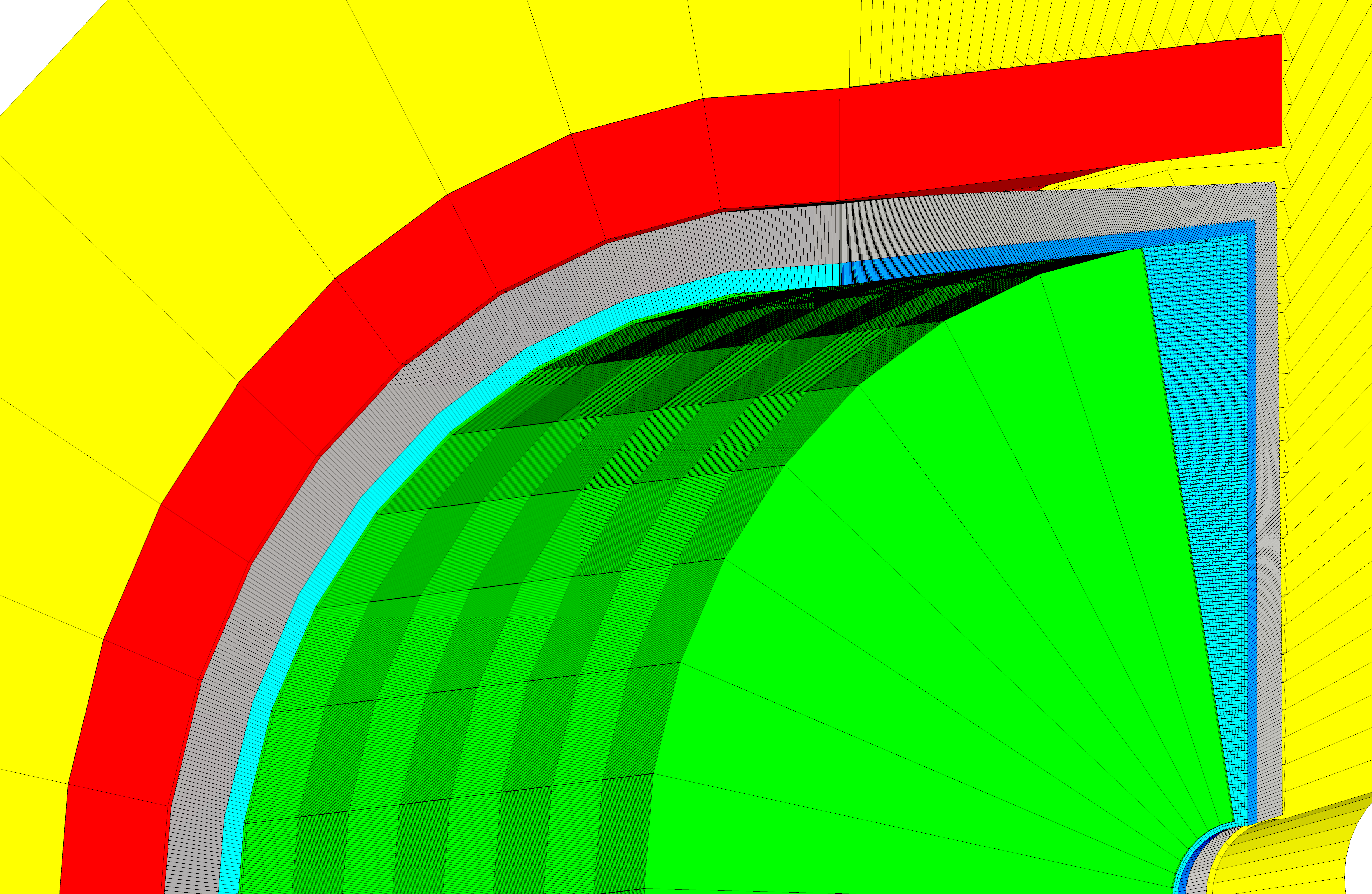}
\caption{
[left] A segmented crystal calorimeter integrated into a collider detector. The precision timing layers (green) are followed by projective crystals longitudinally segmented into front (blue) and rear (purple) compartments. 
[right] SCEPCal embedded in a full detector including  a solenoid (red) in the barrel region and hermetically enclosed by a dual-readout fiber calorimeter (yellow).
}
  \label{fig:scepcal}
\end{figure}

A recent re-examination, based on simulations of the potential of a homogeneous dual-readout EM calorimeter with a spaghetti-type dual-readout HCAL in the light of new technologies, is reported in Ref.~\cite{Lucchini:2020bac}. The ``strawman'' EM calorimeter has two longitudinal sections (E1 and E2) and fine granularity in $\eta-\phi$ space while the HCAL is composed of brass capillary tubes, as shown in Figure~\ref{fig:scepcal}~[left]. 
The use of shorter and narrower crystals reduces the effect of self-absorption of the UV light and enables shower imaging. Smaller crystals are also easier to grow than long crystals and thus would be less expensive. Newly developed wavelength-extended SiPMs open the possibility to efficiently detect \v{C}erenkov light in the red region of the spectrum, above the scintillation peak, where self-absorption is not a problem. In the baseline design, a single SiPM would be used for light readout at the entrance face of E1, while E2 would have two SiPMs at its rear end: one with enhanced UV sensitivity for scintillation light detection with an optical filter passing light with $\lambda<$500 nm, the other with enhanced red sensitivity and a filter passing $\lambda>550$ nm for \v{C}erenkov light detection (for \pwo crystals, other crystals such as BGO or BSO would require slightly different configurations).
Studies in Ref.~\cite{Lucchini:2020bac} show that excellent hadron resolution can be achieved with dual-readout only on E2.
Having no dead material between E1 and E2 reduces the constant term in the resolution. A thin solenoid is placed between the EM calorimeter and the spaghetti dual-readout hadronic calorimeter. The calorimeter could be complemented with a precision m.i.p. timing layer in front of E1, made of {\it e.g.} LYSO crystals, to allow precise particle identification for the $b$ physics program at high intensity running at the $Z$ pole and to remove beam backgrounds at muon colliders. 
A preliminary implementation of the segmented crystal calorimeter geometry with the IDEA detector~\cite{ALY2020162088} is shown in Figure~\ref{fig:scepcal}~[right].

Previous calorimeter systems with a crystal EM calorimeter and a conventional sampling hadron calorimeter, such as the CMS experiment, have had very poor hadronic resolution due to the mismatch between the ratio of their responses to the electromagnetic and  hadronic portions of a hadron-initiated shower (e/h). For crystals without dual readout, e/h is about 2, while for typical sampling calorimeters based on plastic scintillator it is around 1.4~\cite{hcaltestbeam}.  Fluctuations in the portion of the showers in the EM and hadronic calorimeters, coupled with this difference in e/h, produced sampling terms near 100\%$/ \sqrt{E}$. However, with dual readout in the ECAL and HCAL to measure the electromagnetic portion of individual hadronic showers, the responses in each of the two calorimeters can be corrected to an e/h response ratio of 1, eliminating this source of broadening of the measurement resolution.  

\begin{figure}[!htb]
\centering
\includegraphics[width=0.495\linewidth]{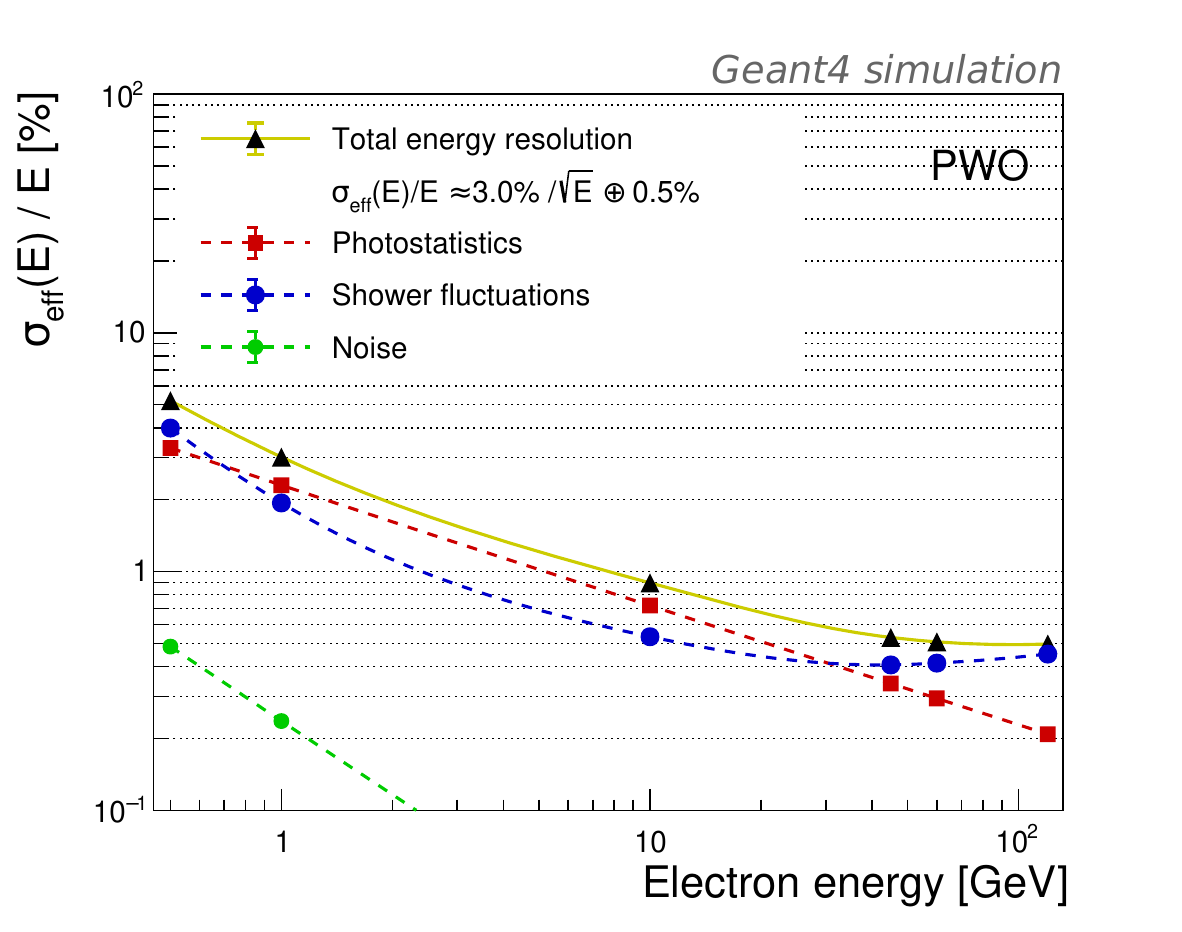}
\includegraphics[width=0.495\linewidth]{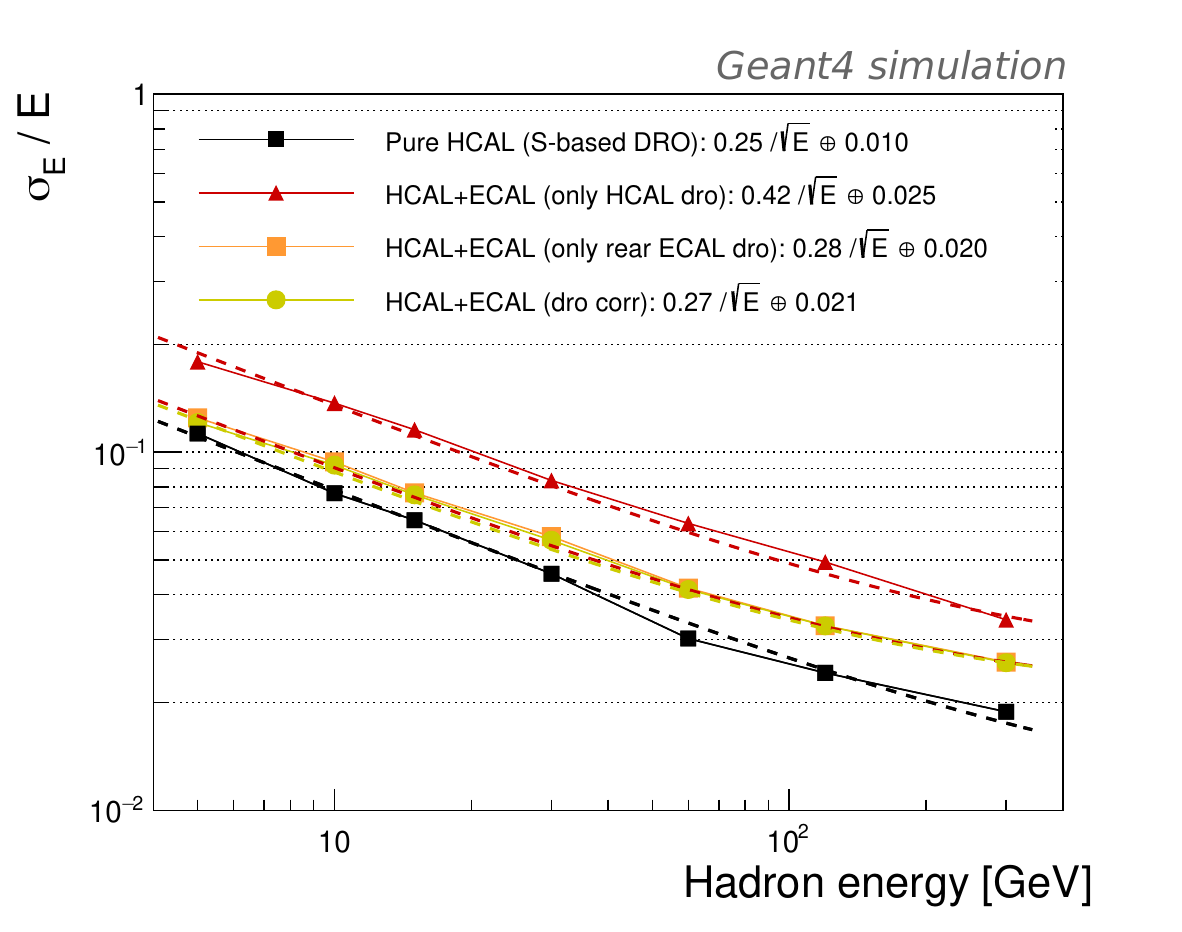}
\caption{Simulated resolutions for a combined dual-readout crystal ECAL and a dual-readout spaghetti HCAL from Ref.~\cite{Lucchini:2020bac}, for a pure dual-readout spaghetti, for that with a conventional crystal EM, and that with a dual-readout crystal EM calorimeter. Note that the energies of particles produced at electron-positron Higgs factories are mostly below 20 GeV, and so this is the most relevant part of the hadronic resolution. The average energy of a charged pion is 3 GeV~\cite{Lucchini:2020bac}.  On average, 13\% of the jet energy is from neutral hadrons~\cite{Lucchini:2020bac}. Shown are EM (left) and hadronic (right) resolutions.}
\label{fig:MarcoRes}. 
\end{figure}

Fig.~\ref{fig:MarcoRes} shows the predicted EM and hadronic resolutions for this calorimeter. The results indicate that the calorimeter system would have an excellent hadron resolution of 25\%$/ \sqrt{E}$. This is much better than that achievable with a high granularity calorimeter and, for particles in jets at Higgs factories in $ZH$ events, whose energies are predominantly less than 20 GeV, the resolution is comparable to a pure spaghetti dual-readout calorimeter. Because it is homogeneous, this calorimeter would allow a typical resolution for EM particles with energy of about half the $Z$ mass of $<$ 1\%, in contrast to current candidate EM calorimeters for future lepton machines which have typical resolutions of several percent, corresponding to a statistical term of the order of 15\%$/\sqrt{E}$.

While the results in Ref.~\cite{Lucchini:2020bac} are encouraging, the  use of red light has not been proven experimentally and should be targeted by dedicated R\&D studies. In addition, by utilizing light both below and above the scintillation peak, the resolution might be further improved.

\begin{figure}[htb]
\centering
\resizebox{0.40\textwidth}{!}{\includegraphics{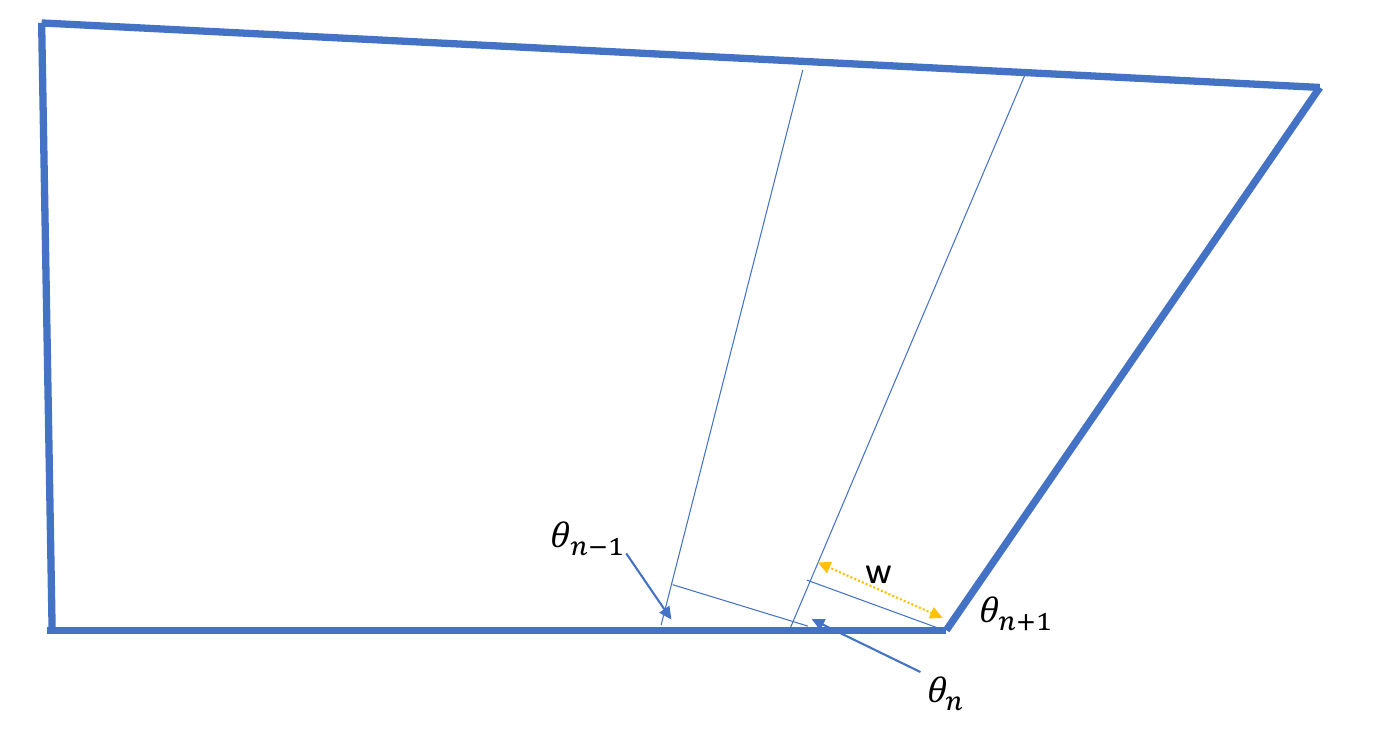}}
\caption{Cartoon of the projective geometry used in the cost estimate for the barrel.}\label{fig:BarrelGeoCartoon}

\end{figure}

Homogeneous calorimeters can be expensive. An estimate for the cost of the
SCEPCAL is given below.
Following the general lines of the IDEA detector design, the costing assumes a \pwo barrel EM calorimeter with an inner radius of 1.8\,m, and a length of 4.7\,m.  
The endcap calorimeter has an outer radius of 1.7\,m and an inner radius of 0.35\,m.  
The crystals are 1.0$\times$1.0$\times$20\,${\rm cm^3}$ and have two depth segments.
The barrel geometry is projective; a cartoon is shown in Fig.~\ref{fig:BarrelGeoCartoon}, with 180 $\theta$ segments in each half barrel and 1130 $\phi$ divisions.

The first depth segment has one SiPM while the second has two. 
The total number of barrel crystal towers is 442,000 with $3 \times 442,000$ SiPMs.  
For the endcap, the number of crystal towers is 174,000, with $3 \times 174,000$ SiPMs.
At a cost of \$8/cm$^3$ (an estimate from the Shanghai Institute of Ceramics), the total cost of the crystals is under 110\,M\$.

Assuming a cost per SiPM of \$5, and a cost per channel of \$4.5 for electronics, power, and monitoring, the per channel cost is \$9.5, corresponding to a total electronics cost of about 18\,M\$.  The total estimated cost of the EM calorimeter is then under 130\,M\$. 
Scaling the cost to the case of \bgo, with a 23\% greater crystal volume and a 14\% lower cost per cm$^3$ according to the SIC cost estimates, and larger crystals with an 11\% larger transverse size to match the \moliere radius, the overall of the EM calorimeter would be essentially the same, within several percent. 

The use of a dedicated electromagnetic crystal calorimeter section eases the design requirements on the spaghetti fiber HCAL. For instance to its sampling fraction can be reduced by increasing the brass tube diameter, resulting in a sizable reduction of the HCAL cost.
The cost for the spaghetti fiber HCAL, when using 2.5\,mm outer diameter brass tubes, as estimated by the IDEA collaboration, is 35\,M\$. (Such a tube diameter is larger than what should be used for a spaghetti-only calorimeter since in that case smaller tubes are needed to achieve the required electromagnetic resolution of around 13\%.) The EM + HCAL cost overall is thus comparable to the cost of alternative types of calorimeters, for instance based on Si-W technology.

\subsection{Particle Flow with dual-readout calorimeters}

\label{sec:lucchini}

The work in Ref.~\cite{lucchini2022particle} proves that advanced particle reconstruction techniques combined with the excellent electromagnetic energy resolution could open new avenues to precise jet measurements, beyond what is possible in a traditional high granularity calorimeter.
Precision EM resolution is essential for the correct assignment of photons from $\pi^0$ decays to jets. For $e^+e^- \rightarrow HZ \rightarrow 6$ jet events, the fraction of events with perfect photon-jet assignment goes from about 50\% for a calorimeter resolution with stochastic term of 30\% to about 70\% for a 3\% term. An EM resolution of a few percent can also significantly improve the missing mass measurement when the $Z$ decays electronically, by exploiting the recovery of bremsstrahlung photons. For an electron momentum of 45 GeV and for a tracker thickness of 0.4~X$_{\rm 0}$ (note: TPC-based trackers for future Higgs factors expect a thickness of 0.12  X$_{\rm 0}$), the resolution is about 0.6\% for an EM stochastic term of 3\%/$\sqrt{E}$, versus 2\% for 30\%$/\sqrt{E}$ (a relative improvement by a factor of more than 3).
In addition, precision EM resolution could aid in $\tau$ reconstruction and be useful for flavor physics studies at high luminosity running at the $Z$ pole.

This work has been extended recently by Princeton (Tully and Lucchini)~\cite{lucchini2022particle} using an implementation in the FCC-ee GEANT4-based simulation framework and a new particle-flow algorithm designed specifically to take advantage of its dual readout, excellent energy resolutions and transverse and longitudinal segmentation. While in the calorimeter design proposed in Ref.~\cite{Lucchini:2020bac,lucchini2022particle}, the longitudinal segmentation is much coarser than typical high granularity Si-W tungsten calorimeters, this deficit is partly compensated by the development of a dedicated particle flow algorithm where the excellent energy resolution for photons and neutral hadrons are emphasized. In addition the dual-readout information provides an additional handle for the algorithm in the step of matching calorimeter hits to charged tracks, as the energy of clustered hits can be corrected using the dual-readout method to yield a linear response to the energy deposits from charged (and neutral) hadrons.

\begin{figure}[!htbp]
\centering
\includegraphics[width=0.495\linewidth]{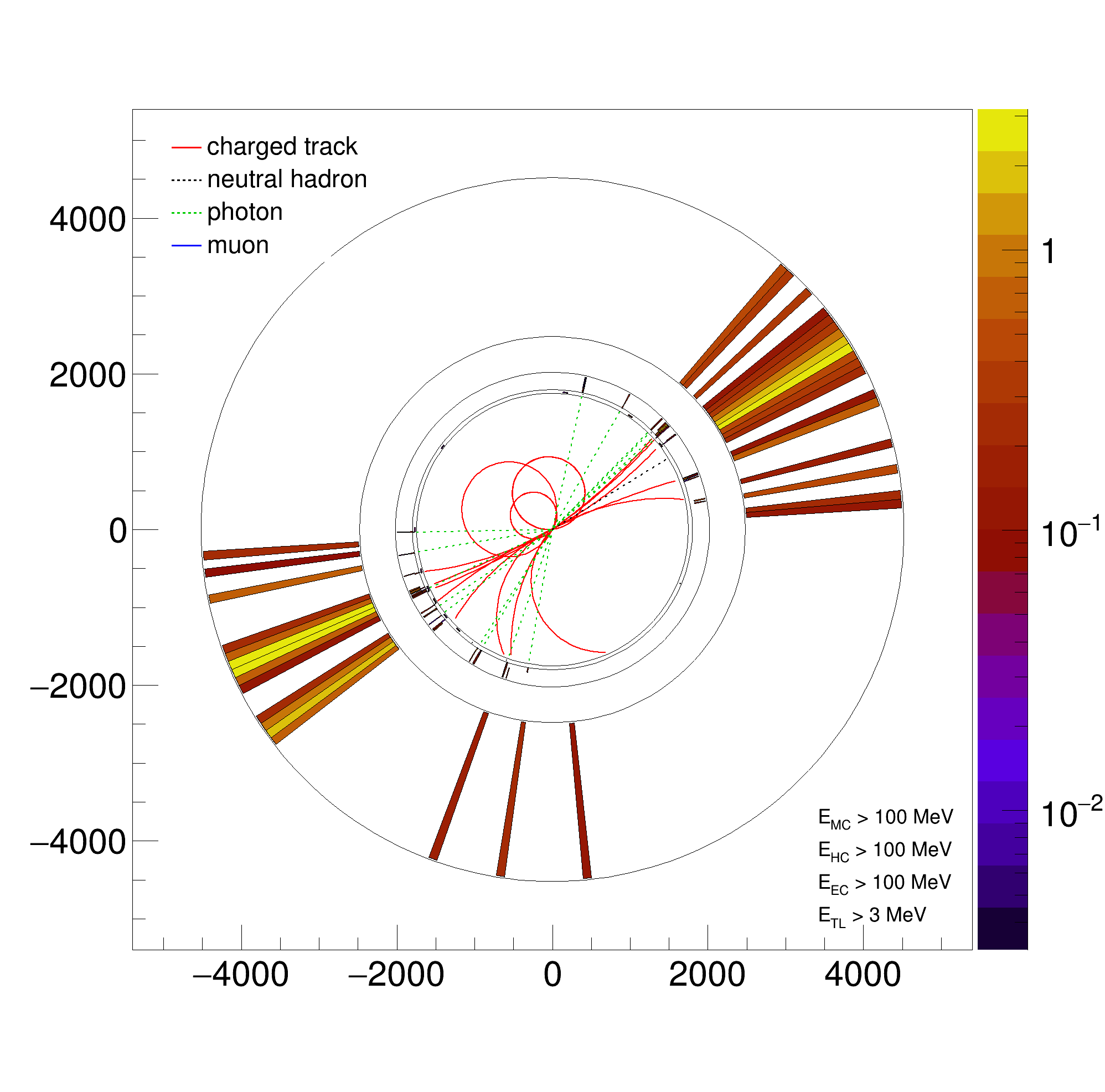}
\includegraphics[width=0.495\linewidth]{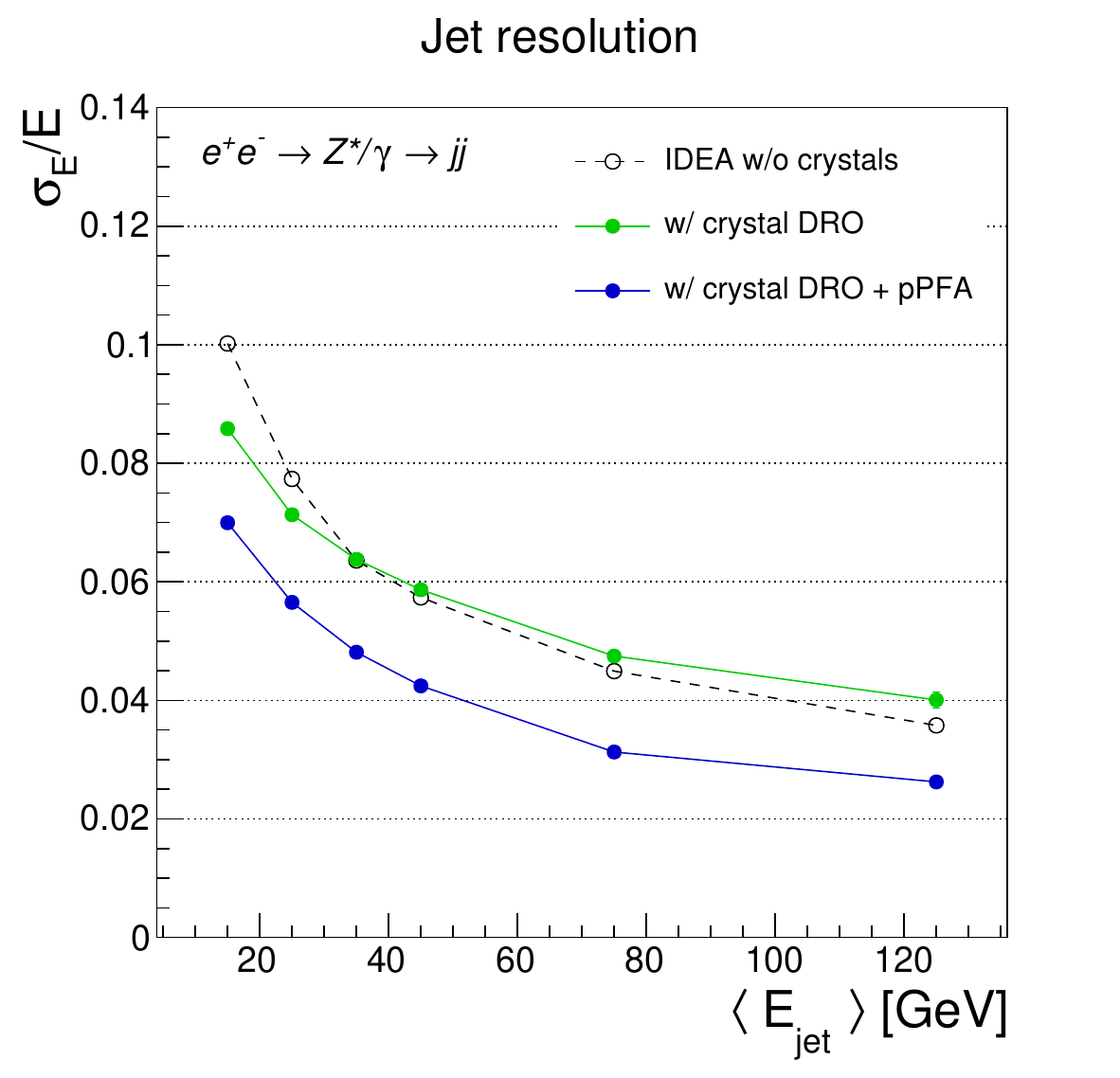}
\caption{ 
[left]
Typical $Z$ to dijet event, showing charged tracks in the tracker, hits in the electromagnetic calorimeter, and then hits in the spaghetti-type dual-readout section.
[right]
Jet energy resolution for a hybrid dual-readout calorimeter (specifically SCEPCal~\cite{Lucchini:2020bac}) with a simple custom particle flow algorithm.}\label{fig:lucchini_pf}
\end{figure}

A $Z\rightarrow jj$ event display  in Figure~\ref{fig:lucchini_pf} [left] shows (from inner to outer radius) generator-level charged particles (with simulated bending in a magnetic field) in a tracker, a crystal dual-readout electromagnetic calorimeter with  SCEPCal~\cite{Lucchini:2020bac} granularity, a dead space for the thin solenoid, and a dual-readout spaghetti fiber calorimeter.
Figure~\ref{fig:lucchini_pf} [right] shows the resulting jet energy resolution, indicating that outstanding jet resolutions may be possible, especially when the particle flow algorithm (pPFA in the figure) and dual readout are combined using a dedicated Dual-Readout proto-Particle Flow Algorithm (DR-pPFA). The jet energy resolutions are Gaussian and thus are reported without truncation\footnote{A caveat: the simulation does not include tracker material in front of the calorimeter (track resolutions are simulated using Gaussian smearing), and so part of our R\&D program is to simulate the results for the benchmark physics processes including a tracker.}. Jet energy resolutions at the level of 3-4\% for jets with energy above 50 GeV are achieved while still maintaining state-of-the-art measurements of electrons and photons.
 

\section{Future R\&D activities}
\label{sec:future}

\subsection{IDEA collaboration prototype plans}
\label{sec:ideafuture}
Unprecedented jet energy resolutions required by future electroweak factories seem to be reachable with a fiber-sampling Dual-Readout (DR) calorimeter, as described in Sec.~\ref{sec:idea}. Although the dual-readout principle has been experimentally proven with several beam tests, many technical problems are still open, and dedicated R\&D is needed to build a hadronic-size prototype in order to check and validate the GEANT4 simulation and experimentally demonstrate the performance.

The DREAM/RD52 project explored methodology for building large fiber-sampling detectors, tweaking the \v{C}erenkov light yield and choosing an appropriate absorber in order to achieve a resolution of about $30\%/\sqrt{E}$ for single hadrons and jets.
However, the effective radius of the DREAM calorimeter was 0.81 $0.81\,\lambda_{int}$, and the RD52 fiber calorimeter laterally contained on average only ~93\% of the hadron shower initiated by a 60\,GeV $\pi^{-}$. 
Fluctuations of the lateral shower leakage were the dominant contribution to the hadronic energy resolution of these detectors. 
 
To overcome this limitation, two complementary projects have been proposed and approved by Italian and Korean National Funding Agencies, respectively.
The projects aim at designing, constructing, and qualifying longitudinally unsegmented, highly granular, fiber-sampling DR calorimeter prototypes to assess:
\begin{itemize}
    \item a stand-alone hadronic resolution around $30\%/\sqrt{E}$ or better, for both single hadrons and jets, while maintaining a resolution for isolated electromagnetic showers close to $10\%/\sqrt{E}$;
    \item a transverse resolution of $O(1  {\text{mrad}})/\sqrt{E}$;
    \item a longitudinal resolution of a few cm through timing measurements;
    \item a modular and scalable construction technique;
    \item an innovative readout architecture based on SiPMs.
\end{itemize}

\subsubsection{HiDRa calorimeter}
The project approved by INFN is called HiDRa: High-Resolution Highly Granular Dual-Readout Demonstrator. The target is to build a Hadronic-Scale Prototype (HSP) made of 16 Highly Granular Modules (HGM), $\sim 13\times13\times200\ {\text{cm}}^3$ each, where the two core modules are equipped with SiPMs and the rest with PMTs. 

The project started January 2022, and it will be completed in three years. The first year will be mainly devoted to both the detector material choice and the construction technique implementation. The following 16 months will be used for the calorimeter construction, while the last 6 months are reserved for the qualification of the prototype with test beams.

Brass or stainless steel are, at present, the baseline options for the absorber material due to their mechanical and shower development properties. Brass capillary tubes were used in 2020 for the construction of an EM-size prototype, $\sim 10\times10\times100\ {\text{cm}}^3$. Commercial 1\,m long capillary tubes with 2\,mm outer diameter have been arranged and glued together to build nine individual modules, assembled in a 3×3 structure that constitutes the final prototype. Thanks to the excellent capillary mechanical precision, the construction procedure is quite simple and fast. The tubes are stacked, one layer after the other, inside a guiding fixture that defines the boundary geometry. This method will be scaled up to build the HGMs forming the HSP.

Solid-state sensors allow for a significant step toward high granularity in fiber-sampling calorimetry. Because of their different properties, S and C signals may be better exploited by using different sensor choices, with either yellow-tuned (for S) or UV-enhanced (for C) sensitivity. A linear response up to thousands of photoelectrons is also required to allow analog signal grouping. Experience with Hamamatsu SiPMs has been highly positive, but progress is needed in order to have sensors with a wider dynamic range (10 or 15 {$\mu$}m pitch), reduced dead area (in a compact SMD package to allow the one-to-one connection with fibers), and affordable cost. Options from different vendors will be investigated.

Plans for studying a long-prospective custom-designed sensor based on a digital SiPM architecture, which would greatly reduce the readout complexity, are also being discussed. The operation of a large number of SiPMs poses a series of system integration challenges due to the reduced space available on the back of the calorimeter, the number of channels, and the cost. The optimal solution would use a custom SiPM with on-board intelligence, motivating the part of the project that deals with the dSiPM evaluation and their possible exploitation.

ASICs integrated into a scalable architecture are needed to investigate system issues while assessing the HSP performance. Both charge integrators and waveform samplers with Feature EXtraction (FEX) are available on the market and the project aims at evaluating several of them.

\subsubsection{Plans of the Dual-Readout Calorimeter R\&D team in Korea}

 {\it High-quality measurements in energy and position}
 
 The project to build a large detector was approved with a research grant of the National Research Foundation of Korea (NRF). 
 The objective is to build a calorimeter prototype with dimensions of $36~{\rm{cm}} \times 36~{\rm{cm}} \times 250~{\rm{cm}}$, consisting of 16 Cu-fiber modules. 
 The size of one module is $9~{\rm{cm}} \times 9~{\rm{cm}} \times 250~{\rm{cm}}$. 
 The lateral containment of the calorimeter is expected to be, on average, 97.5\% of hadron showers initiated by a 60\,GeV $\pi^{-}$. 
 The depth is equivalent to 10\,$\lambda_{int}$, resulting in a longitudinal shower containment above 99\%.
 Moreover, since the \v{C}erenkov light yield affects the hadronic performance, the goal is to collect 100 or more \v{C}erenkov photoelectrons per GeV. 
 SiPMs offer high photon-detection efficiencies. 
 Tests of SiPMs with high sensitivity to the \v{C}erenkov light will be a part of the R\&D plan. 
 Finally, copper is the preferred choice for the absorber material due to its $e/mip$ ratio being close to one, as described in Section~\ref{sec:ideacalorimeter}. 
 This is expected to play an important role in achieving the jet energy resolution of $30\%/\sqrt{E}$ since it has a linear response to low-energy hadrons.
The precise position measurement depends on both the energy resolution and the granularity of the calorimeter. 
SiPMs give the liberty for designing any granularity of the calorimeter because they are manufactured with various sizes and high photon-detection efficiencies. 
A collaboration with a Korean electronics company has been established to fabricate customized readout boards and electronics for highly granular detectors read out with SiPMs.
Recently, the company produced 13 customized readout boards for SiPMs.
An 8~x~4 array of SiPMs is placed on each board.
A total of 416 SiPMs are instrumented to detect scintillation or \v{C}erenkov photons from all individual fibers.
The dimension of each SiPM is $1.3~{\rm{mm}} \times 1.3~{\rm{mm}}$, including photosensor area, cathode and anode.
Signals from SiPMs are fed to the customized electronics, which process waveform and ADC information.
This system is designed to target a time resolution of 50 ps.
The Cu-fiber modules equipped with the readout system will be tested with beams.

{\it Engineering of the module production}

\begin{figure}[hbtp]
\centering
\includegraphics[width=0.6\textwidth]{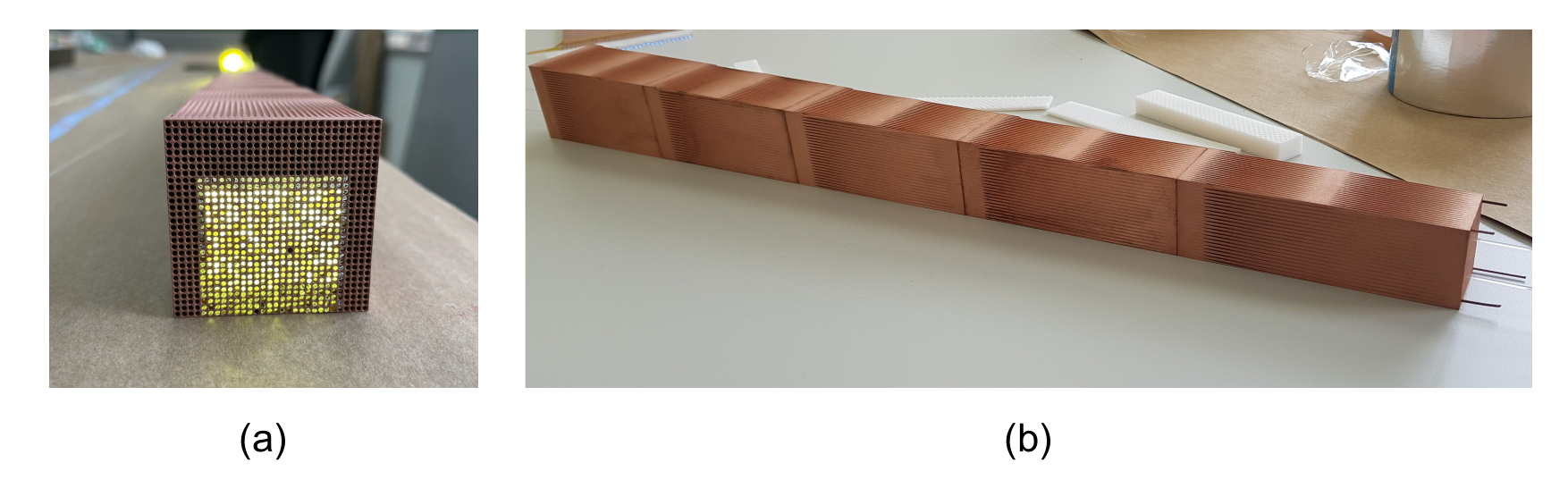}
\caption{The projective geometry of a Cu module whose length is 50 cm. This module was produced with 3D printing.}
  \label{fig:3d_printing}
\end{figure}

Forming the Cu-fiber structure described in Section~\ref{sec:ideacalorimeter} is difficult due to the hardness of copper.
For the RD52 modules, the Cu-fiber structure was done by cutting plates made of 99.5\% Cu with 0.5\% of Te. 
This method worked properly but at the price of ~50\% of the Cu going to waste.
Thus, other endeavors such as 3D printing and skiving are being tested to find more efficient methods. 
3D printing allows the manufacture of copper blocks including holes of 1\,mm diameter interleaved with 0.5\,mm thick absorber, as shown in Fig.~\ref{fig:3d_printing} (a). 
Optical fibers are inserted in these holes. 
In addition, a 50\,cm long module with projective geometry was successfully fabricated, as shown in Fig.~\ref{fig:3d_printing} (b). 
3D printing works nicely, but currently the costs are considerably high. 
To reduce the costs, a different method was proposed, skiving.  
The work has been started, with a skiving company in Korea, to produce the Cu structure with high accuracy and at a relatively low cost. 
The Cu structures produced by skiving are used to investigate efficient fiber stacking methods for mass module production. 
This entails putting fibers and Cu wires or shims alternately between copper sheets. The trials are presently ongoing.

\subsubsection{Machine learning algorithms}

As shown in Section~\ref{sec:ideacalorimeter}, the highly granular fiber dual-readout calorimeter can provide machine-learning-friendly information like scintillation and \v{C}erenkov signals from all individual fibers, signal starting time, analog waveforms, etc. 
Both raw detector signals and the processed higher-level variables can be used to train a machine learning network for particle and jet identification in the longitudinally unsegmented fiber calorimeter. 
Processed higher-level variables may refer to, for example, shower shape, the ratio of the amplitude to the integrated charge of the waveform, and 2D calibrated energy deposit image.
The development of machine learning for particle and jet identification is underway. 
Furthermore, since quantum computing will be an inevitable technology in the future, developing machine learning associated with quantum computing is also in the pipeline.

\subsection{Longer range dual-readout improvements}

In the last two decades, our understanding of processes that are at the heart of calorimetry has vastly improved
\cite{AKCHURIN2005537,Sefkow:2015hna,Sirunyan_2017}, and the avenues for further enhancements have also become clearer.  In this section, we present three ideas that are likely to bring additional capabilities to dual- or multi-readout calorimetry.  First, we consider longitudinal segmentation of fiber calorimeters by timing.  This approach renders otherwise longitudinally unsegmented fiber calorimeters highly segmented in depth and enables the use of powerful features of neural networks in energy reconstruction.  Second, we assess the advantages of structured fibers in spaghetti calorimetry.  The \v{C}herenkov polarization as a discriminant in the separation of scintillation and \v{C}herenkov light is discussed as the third item at the end of this section.

\subsubsection{Longitudinal Segmentation by Timing}
\label{sec:longitudinal}

\begin{figure}[H]
\begin{center}
\resizebox{12cm}{!}{\includegraphics{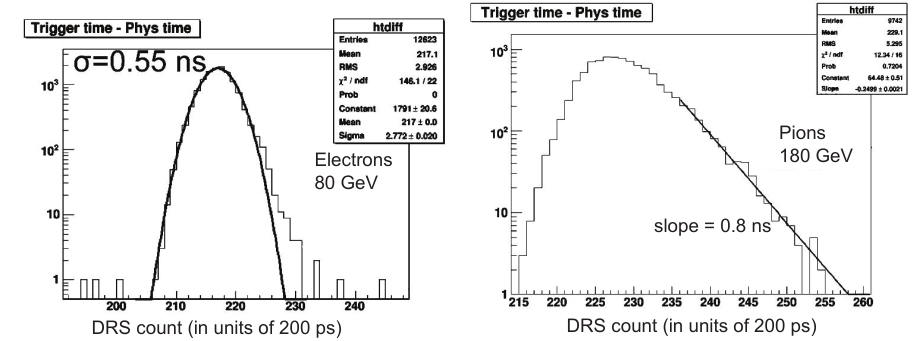}}\vspace{-4mm}
\caption{\small The starting time of signals from 80 GeV electron (left) and 180 GeV pion (right) showers in the DREAM prototype, measured with the DRS readout. The readout was triggered by the passage of a beam particle through the upstream trigger counters.}
\label{fig:DREAMtimingprecision}
\end{center}
\end{figure}

Figure~\ref{fig:DREAMtimingprecision}(a) shows the level of timing precision that was achieved in a 80\,GeV electron beam using the original DREAM prototype: the distribution of the time difference between the calorimeter PMT signals and the trigger counters upstream. This difference was obtained by measuring the start of the calorimeter signals with the Domino Ring Sampler (DRS) chip with a simple threshold, where the waveform sampling was started by the trigger signal.  We observed that a time resolution of 0.55\,ns was achievable with this crude approach. As expected, the same measurement for 180\,GeV pions yielded quite a different time distribution, as shown in Figure~\ref{fig:DREAMtimingprecision}(b). The slope of 0.8\,ns reflected the fact that the light was produced at a depth $z$, which goes like $\exp{(-z/\lambda_{\rm int})}$.  The depth at which the light was produced could be determined with a precision better than 20\,cm since the time resolution was 0.55\,ns and the interaction length was $\sim$25\,cm in this detector.  A factor of 10 improvement (ambitious but likely achievable) would mean a 2\,cm effective longitudinal segmentation.  It is worthwhile to note that there is a significant (measured) difference between the distribution of the starting time of pion and electron induced showers, as shown in Figure~\ref{fig:DREAMtimingprecision}. 

The same idea was also explored in other ways (reported in \cite{Akchurin_2015}) using the information from the scintillation counters enveloping the calorimeter and from the lateral displacement of the center-of-gravity of the calorimeter signals with respect to the coordinates measured in the wire chambers upstream of a calorimeter. Experimental data clearly showed that timing information provides sufficiently precise determination of ``center-of-light'' inside the calorimeter.  

Precisely reconstructing the arrival time of the photon is important for achieving better precision for depth segmentation.   The ``slow'' response time from the photosensor ({\it e.g.} SiPMs for the next generation calorimeters), the overlapping signals from multiple optical photons with different arrival times, and the presence of electronics noise present a challenge.

\begin{figure}[H]
\begin{center}
{\includegraphics[width=0.98\textwidth]{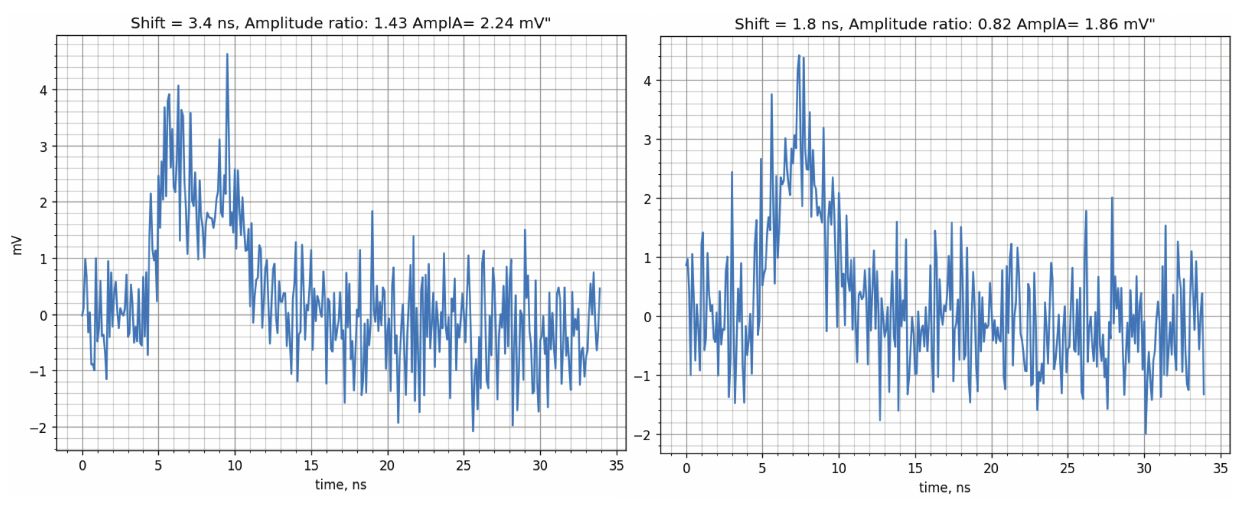}}\vspace{-5mm}
\caption{\small 
Examples of raw signals from the readout electronics where the time separation is 3.4 ns, average amplitude 2.24 mV, and amplitude ratio 1.43 (left). The time separation is 1.8 ns, average amplitude 1.86 mV, and amplitude ratio 0.82 (right). The RMS noise is 0.78 mV.
}
\label{fig:SenSLPulseShapeTwoOfThem}
\end{center}
\end{figure}

We investigated how precisely the arrival time of the photon at the SiPM can be reconstructed based on the raw signal from the readout with the AARDVARC-V3 sampler (running at 10\,GS/s).  We chose recurrent neural networks (RNN) for the signal reconstruction and evaluated the time separation of two low-amplitude signals. Examples of raw signals from the readout electronics are presented in Figure~\ref{fig:SenSLPulseShapeTwoOfThem}.  The single-photon signal amplitude was 2\,mV, and the electronics noise RMS was 0.78\,mV in this configuration. The arrival time difference in the range of 0 to 2\,ns was probed for signals with amplitude between 2 and 40 \,mV.  The amplitude ratio of the two photons was allowed to vary between 0.5 and 2. 

\begin{figure}[H]
\begin{center}
{\includegraphics[width=0.98\textwidth]{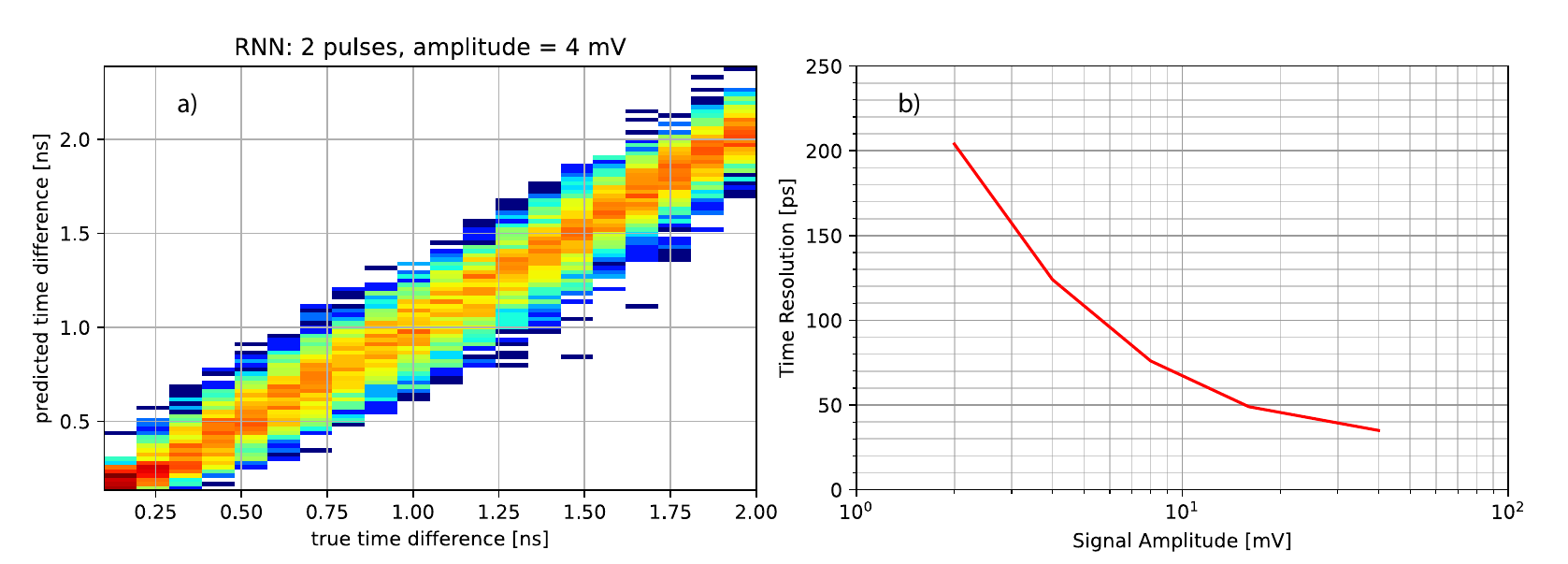}}
\vspace{-5mm}
\caption{\small 
Reconstructed by RNN difference in the arrival time of two signals with amplitude 4\,mV:  (a) predicted {\it vs.} true time separation and (b) the timing resolution for signals between 2 and 40\,mV (1-20 photons equivalent) and a noise level of 0.78\,mV.
}
\label{fig:RNNSummaryPix}
\end{center}
\end{figure}

Long Short-Term Memory (LSTM)~\cite{LSTM} is the particular type of RNN tested here. This particular architecture of the RNN is chosen due to its ability to store information for an arbitrary duration, {\it i.e.} to recognize and memorize characteristic pulse shapes with any length and to be noise-tolerant. 340 time samples were presented to the input of the RNN. Computation in the RNN was performed by a single bi-directional LSTM layer, and the single output was trained to reconstruct the time separation between the two photons. The reconstructed time difference by the RNN as a function of the true time separation is shown in Figure~\ref{fig:RNNSummaryPix}(a) for a signal amplitude of 4\,mV (2 photons equivalent). 

The precision of time reconstruction depends on the signal-to-noise ratio, which changes with the signal amplitude. Figure~\ref{fig:RNNSummaryPix}(b) shows the time resolution for signals between 2 and 40\,mV (1-20 photons equivalent) and a noise level of 0.78\,mV. The RMS is 35\,ps for signals with amplitude 40\,mV, which corresponds to about 2\,cm positional precision in a calorimeter with fibers having a refraction index of 1.50.  This kind of time reconstruction will effectively highly segment a fiber calorimeter in 3D and open up new capabilities, as discussed here.

\subsubsection{Neural Networks for Energy Construction}
\label{sec:NN}

\begin{figure}[ht]
\centering
\includegraphics[width=0.355\textwidth]{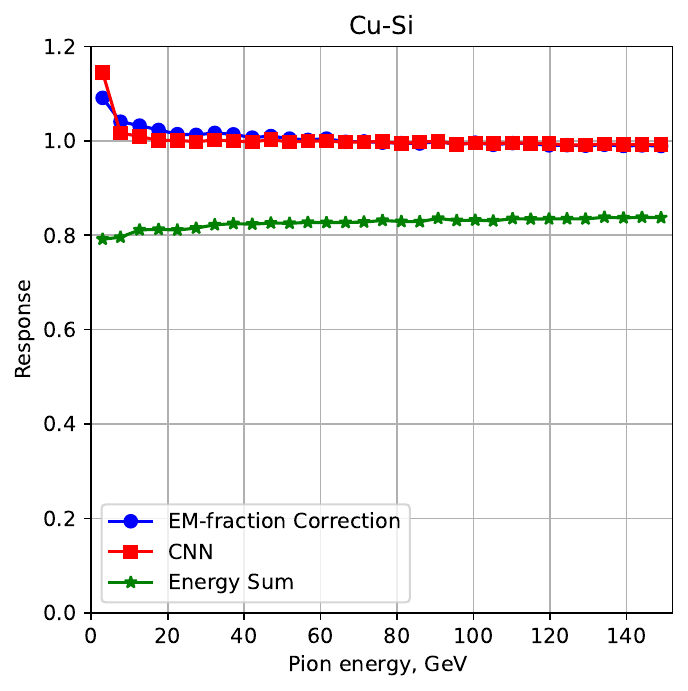}
\includegraphics[width=0.355\textwidth]{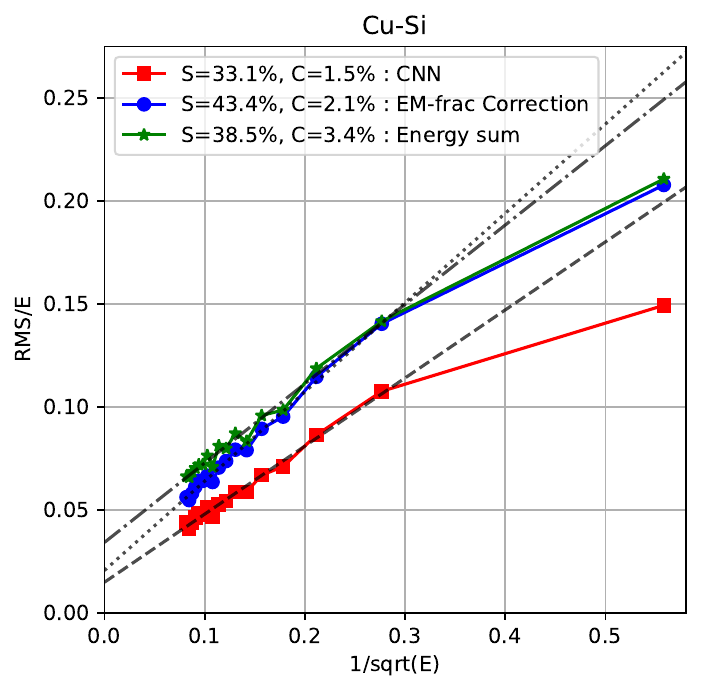}
\caption{The calorimeter response (left) and energy resolution (right) for charged pions are shown. The simple sum over all channels (green), with the $f_{\rm em}$ correction (red) and the CNN regression (blue) show respective energy measurement performance.  The $f_{\rm em}$ correction effectively employs the traditional dual-readout approach~\cite{AKCHURIN2005537}. The energy resolution parameters representing stochastic (S) and constant (C) effects estimated by a linear fit are included in the legend. 
}
\label{fig:CNN_Cu_pions}
\end{figure}

We contrasted ~\cite{Akchurin_2021} the performance of deep neural networks -- Convolutional Neural Network (CNN) and Graph Neural Network (GNN) -- to current state-of-the-art energy regression methods in a finely 3D-segmented calorimeter simulated by GEANT4.  This comparative benchmark gives us some insight about the particular latent signals neural network methods exploit to achieve superior resolution.  A CNN trained solely on a pure sample of pions achieved substantial improvements in the energy resolution, for both single pions and jets, over the conventional approaches and maintained good performance for electron and photon reconstruction.  We also used the GNN with edge convolution to assess the importance of timing information in the shower development for improved energy reconstruction.  In our investigation of energy reconstruction using CNN/GNN techniques, we see promising improvements in resolution beyond what has been demonstrated by dual-, and the proposed triple-, readout calorimeters~\cite{GROOM2007633}.  These investigations represent a stepping stone in the development of state-of-the-art calorimeters for future experiments.

The 3D CNN is trained on a GEANT4-simulated data set with 0.5 to 150\,GeV charged pions. The energy reconstruction performance is then tested on an independent sample in the same energy range.  Figure~\ref{fig:CNN_Cu_pions} shows the reconstruction performance of the CNN compared against the simple energy sum and dual-readout approach.  We have included parametric energy resolution fits of the form $\frac{a}{\sqrt{E}} \oplus b$, where $a$ and $b$ represent the stochastic and constant terms to the comparison plots.  One can see that the CNN outperforms both alternative reconstruction methodologies. Here, the $f_{\rm em}$ correction method represents a dual-readout approach where $f_{\rm em}$ is computed from the energy deposited by the electrons in the shower. 

In the case of showers initiated by photons or electrons, we see that the CNN performance closely resembles, with minimal degradation, the performance of the simple energy sum.  We anticipated this similar performance as the CNN operates as an energy correction and does not find traces from charged hadrons. Thus, the correction to the raw calorimeter response is negligible. As a result, a CNN trained with pion data sets can reconstruct electron/photon energy without introducing a strong, undesired bias.

\begin{figure}[ht]
\centering
\includegraphics[width=0.355\textwidth]{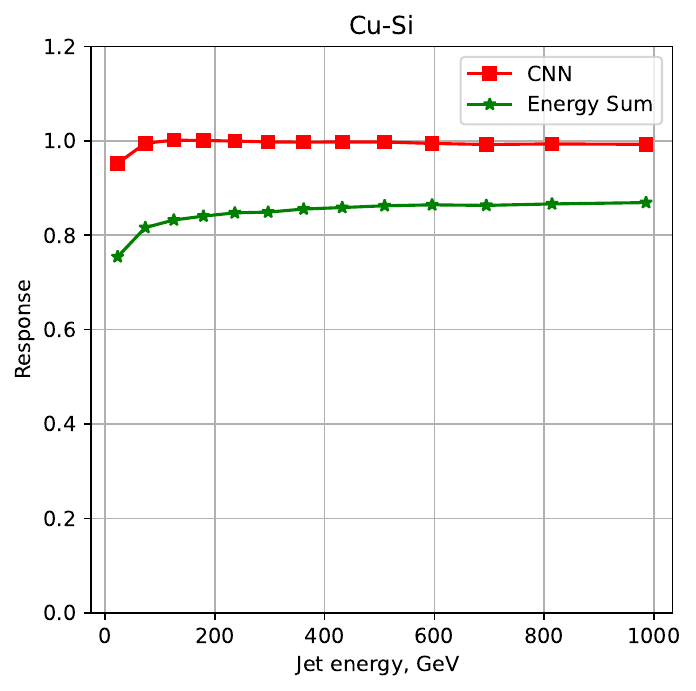}
\includegraphics[width=0.355\textwidth]{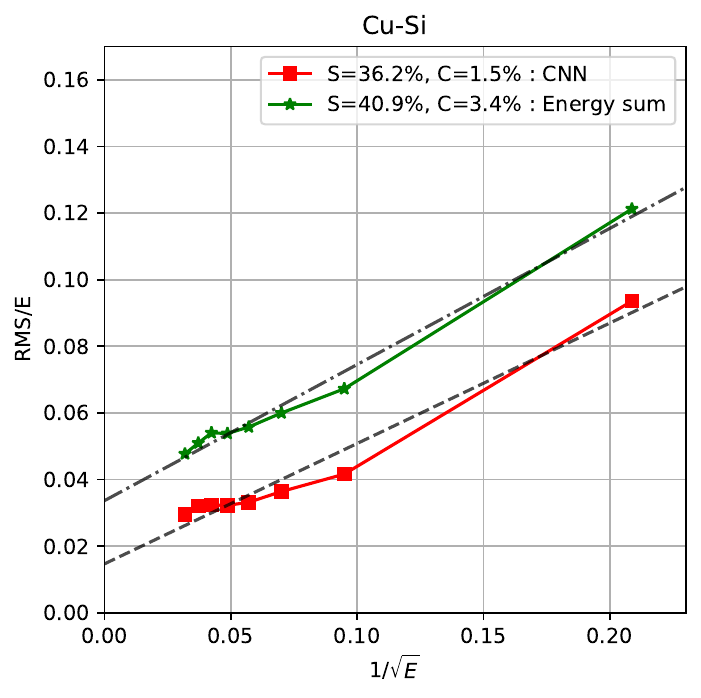}
\caption{The response (left) and energy resolution (right) for jets: the sum over all channels (green) and CNN regression (red).
}
\label{fig:CNN_Cu_jets}
\end{figure}

We further examined the CNN reconstruction performance on jets by using PYTHIA8~\cite{Sjstrand:191756} to simulate $u$-quark jets with energy from 20\,GeV to 1\,TeV. The response linearity and energy resolution are shown in Figure~\ref{fig:CNN_Cu_jets}. The energy scale is preserved without the need for additional corrections. The energy resolution is also significantly improved when compared to the more traditional reconstruction techniques. 

\begin{figure}[ht]
\centering
\includegraphics[width=0.355\textwidth]{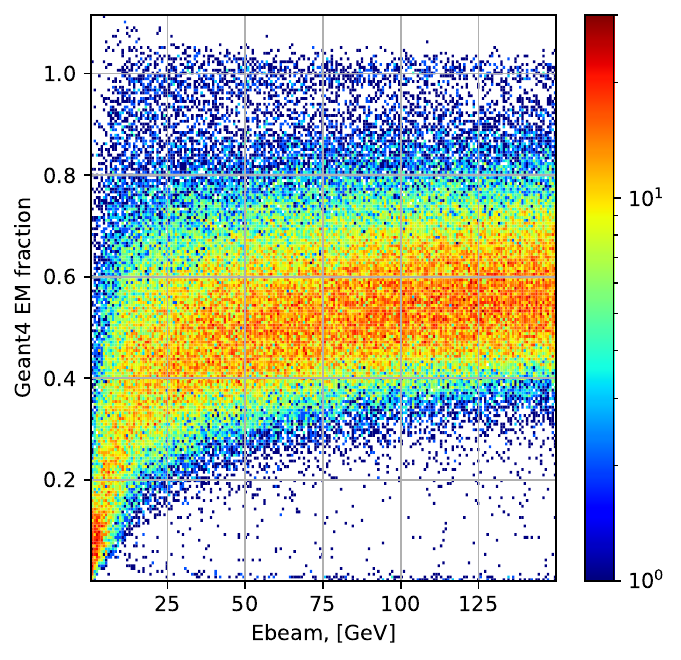}
\includegraphics[width=0.355\textwidth]{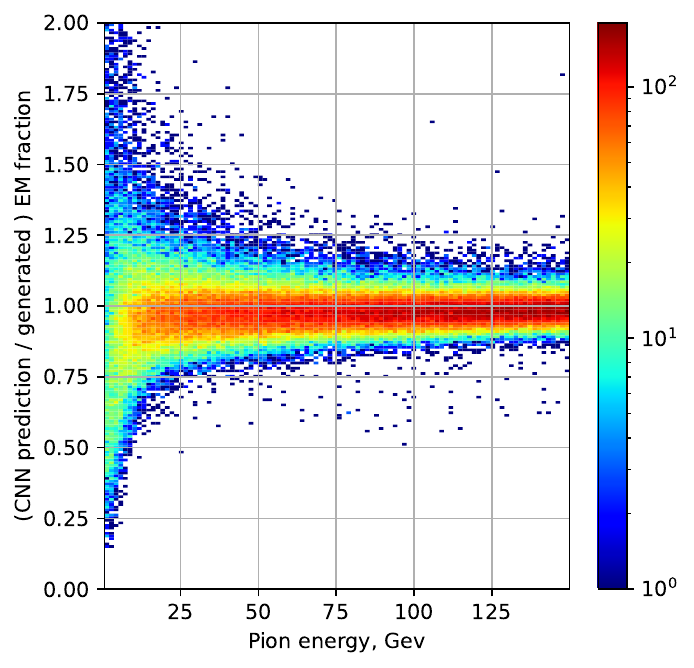}
\caption{The electromagnetic fraction in the energy deposit by pions (left) and the predicted EM fraction by the CNN normalized to the true value (right). 
}
\label{fig:CNN_Cu_pions_EMfrac}
\end{figure}

The large fluctuations in $f_{\rm em}$ in non-compensating calorimeters is the leading source for performance degradation in energy reconstruction. Dual-readout calorimeters are designed to infer $f_{\rm em}$ on an event-by-event basis by using signals from scintillation and \v{C}erenkov light. We have tested an alternative approach for $f_{\rm em}$ reconstruction in a single-readout calorimeter using a CNN that leverages topological information in the shower development. The CNN is trained on simulated charged pions of 0.5--150\,GeV to reconstruct $f_{\rm em}$ and $f_{\rm had}$.  Figure~\ref{fig:CNN_Cu_pions_EMfrac} shows the simulated $f_{\rm em}$ (left) and the ratio of the reconstructed to simulated $f_{\rm em}$ (right) over the range of particle energies.  One can deduce the viability of reconstructing $f_{\rm em}$ in a single-readout highly granular calorimeter from the result illustrated in this figure.

\begin{figure}[ht]
\centering
\includegraphics[width=0.355\textwidth]{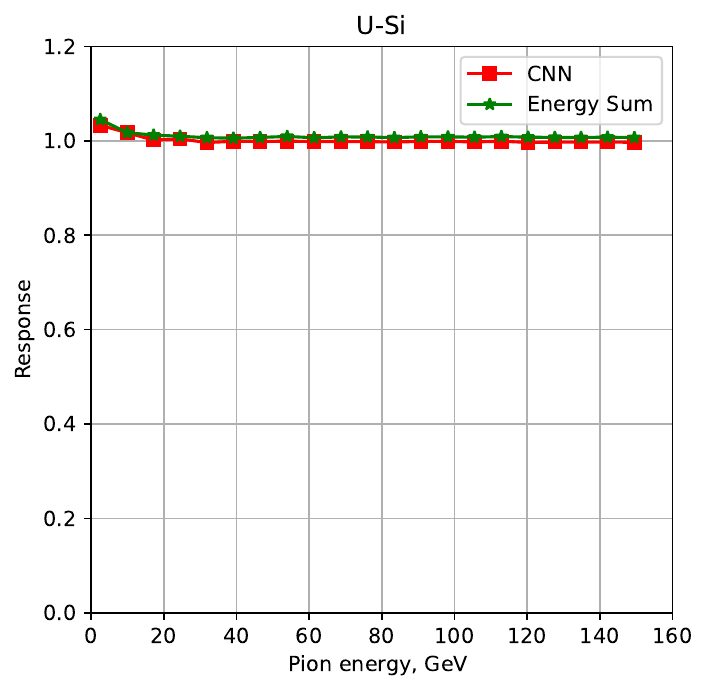}
\includegraphics[width=0.355\textwidth]{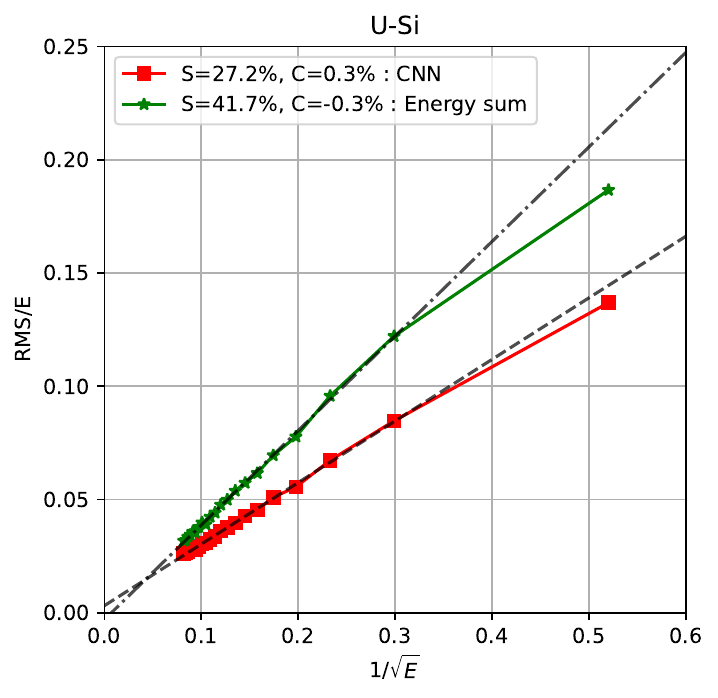}
\caption{The response (left) and energy resolution (right) for pions: the sum over all channels (green) CNN regression (red). 
}
\label{fig:CNN_U_pions}
\end{figure}

In order to gain further insight in the efficacy of the networks described above compared to the traditional dual-read technique, we also studied the energy reconstruction using a compensating ($e/h = 1$) calorimeter where the $f_{\rm em}$ fluctuations are no longer the leading cause for degraded performance.  The simulated U-Si calorimeter shows linear response to pions using the simple sum for energy reconstruction.  The improved reconstruction performance of the CNN over the simple energy sum, as shown in Figure~\ref{fig:CNN_U_pions}, indicates that the CNN exploits the relationship between the invisible energy and the visible signal in the shower, {\it i.e.} the multiplicity and production angle of the secondaries from the hadron interactions (see~\cite{Akchurin_2021} for details).
 
\subsubsection{Structured Fibers for Calorimetry}
\label{sec:newfibers}

\begin{figure}[ht]
  \centering
  \includegraphics[width=0.70\textwidth]{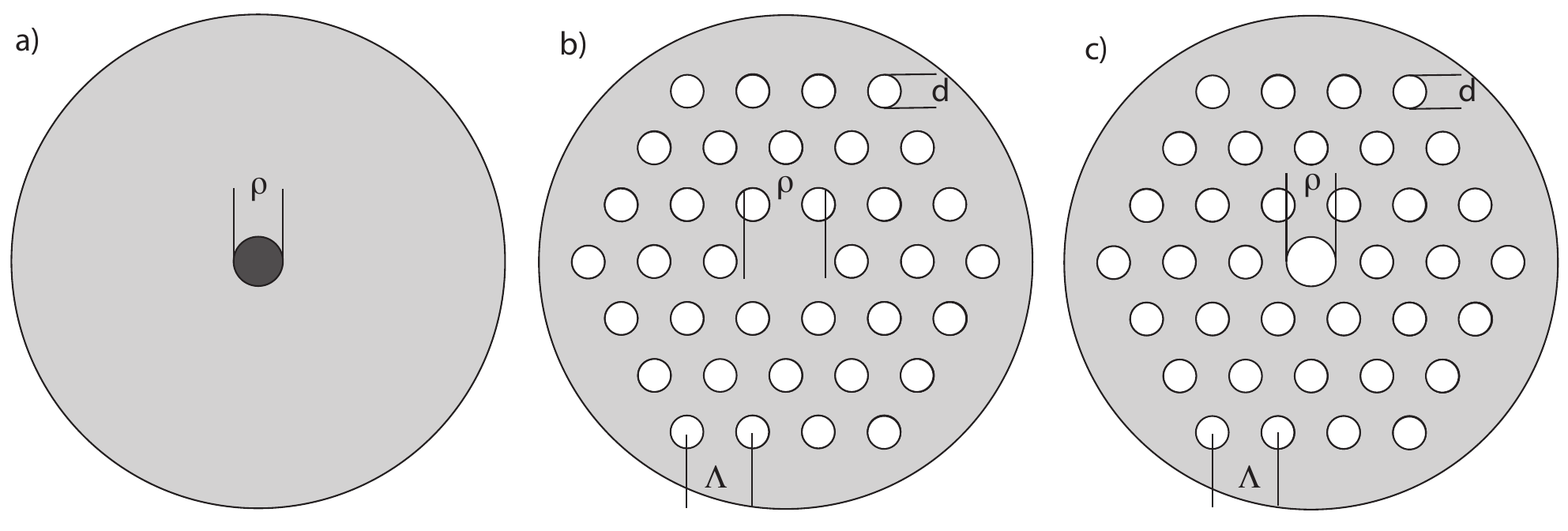}
  \vspace{-0.2cm}
  \caption{\small The cross sections of a) a traditional fiber, b) PCF with a solid core, and c) PCF with a hollow core.  White regions represent air, gray represents silica, and the dark circle represents a higher refractive index core in a traditional fiber a).
  }
  \label{fig:fibertypes}
  \vspace{-0.2cm}
\end{figure}

Traditional optical fibers consist of two coaxial glass cylinders of differing refracting indices.  If the refractive index of the core is higher than that of the outer cylinder, the total internal reflection at the core-cladding surface guides the light along the fiber. The refractive index difference is generally quite small, $\sim 0.1\%$.  Alternatives to total internal reflection to control light flow by embedding micro-structures were suggested in the 1970s, but not until the  1990s were some fibers fabricated and experimentally studied.  Out of this activity, the so-called Photonic Crystal Fibers (PCFs) emerged that confine and guide light through a photonic band-gap effect, which is conceptually similar to how electrons behave in semiconductors.  The PCF's geometry is defined by the micro-structured air hole cladding surrounding the core, which is either solid or hollow (Figure~\ref{fig:fibertypes}).  Depending on geometry, true photonic band-gap guidance occurs only in a hollow core PCF ~\cite{doi:10.1080/09500340.2010.543706}, while in the case of a solid core PCF, the effective refractive index of the central region is higher than the surrounding air hole region (``micro-structured air-cladding'') and the light guidance takes place through modified total internal reflection.  Although the solid core PCFs seem similar to traditional fibers, the additional degrees of freedom offered by the three parameters ($\rho, d$, and $\Lambda$ in Figure~\ref{fig:fibertypes}) open up possibilities that are foreign to traditional fibers.  In what follows, we describe a few ideas and some preliminary studies that are relevant for the next generation of fiber calorimeters.

\begin{figure}[ht]
\centering
       \includegraphics[width=0.7\textwidth]{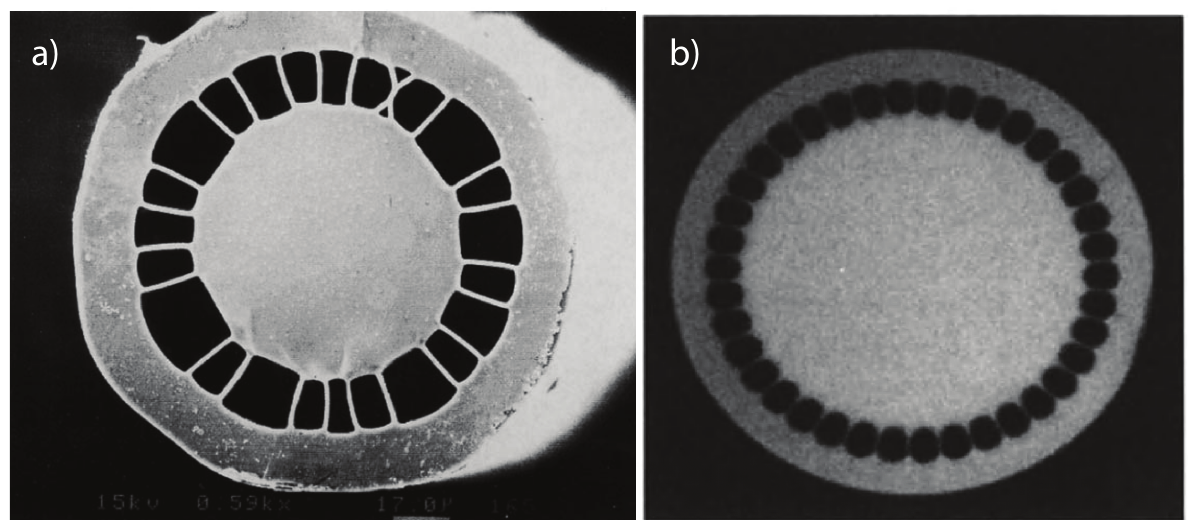}
       \vspace{-0.2cm}
    \caption{\small  Effective air-cladding is accomplished by drawing capillary tubes around a solid core.  The fiber in b) is 150 $\mu$m in diameter with $NA >$0.55 \cite{Eggleton:00,Bjarklev}.
    }
\label{fig:airclad}
\vspace{-0.2cm}
\end{figure}

The limiting factor in (EM) energy resolution in dual-readout calorimetry is the \v{C}erenkov light yield.  Using fused-silica fibers, the original DREAM prototype yielded 8 p.e./GeV, contributing $35\%/\sqrt{E}$ in fluctuations.  A later prototype, using clear plastic fibers, improved the overall \v{C}erenkov light yield to 33 p.e./GeV.  Using super bialkali photocathode PMTs and increased fiber density certainly helped, but the major contribution came from the increased numerical aperture ($NA$) form 0.33 to 0.5 \cite{AKCHURIN2014130}.  A two-fold increase in $NA$  means a four-fold enhancement in light yield.  
     Figure~\ref{fig:airclad} shows two fiber cross-sections from Refs.~\cite{Eggleton:00,Bjarklev}.  The air/silica interface produces a large index difference and makes for extremely high $NA$ values. $NA$s as high as 0.9 have been demonstrated \cite{Bjarklev}. High $NA$ fibers are strongly multi-mode and will capture \v{C}erenkov light very efficiently from a broad range of angles between the fiber axis and charged particle direction; they will also distribute light at the far end of the fiber, a desired feature when SiPMs were used to minimize pixel saturation.  If successful, creating air-clad silica fibers will be a major step towards eliminating the dominant limiting factor in the \v{C}erenkov sector.

\begin{figure}[ht]
\begin{center}
       \includegraphics[width=0.90\textwidth]{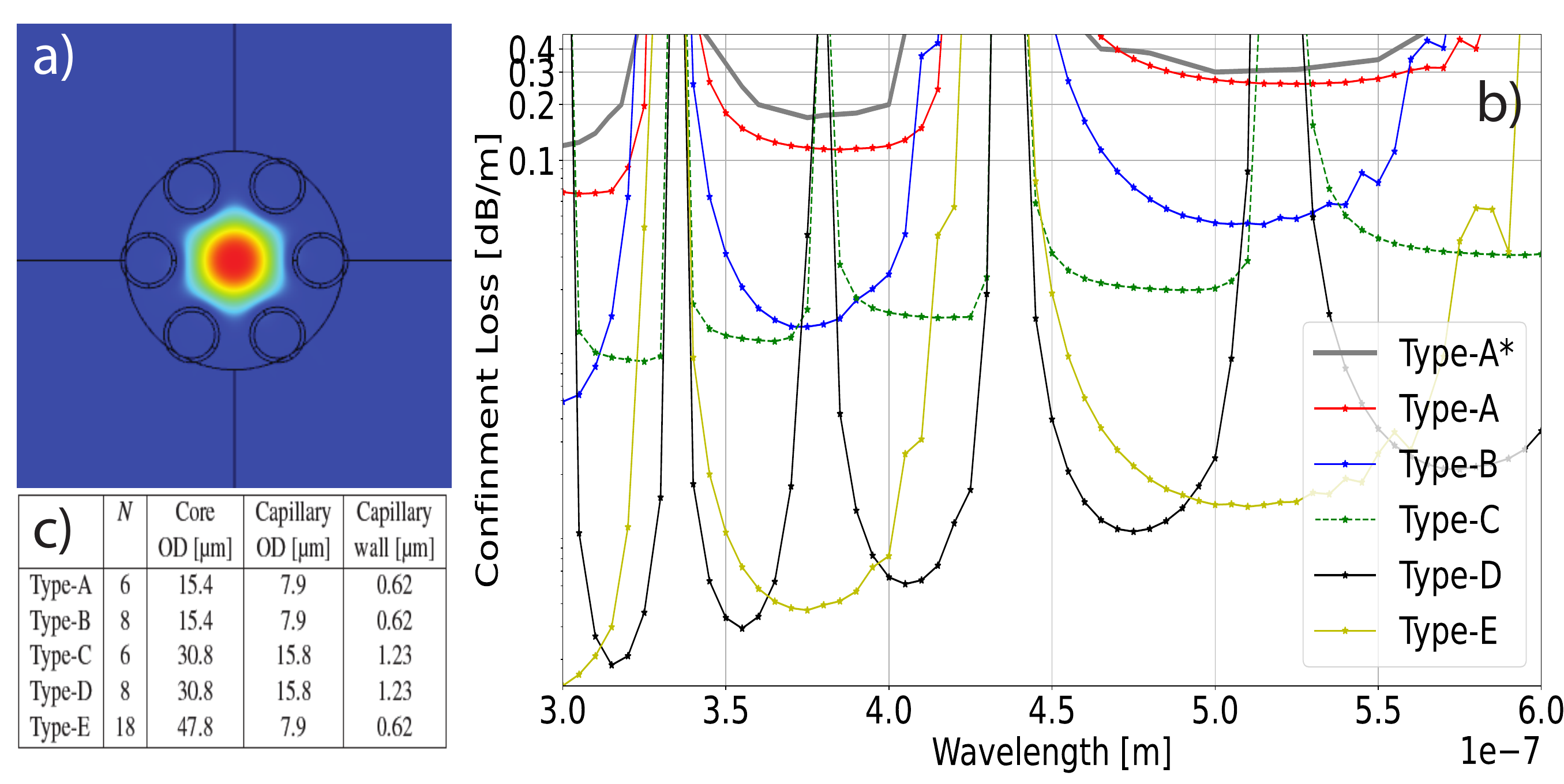}
    \caption{\small The electric field distribution a) in the hollow core of Type-A PCF is well-contained within the central core, b) if the number ($N$), core size, as well as the size and wall-thickness of capillaries are varied, the attenuation properties of the fibers in the wavelength region of interest can be tuned.  For example, eight {\it vs} six capillaries improve the attenuation (Type-B {\it vs} -A), scaling up the dimensions by a factor 2 from Type-A to -C bifurcates the attenuation curves.  Our calculations are in good agreement with published experimental data, indicated by a gray curve, for the Type-A fiber~\cite{HollowCoreNCF}, which gives confidence in the fidelity of computations.
    }
\label{fig:adilfiber}
\end{center}
\vspace{-0.4cm}
\end{figure}

Both types of PCFs lend themselves to exploration for calorimetry. For example, a solid-core PCF might be doped, and a hollow-core PCF might be filled with gas or liquid for scintillation.  The parameter phase space is large.  Using COMSOL, five hollow-core/hollow-capillary configurations (indicated as Type-A to -E in Figure~\ref{fig:adilfiber}) were studied for feasibility. These so-called negative-curvature fibers are somewhat different from the hollow-core PCFs because the curvature of the normal surface in the region around the central hollow-core is negative with respect to the radial unit vector, allowing anti-resonant reflecting in the fiber, and potentially offering additional design flexibility~\cite{219574}.  One of our objectives was to test different configurations for relevant quantities for calorimetry, such as low attenuation in UV and visible wavelengths, wavelength filtering for possible \v{C}erenkov/scintillation discrimination, $NA$, power distribution in lower order modes, and polarization.  Figure~\ref{fig:adilfiber} gives a snapshot.

A 2013 Snowmass whitepaper \cite{winslow2013applications} suggested quantum-dots (QD) or semiconductor nanoparticles for use in particle physics applications, but they have not been widely embraced. 
When semiconductor particles are a few nanometers in size, they behave like individual atoms and demonstrate intriguing optical properties due to quantum confinement.  Much like wavelength shifters, they absorb light at shorter than characteristic wavelengths and re-emit in a narrow band around these wavelengths depending on nanoparticle size: the larger the dot, the longer the emission wavelength. So, quantum-dots are tunable wavelength shifters and good candidates for scintillation- or light-based detector applications ({\it e.g.} \cite{Aberle_2013} for liquid scintillators). 
There is a growing list of quantum-dots. For example,  CdSe (2-8\,nm) emits at 450-650\,nm, and CdS can go down to 380\,nm.  Typically, quantum-dots are suspended in a solvent, but they can also be embedded in a solid or gel or deposited on thin films.     

In the case of traditional fibers, there have been several attempts to fabricate QD-doped glasses, such as co-melting, sol-gel, ion implantation, atomic layer deposition, and so on.  The conventional fiber drawing process 
was not initially successful as QDs clustered due to ion migration at high temperatures to form larger objects than their exciton Bohr radius \cite{10.3389/fmats.2015.00013}.  The melt-extraction technique is better suited but the fiber drawing process at the drawing tower must be modified to include application of cladding and buffer layers in practical fiber production.  Successful fabrication of luminescent traditional fibers  
using a double-crucible approach~\cite{Dong:15},  QD-doped hollow-core PCF ~\cite{Wang_2017} and QD-doped negative-curvature PCF~\cite{nano9060868}, and many variations on the main theme have been reported in the scientific literature.  

\subsubsection{Cherenkov Polarization in Calorimetry}
\label{sec:Cerenkov}

\label{sec:polarization}
\begin{figure}[ht] 
\begin{center}
{\includegraphics[width=0.95\textwidth]{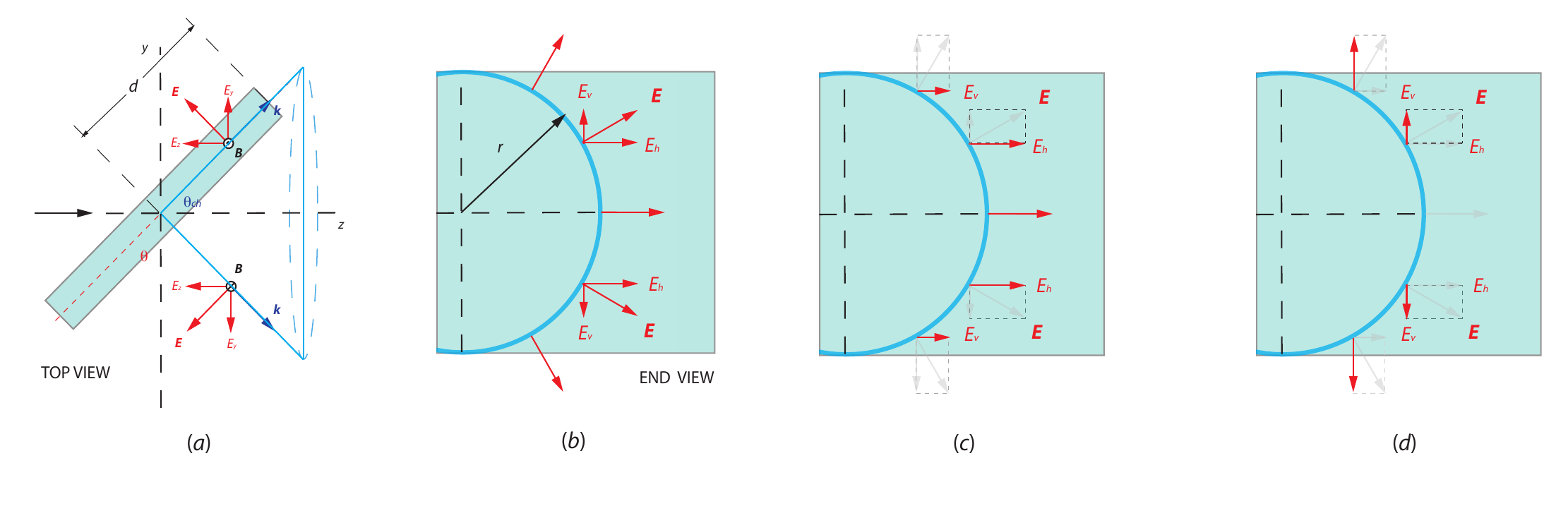}}
\vspace{-8mm}
\caption{\small When viewed from the top (a), a cone section (\v{C}erenkov cone) is developed in a crystal as a charged particle traverses it, indicated by a black arrow on the left.  The end view (b) at the downstream (\v{C}erenkov) sensor end presents an arc or a piece of the \v{C}erenkov cone.  The polarization directions are shown on this exaggerated projection of the cone onto the block end.  The favorable direction is defined when the horizontal components of the electric field vectors $E_h$ are parallel and transmitted through the polarizer (c).  The polarizer is in an unfavorable direction when it is oriented such that the vertical components of the electric field vectors $E_v$ are antiparallel and tend to add to zero (d).}
\label{fig:CherenkovConeSideView}
\end{center}
\end{figure}

An interesting question is if and to what extent polarization of \v{C}erenkov light is maintained in the development of showers.  If it is, can it be used to improve calorimeter performance, especially in separating \v{C}erenkov from scintillation light?  To investigate these questions, the DREAM/RD52 collaboration used a single BSO crystal to analyze the polarization properties of \v{C}erenkov light \cite{AKCHURIN2012333,akchurin2011polarization}.  The BSO crystal was positioned such that the \v{C}erenkov light was directed towards the (\v{C}erenkov) PMT  for a through-going particle ($\theta_{\rm Ch}=61^{\rm o}$).  Figure~\ref{fig:CherenkovConeSideView} shows the relevant quantities for this measurement.

\begin{figure}[hbtp] 
\begin{center}
\resizebox{\textwidth}{!}{\includegraphics{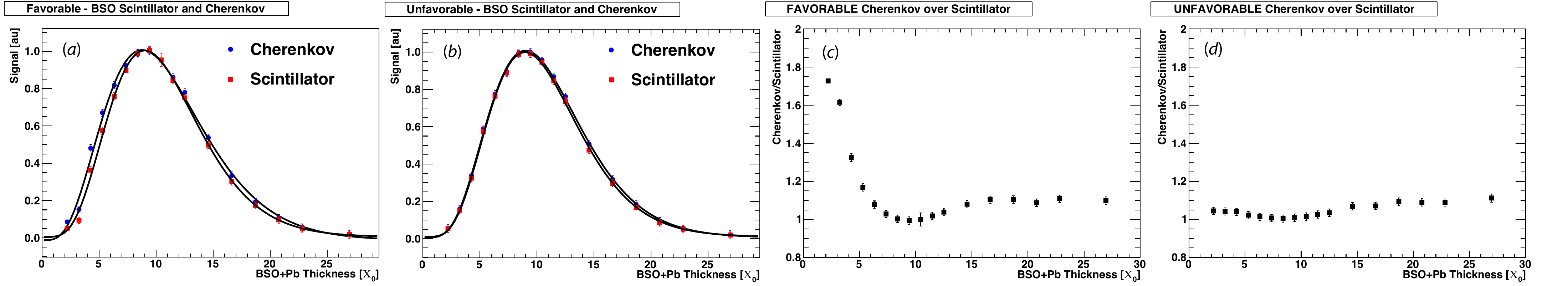}}
\caption{\small The longitudinal shower profiles using 80\,GeV electrons are measured using (a) favorable and (b) unfavorable polarizer orientations at the \v{C}erenkov end.  The solid lines are fit results corresponding to
$t^a e^{-bt}$ parametrization.  The bottom two plots show the $C/S$ ratios for these two cases in which the enhanced \v{C}erenkov signal is clearly visible in (c).  
}
\label{fig:PbProfiles}
\end{center}
\end{figure}

In order to measure the longitudinal polarization profile ($d{\bf P}/dx$) of an electromagnetic shower, we stacked thin lead sheets upstream of the crystal and, using an 80\,GeV electron beam, we measured the \v{C}erenkov signal with favorable and unfavorable polarizer orientations.  Figure~\ref{fig:PbProfiles} shows these normalized profiles such that at the shower maximum, the signals are set to unity.  As the shower develops in the calorimeter, the direction of the secondary particles becomes increasingly random.  Before the shower maximum, the number of secondary shower particles is small, and their directions are strongly aligned with that of the incoming particle's direction.  Therefore, as Figure~\ref{fig:PbProfiles} shows, the \v{C}erenkov polarization direction tends to be maintained.  Once the shower has fully developed, there is no longer a preferred momentum direction among the shower particles, and the polarization averages to zero.

Figure~\ref{fig:PbProfiles} summarizes the results.  It is clear that in the case of favorable polarizer orientation, the \v{C}erenkov signal
appears earlier in depth compared to the scintillation light, Fig.~\ref{fig:PbProfiles}(a).  In the case of the unfavorable polarizer orientation, there is no difference between the \v{C}erenkov and scintillation light profiles because the \v{C}erenkov polarization is suppressed by the polarizer and only randomly polarized \v{C}erenkov light results in a measurable signal, Fig.~\ref{fig:PbProfiles}(b).
Figures~\ref{fig:PbProfiles}(c) and (d) further illustrate the point where the ratio of the $C$ signal to $S$ signal is shown versus depth.  In the favorable case, Fig.~\ref{fig:PbProfiles}(c), $C/S >1$ for $t\lesssim t_{\rm max}$, whereas $C/S \sim 1$ for all $t$ in the unfavorable case of Fig.~\ref{fig:PbProfiles}(d).  There is no significant difference in the exponential tails between the favorable, unfavorable, and/or scintillation cases.  

\v{C}erenkov light polarization adds another unique and discriminating feature that is still to be exploited in the field of calorimetry.  It may be possible to improve energy and direction measurements of high energy particles as we further explore \v{C}erenkov radiation.  Although the timing and spectral characteristics of the \v{C}erenkov light lend themselves to easier utilization for discrimination against scintillation light in the context of multi-readout calorimetry, the full potential of the polarization, or the degree to which the \v{C}erenkov polarization is a discriminant, has yet to be determined.   It may also be possible to ``tag'' the early part of the electromagnetic showers in the front part of a segmented crystal ECAL.  Similarly,  the electromagnetic core of an hadronic shower's (``$\pi^{0}$ component'') may be identified through the polarization information.   In this context, exploration of polarization maintaining fibers ({\it e.g.} PANDA) \cite{1074847,Penninckx:06} and polarization-sensitive single-photon counters \cite{singlephotonpolarization} seems worthwhile.

\subsection{Crystal dual-readout prototypes}
While the simulation results on dual readout in crystals with SiPMs looks promising, and crystals in conjuntion with a fiber HCAL looks even more promising, it is important this be tested via prototypes in conjunction with the IDEA collaboration.  The CalVision collaboration hopes to obtain funding for this purpose. 

\subsubsection{Improvements in dual-readout front-end electronics}
\label{sec:readout}
The challenges of multi-signal extraction with the front-end electronics for dual-readout calorimetry are similar to those recently faced with precision timing detectors.  As with precision timing systems, the sampling rate increases as the pulse waveforms contain important time-domain discriminating information for separating scintillation and \v{C}erenkov components.  The combination of fine-granularity calorimetry, dual readout and precision timing create a flood of raw data at the front-end readout.  However, the dimensionality of the final calorimeter information is much lower than the corresponding raw data.  This situation is ideal for fast, on-detector processing to make major advancements where both digital and analog approaches are relevant to optimize for lower cost, lower power consumption and optimal quality of the extracted information content.

A waveform digitizer with real-time analysis can exploit the distinct signal timing and shape of the \v{C}erenkov and scintillation light to provide a complementary method of C/S signal separation in a dual-readout calorimeter.  Nalu Scientific in collaboration with the University of Hawaii are developing several SoC ASICs that have the potential to meet the needs of the dual-readout calorimetry.  Of particular interest are the AADVARC3, ASoC V3 (an analog-to-digital converter System on a Chip) multi-channel waveform digitizer, partially funded by a DOE SBIR grant~\cite{SBIRASoC,Mostafanezhad, Varner}, and the Phase 2 SBIR development of the HDSoC (High-Density version). Recent measurements have shown that the ASoC V3 meets the requirements for HEP experiments in terms of compactness, low power, high timing resolution, deadtimeless operation, and robustness against pileup~\cite{Mostafanezhad}. SoC ASICs with better rate handling and/or timing and/or channel multiplicity are under development at Nalu, working in partnership with the University of Hawaii and several national labs~\cite{Nalu_CERNSoCWorkshop}.

Analog processing with Field-Programmable Analog Arrays (FPAA) System-on-Chip platforms~\cite{hasler2020} applied to the extraction of the C/S ratio has the potential to greatly reduce processing latency, lower the power consumption, and achieve higher performance than traditional waveform digitizers with digital processing.  As with the fast digitizers, the FPAA would start with the analog values stored from pulse stretching circuits, the most known are the DRS4 developed at PSI and the PSEC4 developed at the University of Chicago.  However, unlike the waveform digitizers, the analog pulse samples would be processed by the FPAA.

Use of FPAAs in the front-end is a novel and important next step for calorimetry.  The strategy of using an FPAA is to provide programmable access to analog pulse samples covering the high C/S regions at the start of the pulse and low C/S regions later in the pulse.  Similarly, a pair of SiPMs, one with dedicated wavelength filtering, would provide an even greater set of C/S information.
A two-layer Artificial Neutral Network (ANN) implemented with an FPAA would be trained with real signals fed into the circuit inputs.  The exact choice of which time samples to feed into the ANN may depend on the crystal, SiPM, filters and front-end circuits.  The weights would be optimized to produce optimal C/S separation with a large C/S ratio producing a large output from the ANN and a low C/S ratio producing a low value.  The output of the ANN would be digitized and recorded with the channel.
The reduction of information would amount to over an order of magnitude from an equivalent of 10 time samples to one ANN output. The analog latency of the circuit is much lower than the clocking circuitry of the waveform digitizer ADCs with real-time analysis.  The power consumption is also expected to be lower with the ANN.  As the need is for an output that provides information on the correction to the S-only energy estimate with a C/S ratio, the precision of the C/S value is not as stringent as the S-only digitization, which again emphasizes that the problem of raw data volume at the front-end can be effectively addressed with dimensional reduction of the information content.\\

Another field of research is related with the development of fully digital CMOS sensors (digital SiPMs), that could provide an even more appealing and innovative solution. Several implementations have been proposed in the last 10 years and the interesting one, in this context, would be based on an array of Single-Photon Avalanche Diodes (SPADs), with pixel-level signal digitisation and on-chip counting and timing functions. High dynamic range and a PDE enhanced in the blue region would also be required. On the other hand, for this R\&D, funds have not yet been secured.

\subsubsection{Beyond just scintillation and Cherenkov: multi-readout}
\label{sec:moreInfo}

\begin{figure}[hbtp]
\centering
\resizebox{0.48\textwidth}{!}{\includegraphics{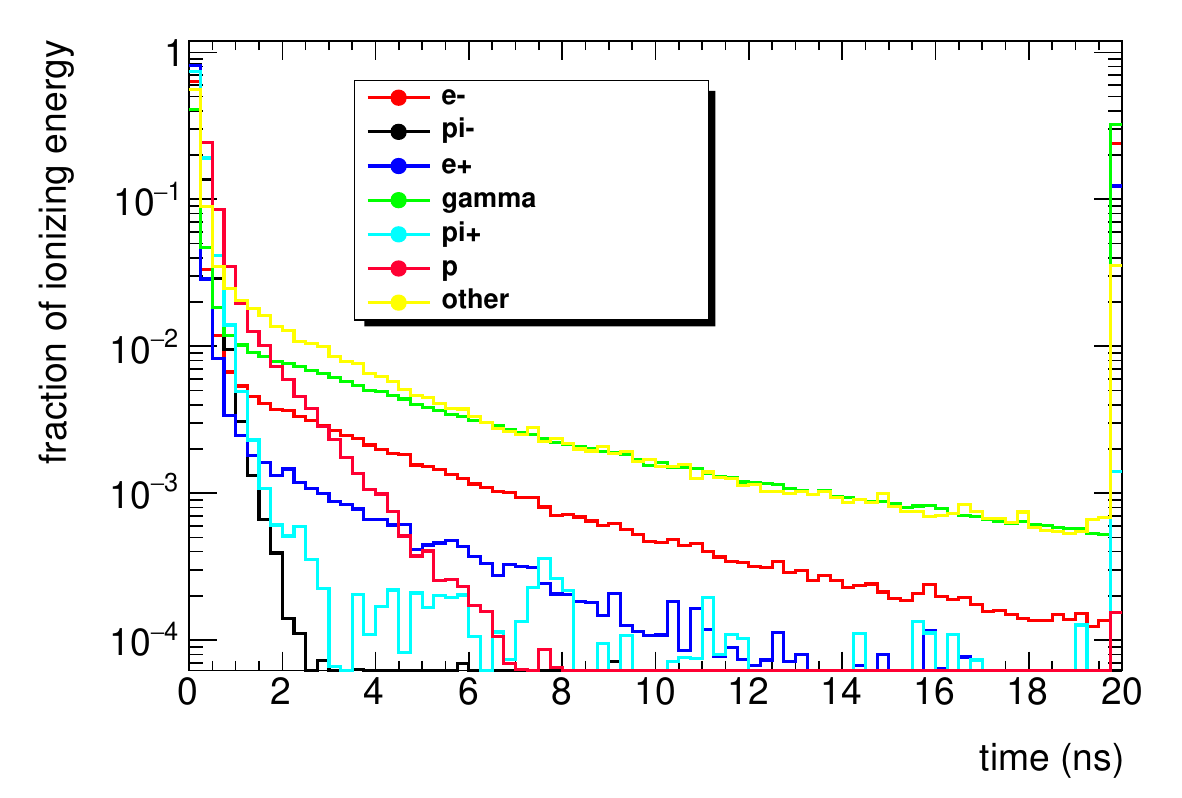}}
\resizebox{0.48\textwidth}{!}{\includegraphics{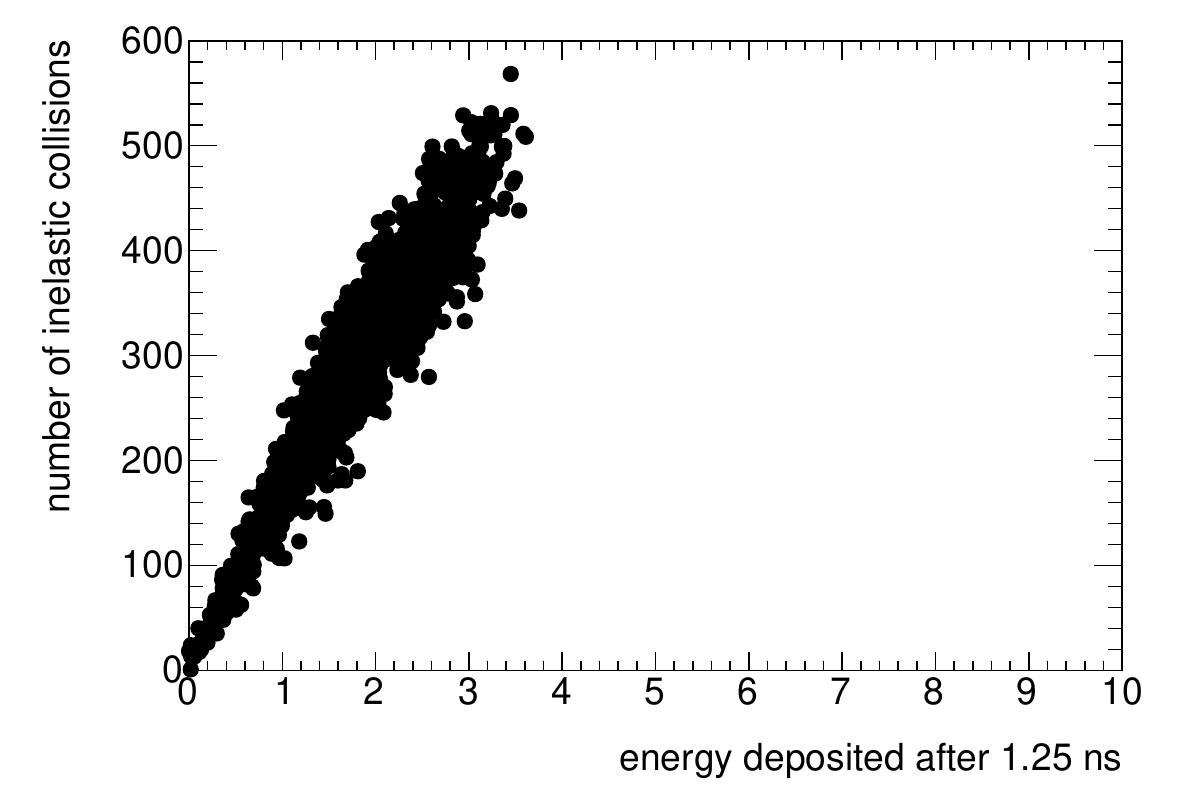}}
\caption{ 
[left] Time of energy deposition after subtraction of time of flight at the speed of light
to deposition position for various shower particle types.
[right] Correlation between energy deposited after 1.25\,ns and the number of inelastic interactions in a GEANT4-simulated 10~\GeV  pion shower in a large block of PbWO4.  
}\label{fig:timingcor}
\end{figure}

As shown in Fig.~\ref{fig:beta}, the energy deposited by slow particles (below the \v{C}erenkov threshold speed) is correlated with the
total number of inelastic collisions in the shower. However, there are proxies that could
potentially have a better correlation.
As shown in Fig.~\ref{fig:timingcor} [left], the energy deposits from protons
produced in inelastic collisions tends to occur at late times.
Comparing Fig.~\ref{fig:timingcor}
with Fig.~\ref{fig:beta} shows that timing measurements have the potential
to be a better proxy for the missing energy than the particle velocity measurement.
The challenge is the optimization of the timing window.  At short times, the separation will be diluted due to the scintillator decay time
and the light propogation time within the crystal.  
For windows at longer times,
the energy deposits are small and start to lose the tight correlation.
Simulations are needed to understand the best way to use this information.
For other studies on the use of timing as a dual-readout proxy, see Refs.~\cite{AkchurinNN,7454834}.

\subsubsection{Novel materials and homogeneous dual-readout hadron calorimetry}
\label{sec:bluesky}
In the long term, the development of precision calorimetry is being pursued with two thrusts: development of cost-effective materials that allow a larger fraction of the calorimeter to be homogeneous instead of sampling, and development of novel wavelength shifters and light control mechanisms that improve {\it e.g.} the \v{C}erenkov light collection

\begin{figure}[htb]
    \centering
     \includegraphics[width=0.52\textwidth]{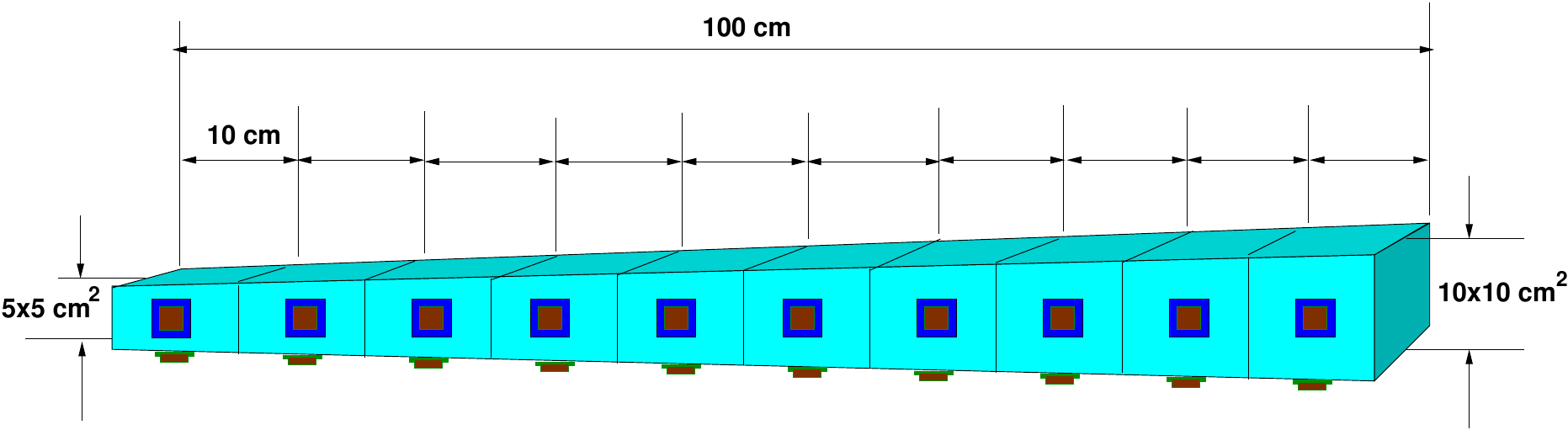}
    \caption{A typical HHCAL cell with pointing geometry~\protect{\cite{zhu2008hhcal}}.  }
    \label{fig:hhcal}
\end{figure}

A hadron calorimeter that was entirely made of highly segmented crystals using maximal information (HHCAL) could represent the ultimate in calorimetry. Figure~\ref{fig:hhcal} 
shows a potential HHCAL cell geometry~\cite{zhu2008hhcal}.
In this section, we discuss past research for a cost effective material and the potential resolutions enabled if it is found.
This goal has long been pursued in the US by physicists at FNAL, Caltech, Argonne, and Oak Ridge.
 Because of the unprecedented volume (70 to 100 m$^3$) required for an HHCAL, development of inorganic scintillators with a low unit cost of less than \$1/cm$^3$ is crucial for this detector concept~\cite{Mao2012HHCAL_IEEE}. 
 The material of choice must be dense and UV transparent (for effective collection of the \v{C}erenkov light) to allow for a clear discrimination between the \v{C}erenkov and scintillation light by wavelengths.
 The decay time should be long enough to 
 allow for a clear discrimination between \v{C}erenkov and scintillation with timing information, while short enough to minimize out-of-time pileup at colliders.
Current material costs are too high for practical calorimeters.  Two thrusts are being pursued.  The first uses Kyropoulos growth technology to lower the cost of growing crystals.  
This technique is low cost but it needs to be evolved to allow higher crystal densities and shorter radiation lengths.
The second is cerium doped DSB:Ce glasses  (a mixture of barium and silicon oxides)~\cite{Auffrayetal2015} and ceramics. The performance and suitability of large samples of these glasses and ceramics were studied by Novotny {\it  et al.}~\cite{Novotnyetal2017,Novotnyetal2019}, including large DSB:Ce blocks of $23 \times 23 \times 125\,$mm$^3$ loaded with up to 20\% of gadolinium oxide, with promising results. Heavily doped samples were shown to be sufficiently fast (a slow component with 400 ns decay time), and have a suitable radiation length (2.2\,cm), wavelength cutoff (318 nm), and scintillation emission wavelength (440 and 460\,nm). 
A recent DOE SBIR award was issued for the development of a dense scintillating glass with a mass production cost of less than \$2/cm$^3$~\cite{sbir2020}. Major development objectives 
include: large block dimensions, low levels of macro defects, high light output, fast decay time, radiation resistance, and overall performance comparable to \pwo at a much lower cost.

Another useful development area relates to a better handling of light.  Some possibilities are
 using quantum dot waveshifters, using interference filters, using waveform digitizers and using engineered diffraction filters.

\section{Executive Summary}

Future collider detectors are tasked with resolving the information content of Higgs boson events and a wealth of potential new physics processes at higher precision with the possibility to shed light on the nature of the fundamental interactions and on the origin of the dark matter.  Calorimeters have a central role in bringing together electromagnetic, charged and neutral hadron information with the central tracker and muon systems to form jets, isolated particles and missing energy. Calorimeter R\&D has the unique opportunity to maximize the information content through intrinsic multi-readout methods for improving energy resolution and response, and providing precision timing and particle identification.  The quality of the information provided to particle flow underpins its overall performance, especially in the presence of high-luminosity beam backgrounds.  The technique of hybrid dual-readout calorimetry, described here, delivers the highest intrinsic electromagnetic and hadron resolutions, precision timing, and matches the highest performance particle-flow calorimeters for jets.  Innovations through hybrid dual-readout R\&D will enable massive on-detector raw data reductions while achieving the highest quality information content per particle shower of any previous calorimeter.

Dual-readout calorimetry is a proven technique for improving calorimeter resolutions and yet its full potential remains to be explored.  As inelastic collisions produced in a hadronic shower are associated with a lower response due to energy lost to binding energies, neutrons migrating far from the shower, neutrinos produced in pion decays, and  other sources, the method uses proxies to estimate their number.  Since fluctuations in the average response due to fluctuations in the number of inelastic collisions in the shower dominates the hadronic resolution, it can be greatly improved using an energy scale correction based on this proxy.  In the classic method, the total energy of all ionizing particles is estimated via scintillation light, and the energy depositions of protons produced in inelastic collisions (often via neutron interactions) are estimated via the fraction of the shower particles with a velocity too low to produce \v{C}erenkov light.
Current state-of-the-art calorimeter resolutions are the result of pioneering work by the DREAM/RD52 and IDEA collaborations,

In the near-term future, a prototype spaghetti calorimeter, with scintillating fibers for scintillation light and clear plastic fibers for \v{C}erenkov light, is being constructed by physicists from Italy and S. Korea. Because of their efforts, if a circular Higgs factory is constructed, such a calorimeter is likely to be part of the detector at one of the interaction regions.
The US (especially the Texas Tech University) has long had an intellectual impact in spaghetti dual-readout calorimetry.  Refunding experimental work in this area would result in a leadership role for the US in this detector.

In addition, the US has also shown intellectual leadership in revitalizing an old but abandoned idea: using the dual-readout technique with a homogeneous crystal electromagnetic calorimeter with a backing spaghetti-type hadron calorimeter.  Simulation studies show that new photodetector technologies make this possible now, and could lead to the first calorimeters with both excellent resolution for photons/electrons and state-of-the-art jet resolutions.  Through funding in this area, the US could maintain our leadership through the construction of prototypes and advocacy for inclusion of a dual-readout electromagnetic calorimeter at a Higgs factory experiment.

US physicists are also creating new types of particle flow algorithms appropriate for this type of detector that make use of its excellent energy resolution.  Funding for this type of research will certainly improve their efficacy and allow US leadership.  

US physicists are also looking for inexpensive materials that could allow a homogeoneous hadron calorimeter, and for proxies beyond timing and \v{C}erenkov radiation to improve the estimate of the number of inelastic collisions.
US groups would like to investigate using the time distribution of the energy deposits as a proxy for the number of inelastic collisions.
In addition, such calorimeters can be improved via improved readout systems, dedicated ASICs for time and energy estimation, by light-weight mounting systems, and by sophisticated machine learning techniques.  However, funding is necessary to allow this work.

The world is coming to the consensus that the next highest priority for the field, after the HL-LHC, is a Higgs factory~\cite{EuropeanStrag}.  The US has an opportunity to play a leading role in the IDEA collaboration through work in dual-readout calorimetry if funding is given before it's too late.

\section{Acknowledgments}

We acknowledge support from the following funding agencies: DOE and NSF (USA).
This research was supported by the National Research Foundation of Korea (NRF) grants 2020R1A2C3013540, 2021K1A3A1A79097711, and 2018R1A6A1A06024970, and by INFN funding to the RD-FCC and HiDRa projects.
\bibliographystyle{elsarticle-num}
\bibliography{theBIB}

\end{document}